\documentclass[pteplogo]{ptephy_v1}
\usepackage{ulem}

\preprintnumber{XXXX-XXXX}

\def\vector#1{\mbox{\boldmath $#1$}}

\begin{document}

\title{Vacuum Polarization and Photon Propagation in an Electromagnetic Plane Wave}

\author[1,*]{Akihiro Yatabe}
\affil{Advanced Research Institute for Science and Engineering, Waseda University, 3-4-1, Okubo, Shinjuku, Tokyo 169-8555, Japan\email{yatabe@heap.phys.waseda.ac.jp}}

\author[1]{Shoichi Yamada}

\begin{abstract}
The QED vacuum polarization in external monochromatic plane-wave electromagnetic fields is calculated with spatial and temporal variations of the external fields being taken into account. We develop a perturbation theory to calculate the induced electromagnetic current that appears in the Maxwell equations, based on Schwinger's proper-time method, and combine it with the so-called gradient expansion to handle the variation of external fields perturbatively. The crossed field, i.e., the long wavelength limit of the electromagnetic wave is first considered. The eigenmodes and the refractive indices as the eigenvalues associated with the eigenmodes are computed numerically for the probe photon propagating in some particular directions. In so doing, no limitation is imposed on the field strength and the photon energy unlike previous studies. It is shown that the real part of the refractive index becomes less than unity for strong fields, the phenomenon that has been known to occur for high-energy probe photons. We then evaluate numerically the lowest-order corrections to the crossed-field resulting from the field variations in space and time. It is demonstrated that the corrections occur mainly in the imaginary part of the refractive index.
 \end{abstract}

\subjectindex{B39}

\maketitle

\section{Introduction}
In the quantum vacuum, virtual particles and anti-particles are produced and annihilated repeatedly in very short times as intuitively represented by bubble Feynman diagrams. When an external field is applied, even these virtual particles are affected, leading to modifications of the property of quantum vacuum. One of the interesting consequences is a deviation of the refractive index from unity accompanied by a birefringence, i.e., distinct refractive indices for different polarization modes of photon\footnote{Interestingly, this does not occur for the nonlinear electrodynamics theory by Born and Infeld~\cite{Bialynicki-Birula1983}.}. It is a purely quantum effect that becomes remarkable when the strength of the external field approaches or even exceeds the critical value, $f_c = m^2/e$ with $m$ and $e$ being the electron mass and the elementary charge, respectively, whereas, the deviation of the refractive index from unity is proportional to the field-strength squared for much weaker fields. Photon splitting, which is another phenomenon in external fields, has been also considered~\cite{1987PhRvL..59.1065A}.

Such strong electromagnetic fields are not unrealistic these days. In fact, the astronomical objects called magnetars are a subclass of neutron stars, which are believed to have dipole magnetic fields of $\sim 10^{14-15}$G~\cite{mcgill}\footnote{The online catalog of magnetars is found at (http://www.physics.mcgill.ca/~pulsar/magnetar/main.html).}. Although the origin of such strong magnetic fields is still unknown, they are supposed to have implications for various activities of magnetars such as giant flares and X-ray emissions~\cite{2008A&ARv..15..225M}. In fact, their strong magnetic fields are thought to affect the polarization properties of surface emissions from neutron stars by the quantum effect~\cite{2002PhRvD..66b3002H,2015MNRAS.454.3254T}. This phenomenon may have indeed been detected in a recent optical polarimetric observation~\cite{2017MNRAS.465..492M}. The quantum correction may also play an important role through the so-called resonant mode conversions~\cite{1979PhRvD..19.3565M,2003PhRvL..91g1101L,2017ApJ...850..185Y}. On the other hand, the progress in the high-field laser is very fast. Although the highest intensity realized so far by Hercules laser at CUOS~\cite{hercules} is still sub-critical ($2 \times 10^{22} \mathrm{W/cm^2}$) for the moment, we may justifiably expect that the laser intensity will reach the critical value in not-so-far a future. Some theoretical studies on the vacuum polarization are meant for the experimental setups in the high-field laser~\cite{2006OptCo.267..318H,2014PhRvD..89l5003D,2014PhRvD..90d5025D,karbstein1,KingHeinzl2016}.

The study of the vacuum polarization in strong-field QED has a long history. It was pioneered by Toll in 1952~\cite{1952PhDT........21T}. He studied in his dissertation the polarization of vacuum in stationary and homogeneous magnetic fields in detail and many authors followed with different methods, both analytic and numerical~\cite{Baier1967a,1971PhRvD...3..618B,Adler71,tsaierber74,tsaierber75,kohriyamada,2007NuPhB.778..219S,hattoriitakura1,hattoriitakura2,kishikawa,karbstein2013}, and obtained the refractive indices. The vacuum polarization for mixtures of constant electric and magnetic fields was also investigated~\cite{1970PhRvD...2.2341B,batalin_shabad_1971,urrutia1978vacuum,artimovich_1990,DitGies,2000NuPhB.585..407S}. Note that such fields can be brought to either a purely magnetic or a purely electric field by an appropriate Lorentz transformation, with so-called crossed fields being an exception. 

In~\cite{1952PhDT........21T}, the polarization in the crossed field was also discussed. The crossed field may be regarded as a long wavelength limit of electromagnetic waves, having mutually orthogonal electric and magnetic fields of the same amplitude. Toll first calculated the imaginary part of the refractive index from the amplitude of pair creations and then evaluated the real part of refractive index via the Kramers-Kronig relation. Although there was no limitation to the probe-photon energy, the external-field strength was restricted to small values (weak-field limit) because he ignored the modification of the dispersion relation of the probe photon. Baier and Breitenlohner \cite{Baier1967b} obtained the refractive index for the crossed field in two different ways: they first employed the polarization tensor that had been inferred in~\cite{Baier1967a} from the 1-loop calculation for the external magnetic fields, and utilized in the second method the expansion of the Euler-Heisenberg Lagrangian to the lowest order of field strength. Note that both approaches are valid only for weak fields or low-energy probe photons.

The expression of the polarization tensor to the full order of field strength for the external crossed field was obtained from the 1-loop calculation with the electron propagator derived either with Schwinger's proper-time method~\cite{narozhnyi69} or with Volkov's solution~\cite{ritus72}. In~\cite{narozhnyi69}, the general expressions for the dispersion relations and the refractive indices of the two eigenmodes were obtained. Note, however, that the refractive indices were evaluated only in the limit of the weak-field and strong-field\footnote{Although these limits are referred to as ''weak-field limit'' and ``strong-field limit'' in the literature, they may be better called ``weak-field or low-energy limit'' and ``strong-field and high-energy limit'', respectively. See Fig.~\ref{region} for the actual parameter region.}. On the other hand, another expression of polarization tensor was obtained and its asymptotic limit was derived in~\cite{ritus72} although the refractive index was not considered.

The evaluation of the refractive index based on the polarization tensor of~\cite{ritus72} was attempted by Heinzl and Schr\"oder~\cite{heinzl} in two different ways: the first one is based on the hypothesized expression of the polarization tensor in the so-called large-order expansion with respect to the probe-photon energy; in the evaluation of the real part of the refractive index, the external crossed field was taken into account only to the lowest order of the field strength in each term of the expansion and the imaginary part was estimated from the hypothesized integral representation; in the second approach, the polarization tensor was expanded with respect to the product of the external-field strength and the probe-photon energy, and the refractive index was evaluated; the imaginary part was calculated consistently to the leading order and the anomalous dispersion for high-energy probe photons, which had been demonstrated by Toll~\cite{1952PhDT........21T}, was confirmed. Note that in these evaluations of the refractive index in the crossed field, the modification of the dispersion relation for the probe photon was again ignored as in~\cite{1952PhDT........21T} and hence the results cannot be applied to super-critical fields.

It should be now clear that the vacuum polarization and the refractive index have not been fully evaluated for supra-critical field strengths even in the crossed field. One of our goals is hence to do just that.

It is understandable, on the other hand, that the evaluation of the refractive index in the external electromagnetic plane-wave is more involved because of its non-uniformity. In fact, the refractive index has not been obtained except for some limiting cases. The polarization tensor and the refractive index in the external plane-wave were first discussed by Becker and Mitter~\cite{1975JPhA....8.1638B}. They derived the polarization tensor in momentum space from the 1-loop calculation with the electron propagator obtained by Mitter~\cite{Mitter1975}, which is actually Volkov's propagator represented in momentum space. Although the formulation is complete, the integrations were performed only for circularly polarized plane-waves as the background. The refractive indices were then evaluated at very high energies ($\gg m$) of the probe photon.

Baier et al.~\cite{baier75} calculated scattering amplitudes of a probe photon again by the circularly-polarized external plane-wave to the 1-loop order, employing the electron propagator expressed with the proper-time integral. The general expression of the dispersion relation was obtained but evaluated only in the weak-field and low-energy limit. The refractive indices for the eigenmodes of probe photons were also calculated in this limit alone. Affleck~\cite{1988JPhA...21..693A} treated this problem by expanding the Euler-Heisenberg Lagrangian to the lowest order of the field strength, assuming that the external field varies slowly in time and space. The refractive index was evaluated only in the weak-field limit again. Recently, yet another representation of the polarization tensor in the external plane-wave was obtained from the calculation of the 1-loop diagram with Volkov's electron propagator~\cite{meuren}. Only the expression of the polarization tensor was obtained, however, and no attempt was made to evaluate it in this study.

In their paper~\cite{2014PhRvD..89l5003D}, Dinu et al. employed the light front field theory, one of the most mathematically sophisticated formulations, to derive the amplitude of photon-photon scatterings, from which the refractive index integrated over the photon path was obtained. They calculated it for a wide range of the probe-photon energy and field strength. Although they gave the expression for the local refractive index, it was not evaluated. The eigenmodes of probe photons were not calculated, either.

In this paper, we also derive the expression of the polarization tensor and the refractive index for the external electromagnetic plane-wave, developing a perturbation theory for the induced electromagnetic current based on the proper-time method. It is similar to Adler's formulation~\cite{Adler71} but is more general, based on the interaction picture, or Furry's picture, and not restricted to a particular field configuration. Combining it with the so-called gradient expansion, we calculate the lowest-order correction from temporal and spatial field variations to the induced electromagnetic current, and hence to the vacuum polarization tensor also, for the crossed fields. This is nothing but the WKB approximation and, as such, may be applicable not only to the electromagnetic wave but also to any slowly-varying background electromagnetic fields. We then evaluate numerically the refractive indices for eigenmodes of the Maxwell equations with the modification of the dispersion relation being fully taken into account. Note that unlike~\cite{2014PhRvD..89l5003D} our results are not integrated over the photon path but local, being obtained at each point in the plane wave.

The paper is organized as follows: we first review Schwinger's proper time method briefly and then outline the perturbation theory based on the Furry picture to obtain the induced electromagnetic current to the linear order of the field strength of the probe photon in Sec.~2. This is not a new stuff. We then apply it to the plane-wave background in Sec.~3; in so doing, we also appeal to the so-called gradient expansion of the background electromagnetic wave around the crossed field. Technical details are given in Appendices.
Numerical evaluations are performed both for the crossed fields and for the first-order corrections in Sec.~4;
we summarize the results and conclude the paper in Sec.~5.

\section{Perturbation Theory in Proper-Time Method}
In this section, we briefly summarize Schwinger's proper-time method and outline
its perturbation theory, which will be applied to monochromatic plane-waves in the next section.

\subsection{Schwinger's Proper-Time Method}
The effective action of electromagnetic fields is represented as
\begin{eqnarray}
\Gamma = \Gamma _{\mathrm{cl}} + \Gamma _{\mathrm{q}} ,
\end{eqnarray}
where $\Gamma _{\mathrm{cl}}$ is the classical action and 
$\Gamma _{\mathrm{q}}$ is the quantum correction, which satisfies the following relation:
\begin{eqnarray}
 \frac{\delta \Gamma  _{\mathrm{q}}}{\delta A_{\mu}} \equiv \langle j^{\mu} (x) \rangle = ie \, \mathrm{tr} [ \gamma ^{\mu} G (x,x) ] . \label{currentterm}
\end{eqnarray}
Then, the vacuum Maxwell equation is modified as
\begin{eqnarray}
 - \Box A_{\mu} + \partial^{\nu} \partial_{\mu} A_{\nu}  - \langle j _{\mu} \rangle = 0. \label{normal_maxwell_eq}
\end{eqnarray}
Although there is no electromagnetic current generated by real charged particles in the vacuum,
$\langle j _{\mu} \rangle$ defined in this way is referred to as the induced electromagnetic current~\cite{DitGies}.
This term can be written with the electron propagator $G(x,y)$~\cite{schwinger51}
with $\mathrm{tr}$ in Eq.~(\ref{currentterm}) being the trace, or the diagonal sum on spinor indices;
$\gamma ^{\mu}$'s are the gamma matrices.
In this paper, the Greek indices run over 0 through 3 and the Minkowski metric is
assumed to be $\eta = \mathrm{diag} (+,-,-,-)$.

The electron propagator $G$ in the external electromagnetic field
is different from the ordinary one in the vacuum and the
modification by the external field,
the strength of which is close to or even exceeds the critical value $f_c$,
cannot be treated perturbatively.
The proper-time method is a powerful tool to handle such situations. 
The electron propagator satisfies the Dirac equation in the external electromagnetic field $A^{\mu}$:
\begin{eqnarray}
( i \gamma ^{\mu} \partial_{\mu}  - e \gamma ^{\mu} A_{\mu} (x) - m ) G (x,y) = \delta ^4 (x-y). \label{diracequation}
\end{eqnarray}

It is supposed in the proper-time method that there exists an operator $\hat{G}$,
the $x$-representation of which gives the propagator as $\langle x| \hat{G} |y \rangle = G(x,y)$.
Then Eq.~(\ref{diracequation}) can be cast into the following equation for the operators:
\begin{eqnarray}
( \gamma ^{\mu} \hat{\Pi} _{\mu} -m ) \hat{G} = \hat{{\bf 1}} ,
\end{eqnarray}
where $\hat{\bf 1}$ is the unit operator and $\hat{\Pi}_{\mu} = i \partial _{\mu} - e A _{\mu}$.
Here we used $\delta ^4 (x-y) = \langle x|y \rangle$. 
From this equation, the operator $\hat{G}$ is formally solved as
\begin{eqnarray}
\hat{G} = \frac{\hat{\bf 1}}{\gamma ^{\mu} \hat{\Pi}_{\mu} -m} ,
\end{eqnarray}
which can be cast into the following integral form:
\begin{eqnarray}
\hat{G} = i ( - \gamma ^{\mu} \hat{\Pi} _{\mu} -m) \int ^{\infty} _{0} ds \exp \left[ -i ( m^2 - ( \gamma ^{\mu} \hat{\Pi} _{\mu} )^2 - i \varepsilon ) s \right]. \label{propertime_operator_green_function}
\end{eqnarray}
In the above expression, the parameter $s$ is called the proper-time
and $-i \varepsilon$ is introduced to make the integration convergent as usual and will be dropped hereafter for brevity.
The electron propagator, being an $x$-representation of this operator, is obtained as 
\begin{eqnarray}
G (x, y) = i \int ^{\infty} _{0} ds e^{-im^2 s} \left[ \langle x| - \gamma ^{\mu} \hat{\Pi} _{\mu} e^{-i ( - ( \gamma ^{\nu} \hat{\Pi} _{\nu} )^2 ) s } |y \rangle -  \langle x| m e^{-i ( - ( \gamma ^{\mu} \hat{\Pi} _{\mu} )^2 ) s } |y \rangle \right] . \label{propertime_function_green_function}
\end{eqnarray}

Here we had better comment on the boundary condition for the electron propagator, or the causal Green function, in the electromagnetic wave. This issue may be addressed most conveniently for finite wave trains in the so-called light front formulation (e.g.~\cite{kogut1970quantum,neville1971quantum}), in which double null coordinates are employed. This is because the asymptotic states in the remote past and future (in the null-coordinate sense) are unambiguously defined~\cite{neville1971quantum}, which is crucially important particularly when one calculates $S$-matrix elements~\cite{2014PhRvD..89l5003D,neville1971quantum}; it is also important that the translational symmetry is manifest in one of the null coordinates. Then the causal Green function is obtained in the usual way, i.e., by the appropriate linear combination of the homogeneous Green functions with positive- and negative-energies according to the time ordering in the null coordinate~\cite{kogut1970quantum,neville1971quantum}. On the other hand, it is a well-known fact that the Dirac equation can be solved in a closed form for an arbitrary plane wave~\cite{Mitter1975,volkov1935class}. It is then possible to construct the same causal Green function with these Volkov solutions~\cite{ritus72,Mitter1975}. According to Ritus~\cite{ritus72}, all that is needed is a well-known $-i \epsilon$ prescription, i.e., the introduction of an infinitesimal negative imaginary mass. It was pointed out by Mitter~\cite{Mitter1975} then that this is equivalent to the same prescription in the proper-time method of Schwinger, that is, the formulation we adopt in this paper (see Eq.~(\ref{propertime_operator_green_function})). In this sense, the propagator we employ in this paper is the causal Green function thus obtained in the limit of the infinite wave train. As will become clear later (see Eq.~(\ref{maxwell_equation_wigner_representation_fourier}) in Section~\ref{Application_to_the_Single_Plane-Wave}), since we employ the gradient expansion in the local approximation, the distinction between the finite or infinite wave train will not be important in our formulation.

Returning to Eq.~(\ref{propertime_function_green_function}) and interpreting the operator $ e^{-i ( - ( \gamma ^{\mu} \hat{\Pi} _{\mu} )^2 ) s }$ as the evolution operator in the proper-time,
one can reduce the original field-theoretic problem to the one in quantum mechanics 
for the Hamiltonian $H = - ( \gamma ^{\mu} \hat{\Pi} _{\mu} )^2$.
Then the transformation amplitude is given as
\begin{eqnarray}
\langle x| e^{-i ( - ( \gamma ^{\mu} \hat{\Pi} _{\mu} )^2 ) s } |y \rangle  &=& \langle x| e^{-i H s } |y \rangle  \nonumber \\
&=& \langle x(s)|y(0) \rangle . 
\end{eqnarray}
Here the state $| x(s) \rangle$ is defined as the eigenstate for the operator $\hat{x}$ in the Heisenberg picture:
\begin{eqnarray}
 |x (s) \rangle \equiv e^{i Hs} |x \rangle .
\end{eqnarray}

The Hamiltonian $H$ is expressed as
\begin{eqnarray}
H = - \hat{\Pi}^2 + \frac{1}{2} e \sigma ^{\mu \nu} F_{\mu \nu} , \label{propertimeHamiltonian}
\end{eqnarray}
where we used the Clifford algebra for the gamma matrices
$\{ \gamma ^{\mu} , \gamma ^{\nu} \} = 2 \eta ^{\mu \nu}$
and the commutation relation $[ \Pi ^{\mu} , \Pi ^{\nu} ] = -ie F^{\mu \nu}$
to obtain $\hat{\Pi} ^2 = \hat{\Pi} _{\mu} \hat{\Pi}^{\mu}$
and $\sigma ^{\mu \nu} = \frac{i}{2} [ \gamma ^{\mu}, \gamma ^{\nu}]$;
$A_{\mu}$ and $F_{\mu \nu}$ are the vector potential and the field tensor
for the external electromagnetic field, respectively.
The proper-time evolutions of the 
operators $\hat{x}$ and $\hat{\Pi}$ are given by the Heisenberg equations:
\begin{eqnarray}
\frac{d \hat{x}^{\mu} (s)}{ds} &=& 2 \hat{\Pi} ^{\mu} (s) , \label{heisenberg_x} \\
\frac{d \hat{\Pi} ^{\mu} (s) }{ds} &=& 2e {F^{\mu}}_{\nu}  \hat{\Pi} ^{\nu} (s) + e^{i H s} ie \frac{\partial  {F^{\mu} }_{\nu} }{\partial x_{\nu} } e^{-i H s} + e^{i H s} \frac{1}{2} e \sigma ^{\nu \lambda}  \frac{\partial F_{\nu \lambda}}{\partial x _{\mu}} e^{-i H s} . \label{heisenberg_pi} 
\end{eqnarray}
Then, the induced electromagnetic current $\langle j^{\mu} \rangle$ in Eq.~(\ref{currentterm}) is represented as follows~\cite{Adler71}:
\begin{eqnarray}
\langle j^{\mu} (x) \rangle = \frac{e}{2} \int ^{\infty} _0 ds \: e^{-im^2 s}  \mathrm{tr} \left[ \langle x(s) | \hat{\Pi} ^{\mu} (s) + \hat{\Pi} ^{\mu} (0) | x(0) \rangle  - i \sigma ^{\mu \nu} \langle x(s) | \hat{\Pi} _{\nu} (s) - \hat{\Pi} _{\nu} (0) | x(0) \rangle  \right] . \ \  \label{plcurrent}
\end{eqnarray}
Note that this is equivalent to the 1-loop approximation with the external field
being fully taken into account.

In order to obtain the refractive index of the vacuum in the presence of an external electromagnetic field, we have to consider a probe photon in addition to the background electromagnetic field and apply Eq.~(\ref{normal_maxwell_eq}) to the amplitude of the probe photon. In so doing, the induced electromagnetic current $\langle j_{\mu} \rangle$ needs to be evaluated to the linear order of the amplitude of the probe photon and the perturbation theory is required at this point~\cite{DitGies}.
The Heisenberg equations given above can be solved analytically for some limited cases such as
time-independent homogeneous electric or magnetic fields and
single electromagnetic plane-waves~\cite{schwinger51}.
We will employ the latter as an unperturbed solution in the perturbative calculations in Section 3.
It is stressed that calculating the effective action for a given plane-wave background and taking its derivative with respective to the field strength is not sufficient for the evaluation of the refractive index, since the probe photon in general has a different wavelength and propagates in a different direction from those of the background electromagnetic wave. We hence need to take these differences fully into account in the perturbative calculations of the induced electromagnetic current. This was essentially done by~\cite{1975JPhA....8.1638B} in a different framework, i.e., performing 1-loop calculations in momentum space. In this paper we assume that the background wave has a long wavelength and calculate the refractive index locally in the sense of the WKB-approximation. In so doing, we appeal to the gradient expansion of the background plane wave as explained in Section 3.

\subsection{Outline of Perturbation Theory}
We now consider the perturbation theory in the proper time method. The purpose is to evaluate the induced electromagnetic current Eq.~(\ref{plcurrent}) up to the linear order of the amplitude of the probe photon, which is supposed to propagate in an external electromagnetic field.
It is then plugged into Eq.~(\ref{normal_maxwell_eq}) to derive
the refractive indices.
The Heisenberg equations (\ref{heisenberg_x}), (\ref{heisenberg_pi}) can be
analytically solved for a single monochromatic electromagnetic plane-wave~\cite{schwinger51}.
The calculation of the first order corrections to this solution is
the main achievement in this paper.
As explained in the next section, we employ further the gradient expansion of the background electromagnetic wave, which in turn enables us to obtain the refractive indices locally in the WKB sense.
In this section, we give the outline of the generic part of this perturbation theory,
which is not limited to the plane-wave background.
We will then proceed to its application to the monochromatic
plane-wave background in the next section.  

In the perturbation theory, the external electromagnetic fields are divided into two pieces:
the background $A^{\mu}$ and the perturbation $b^{\mu}$.
The corresponding field strengths are denoted by $F_{\mu \nu}$ and $g_{\mu \nu}$, respectively.
We take the latter into account only to the first order.
Then, the Hamiltonian given in Eq.~(\ref{propertimeHamiltonian}) can be written as
\begin{eqnarray}
H &=& - \left( i \partial _{\mu} - e A_{\mu} ( \hat{x} ) - eb_{\mu} (\hat{x} ) \right) ^2 + \frac{1}{2} e \sigma ^{\mu \nu} \left( F_{\mu \nu} (\hat{x}) + g_{\mu \nu} ( \hat{x} ) \right) \nonumber \\
&=& H^{(0)} + \delta H.
\end{eqnarray}
In this expression, $H^{(0)}$ is the unperturbed Hamiltonian, for which we assume that the proper-time evolution is
known, preferably analytically as in the time-independent homogeneous electric or magnetic fields and the single plane-wave.
$\delta H$ is the perturbation to the Hamiltonian.
It is evaluated to the first order of $b_{\mu}$ and expressed with $\delta \Pi_{\mu}=-e b_{\mu}$ as
\begin{eqnarray}
\delta H = - \hat{\Pi} ^{(0)} _{\mu} \delta \hat{\Pi} ^{\mu} - \delta \hat{\Pi} ^{\mu} \hat{\Pi} ^{(0)} _{\mu} + \frac{1}{2} e \sigma ^{\mu \nu} g_{\mu \nu} . 
\end{eqnarray}

In the proper-time method, the amplitudes of operators such as $\langle x(s) | \hat{\Pi} _{\mu} (s) | x(0) \rangle$ are evaluated very frequently
and in the perturbation theory they need to be calculated with perturbations to 
both the operators and the states being properly taken into account.
In so doing, we employ the interaction picture, which is also referred to
as the Furry picture~\cite{neville1971quantum} in the current case,
rather than making full use of the properties of particular field configurations
as in~\cite{Adler71}. 
The relation between the operator in the Heisenberg 
picture $\hat{A}_H (u)$ and that in the interaction picture $\hat{A}_I (u)$
is then given by the transformation:
$\hat{A} _H (u)= U^{-1} (u) \hat{A} _I (u) U (u)$,
where the operator $U(u)$ is written as $U(u) = e^{iH^{(0)} u} e^{-iH u}$.
It also satisfies the following equation:
$i \frac{\partial}{\partial u} U (u) = \delta H_I (u) U(u)$.
Here the perturbation Hamiltonian in the interaction picture $\delta H_{I}$ is 
given as 
$ \delta H_{I} (u) \equiv e^{i H^{(0)} u} \delta H e^{-i H^{(0)} u}$.
The equation of $U(u)$ can be solved iteratively as
\begin{eqnarray}
U(u) &=& 1 + (-i) \int ^u _0 d u_1 \delta H_I (u_1) \nonumber \\
& & + ( -i )^2 \int ^u _0 d u_1 \int ^{u_1} _0 d u_2 \delta H_{I} (u_1) \delta H_{I} (u_2) + \cdots \nonumber \\
& & + (-i )^n \int ^u _0 d u_1 \cdots \int ^{u_{n-1}} _0 d u_n \delta H_I (u_1 ) \cdots \delta H_I (u_n) \nonumber \\
& & + \cdots . \label{propertime_evolution}
\end{eqnarray}
Note that the right hand side of this equation includes only unperturbed quantities,
since the operators obey the free Heisenberg equations in the interaction picture.

The transformation amplitude $\langle x(s) | x(0) \rangle$ is also expressed
with the unperturbed operators and states as
\begin{eqnarray}
& & \langle x(s)| x(0) \rangle \simeq  \langle x^{(0)} (s)| \left[ 1 - i \int ^s _0 d u \delta H_I (u) \right] |x(0) \rangle . \label{perturbedamplitude}
\end{eqnarray}
In this expression, the index $(0)$ attached to the state indicates its proper-time evolution by $H^{(0)}$: 
\begin{eqnarray}
| x^{(0)} (s) \rangle =  e^{iH^{(0)} s} | x(0) \rangle .
\end{eqnarray}
Since we assume that the interaction and Heisenberg pictures are coincident with each other at $u=0$, we have
\begin{eqnarray}
 |x ^{(0)} (0) \rangle  = |x (0) \rangle .   
\end{eqnarray}
The operators $\hat{\Pi} (u)$ and the states $|x(u) \rangle$ 
at an arbitrary proper-time $u$ are expressed with the unperturbed counterparts $\hat{\Pi}^{(0)}$,
$\hat{x}^{(0)}$ and $|x^{(0)} \rangle$ via the operator $U(u)$ given in Eq.~(\ref{propertime_evolution}) in a similar way.

The amplitudes that appear in Eq.~(\ref{plcurrent}) for $\langle j^{\mu}(x) \rangle$ can be represented as
\begin{eqnarray}
& & \langle x(s)| \hat{\Pi}^{\mu} (s)|x(0) \rangle = \langle x^{(0)}(s)| \hat{\Pi}^{\mu}_I (s) U (s) |x (0) \rangle , \label{Hetoin1}  \\
 & & \langle x(s)| \hat{\Pi}^{\mu} (0) |x(0) \rangle = \langle x^{(0)} (s)| U(s) \hat{\Pi}^{\mu}_I (0) |x(0) \rangle , \label{Hetoin2}
\end{eqnarray}
with the operators and states in the interaction picture.
The calculations of these amplitudes are accomplished by the permutations of operators
$\hat{x} (s)$ and $\hat{x} (0)$ with the employment of their commutation relations
so that $\hat{x} (s)$ should sit always to the left of $\hat{x} (0)$.

\section{Application to the Single Plane-Wave} \label{Application_to_the_Single_Plane-Wave}
In this section, we apply the perturbation theory outlined above to the calculation of the induced electromagnetic current in the monochromatic plane-wave.
It is stressed that the distinction between the electromagnetic wave train having a finite or infinite length is not important in our calculations, since they employ only local information of the electromagnetic wave in the background thanks to the gradient expansion.
We first summarize the well-known results for the unperturbed background~\cite{schwinger51}.
The plane wave is represented as
\begin{eqnarray}
F_{\mu \nu} = f_{\mu \nu} F( \Omega \xi ) , 
\end{eqnarray}
with $f_{\mu \nu}$ being a constant tensor that sets the typical amplitude of the wave and $F(\Omega \xi)$ being an arbitrary function of $\Omega \xi = \Omega n_\mu x^\mu$; $\Omega$ is a frequency of the wave and
$n^{\mu}$ is a null vector that specifies the direction of wave propagation. 
The Heisenberg equations are written in this case as
\begin{eqnarray}
\frac{d \hat{x} ^{\mu} (s)}{ds} &=& 2 \hat{\Pi} ^{\mu} (s) , \label{plx} \\
\frac{d \hat{\Pi} ^{\mu} (s)}{ds} &=& 2 e {F^{\mu}}_{\nu} (s) \hat{\Pi}^{\nu} (s) + \frac{e}{2} n^{\mu}  f_{\nu \lambda} \sigma ^{\nu \lambda} \frac{d F ( \Omega \xi (s))}{d \xi (s)} . \label{plPi} 
\end{eqnarray}
Note that the phase $\xi (s) = n_{\mu} \hat{x}^{\mu} (s)$ in these equations is an operator and a function of the proper time $s$.
The term that contains $\partial ^{\nu} {F^{\mu}}_{\nu}$ vanishes in the equation of $\hat{\Pi}$ because
it is written as
$\partial _{\nu} {F_{\mu}}^{\nu} = {f_{\mu}}^{\nu} n_{\nu} [ d F ( \Omega \xi (s)) /d \xi (s) ]$ and
the following relation ${f_{\mu}}^{\nu} n_{\nu} = 0$ holds for the plane-wave.

To solve these equations, one introduces
\begin{eqnarray}
C^{\mu} = {f^{\mu} }_{\nu} \hat{\Pi} ^{\nu} (s) - e f^2 n^{\mu} A ( \Omega \xi (s)), 
\end{eqnarray}
which one can show is a constant of motion.
In this expression, $f^2 = f_{\mu \nu} {f^{\nu}}_{\lambda} / n_{\mu} n_{\lambda}$ is the amplitude squared of the plane-wave and $A( \Omega \xi (s))$ is defined as a quantity that satisfies the following relation: $F ( \Omega \xi (s) ) = dA ( \Omega \xi (s)) / d \xi (s) $.
Then, the operators $\hat{\Pi}^{\mu} (s)$ and $\hat{\Pi}^{\mu} (0)$ are obtained as follows:
\begin{eqnarray}
 & & \hat{\Pi} ^{\mu} (s) = \frac{\hat{x} ^{\mu} (s) - \hat{x} ^{\mu}(0)}{2s} \nonumber \\
 & & \hspace{0.3cm} + \frac{s}{\xi (s) - \xi(0)} \biggl[ 2 C^{\mu} eA ( \Omega \xi (s)) + n^{\mu} e^2 f^2 A^2 ( \Omega \xi (s)) + \frac{1}{2} e \sigma ^{\nu \lambda} f_{\nu \lambda} n^{\mu} F ( \Omega \xi (s)) \biggr] \nonumber \\
& & \hspace{0.3cm} - \frac{s}{ \left( \xi (s) - \xi (0) \right) ^2} \int ^{\xi (s)} _{\xi (0)} d \xi (u) \biggl[ 2 C^{\mu} e A ( \Omega \xi (u)) + n^{\mu} e^2 f^2 A^2 ( \Omega \xi (u)) + \frac{1}{2} e \sigma ^{\nu \lambda} f_{\nu \lambda} n^{\mu} F ( \Omega \xi (u)) \biggr] , \hspace{0.8cm} \label{Pis}
\end{eqnarray}
\begin{eqnarray}
& & \hat{\Pi}^{\mu} (0) = \frac{\hat{x}^{\mu} (s) - \hat{x} ^{\mu} (0)}{2s} \nonumber \\
& & \hspace{0.3cm} + \frac{s}{\xi (s) - \xi (0)} \biggl[ 2C^{\mu} e A( \Omega \xi (0)) + n^{\mu} e^2 f^2 A^2 ( \Omega \xi (0)) + \frac{1}{2} e \sigma ^{\nu \lambda} f_{\nu \lambda} n^{\mu} F ( \Omega \xi (0)) \biggr] \nonumber \\
& & \hspace{0.3cm} - \frac{s}{( \xi (s) - \xi (0) )^2} \int ^{\xi (s)} _{\xi (0)} d \xi (u) \biggl[ 2 C^{\mu} e A ( \Omega \xi (u)) + n^{\mu} e ^2 f^2 A^2 ( \Omega \xi (u)) + \frac{1}{2} e \sigma _{\nu \lambda} f^{\nu \lambda} n^{\mu} F ( \Omega \xi (u)) \biggr]. \hspace{0.8cm} \label{Pi0}
\end{eqnarray}
$C^{\mu}$ is also expressed as
\begin{eqnarray}
 C^{\mu} &=& \frac{{f^{\mu}}_{\nu} ( \hat{x}^{\nu} (s) - \hat{x}^{\nu} (0))}{2s} - \frac{1}{\xi (s) - \xi (0)} \int ^{\xi (s)} _{\xi (0)} d \xi (u) n^{\mu} f^2 e A ( \Omega \xi (u)) .
\end{eqnarray}
The amplitude $\langle x'(s)|x''(0) \rangle$ is given, on the other hand, as
\begin{eqnarray}
& & \langle x'(s)|x''(0) \rangle \nonumber \\
& & = \frac{1}{i (4 \pi )^2} \exp \left[ - \int ^{x'} _{x''} dx_{\mu} e A^{\mu} (x) \right] \frac{1}{s^2} \exp \Biggl\{ - \frac{i}{4s} (x' - x'' )^2  + \frac{i f^2 s}{( \xi' - \xi'' )^2} \left[ \int ^{\xi'} _{\xi''} e A ( \Omega \xi) d \xi \right] ^2 \nonumber \\
& & \hspace{0.5cm} - \frac{is}{\xi' - \xi''} \int ^{\xi'} _{\xi''} d \xi \left[ e^2 f^2 A^2 ( \Omega \xi ) + \frac{1}{2} e \sigma _{\rho \lambda} f^{\rho \lambda} F ( \Omega \xi ) \right] \Biggr\} . \label{plamp}
\end{eqnarray}
To derive the induced electromagnetic current $\langle j^{\mu} \rangle$, we use the amplitude $\langle x(s)| x(0) \rangle$, which is immediately obtained from the above equation as 
\begin{eqnarray}
& & \langle x(s) | x(0) \rangle \nonumber \\
&=& \frac{1}{i (4 \pi )^2 s^2} \exp \left[ - \frac{i}{2} e \sigma ^{\alpha \beta} f_{\alpha \beta} F (\Omega \xi ) s \right] \nonumber \\
&=& \frac{1}{i (4 \pi )^2 s^2} \left( {\bf 1} - \frac{ies}{2} F ( \Omega \xi ) \sigma ^{\alpha \beta} f_{\alpha \beta} \right) . \label{plac} 
\end{eqnarray}
We then find from Eqs.~(\ref{Pis}) through~(\ref{plamp}) that $\langle j^{\mu} \rangle$ in Eq.~(\ref{plcurrent}) is vanishing as pointed out first by Schwinger in his seminal paper~\cite{schwinger51}.
This situation changes, however, if another plane wave is added.

In this paper, we consider the propagation of a probe photon through the external monochromatic
plane-wave with the former being treated as a perturbation to the latter as usual.
We have in mind its application to high-field lasers.
Since the wavelengths of these lasers are close to optical wavelengths,
we assume in the following that the wavelength of the unperturbed monochromatic
plane-wave is much longer than the electron's Compton wavelength, or 
$\Omega_0 /m \ll 1$ for the wave frequency $\Omega _0$.
It may be then sufficient to consider temporal and spatial variations of the unperturbed 
plane-wave to the first order of $\Omega _0$.
This is equivalent to the so-called gradient expansion of the unperturbed field to the first order, which can be expressed generically as $F_{\mu \nu}  \simeq f_{\mu \nu} (1 + \Omega \xi)$. In fact, the plane-wave field given as $F_{\mu \nu} = f_{0 \mu \nu} \sin ( \Omega _0 n_{\alpha} x^{\alpha} )$ is Taylor-expanded at a spacetime point $x_0 ^{\mu}$
as $f_{0 \mu \nu} \sin ( \Omega _0 n_{\alpha} x_0 ^{\alpha} ) ( 1 + \cos ( \Omega _0 n_{\beta} x_0 ^{\beta} ) \Omega _0 n_{\gamma} ( x^{\gamma} - x_0 ^{\gamma}) )$ to the first order. This can be recast into $F_{\mu \nu} \simeq f_{\mu \nu} ( 1 + \Omega n_{\alpha} x^{\alpha})$ after shifting coordinates by $x_0 ^{\alpha}$ and employing the local amplitude and the gradient of the background field at $x^\alpha$ as $f_{\mu \nu} = f_{0 \mu \nu}  \sin ( \Omega _0 n_{\alpha} x_0 ^{\alpha} )$ and $\Omega =  \cos ( \Omega _0 n_{\beta} x_0 ^{\beta} ) \Omega _0$, respectively. Note that the above assumption on $\Omega_0$ implies $\Omega /m \ll 1$.

Gusynin and Shovkovy~\cite{gusynin1999derivative} developed a covariant formulation to derive the gradient expansion of the QED effective Lagrangian, employing the world-line formalism under the Fock-Schwinger gauge. Although their method is systematic and elegant indeed, the results obtained in their paper cannot be applied to the problem of our current interest, since the actual calculations were done only for the following field configurations: $F_{\mu \nu} = \Phi (x_{\alpha}) f_{\mu \nu}$, where $\Phi (x_{\alpha})$ is an arbitrary slowly-varying function of $x_{\alpha}$ while $f_{\mu \nu}$ is a constant tensor; the former gives a field variation in space and time and the latter specifies a field configuration. Although it appears quite generic, it does not include the configurations of our concern, i.e., those consisting of two electromagnetic waves propagating in different directions, unfortunately. Note that if the background and probe plane-waves are both traveling in the same direction and having the identical polarization, then one may regard the sum of their amplitudes as $\Phi$ and apply the gradient expansion of Gusynin and Shovkovy~\cite{gusynin1999derivative} to them; in this case, however, Schwinger~\cite{schwinger51} already showed that there is no quantum correction to the effective Lagrangian.

In our method, the probe photon, which is also treated as a classical electromagnetic wave,
is assumed to be monochromatic locally.
Strictly speaking, it has neither a constant amplitude nor a constant frequency because
the external field changes temporally and spatially.
As long as the wavelength of the external field is much longer than that
of the probe photon, which we assume in the following, the above assumption
that the probe field can be regarded as monochromatic locally may be justified.
We need to elaborate on this issue a bit further, though. As Becker and Mitter developed in their paper~\cite{1975JPhA....8.1638B}, the polarization tensor $\Pi^{\mu\nu} (x_1, x_2)$ depends not only on the difference of the two coordinates $x_1 - x_2$ but also on each of them separately and, as a result, its Fourier transform has two momenta corresponding to these coordinates. If the electromagnetic wave in the background is monochromatic, then the Floquet theorem dictates that the difference between them should be equal to some multiple of the wave vector of the electromagnetic wave in the background~\cite{zel1967quasienergy}. It follows then that eigenmodes of the probe photon are not diagonal in momentum in general. In fact, they should satisfy the following Maxwell equation:
\begin{eqnarray}
 \Box _{x_1} b^{\mu} (x_1, x_2) - \partial ^{\nu} _{x_1} \partial ^{\mu} _{x_1} b_{\nu} (x_1,x_2) = \int dx' \Pi ^{\mu \nu} (x_1,x') b_{\nu} (x', x_2) \label{maxwell_equation_discussion_Wigner_representation}
\end{eqnarray}
Becker and Mitter Fourier-transformed this equation and attempted to solve it in momentum space. Although they showed analytically that the momenta of probe-photon were indeed mixed in the expected way, they ignored the mixing in actual evaluations of the refractive index, since the effect is of higher order in the coupling constant.

We take another approach in this paper. Assuming, as mentioned above, that the electromagnetic wave in the background varies slowly in time and space and hence the probe photon can "see" the local field strength and its gradient alone, we expand the above equation in the small gradient. In so doing, we employ the Wigner representations of variables:
\begin{eqnarray}
 & & \Pi ^{\mu \nu} (x_1 - x_2 : X) = \int \frac{d^4 p}{(2 \pi )^4} \tilde{\Pi}^{\mu \nu} (p, X) e^{ip ( x_1 - x_2)}, \\
 & & b^{\mu} (x_1 - x_2 : X) = \int \frac{d^4 p}{(2 \pi )^4} \tilde{b}^{\mu} (p, X) e^{i p ( x_1 - x_2)} ,
\end{eqnarray}
where $X = (x_1 + x_2)/2$ is the center-of-mass coordinates and $\Pi ^{\mu \nu}$ and $b^{\mu}$ are regarded in these equations as functions of the relative coordinates $x_1-x_2$ and $X$ instead of $x_1$ and $x_2$.
Inserting these expressions into the right hand side of Eq.~(\ref{maxwell_equation_discussion_Wigner_representation}) and Fourier-transforming it with respective to the relative coordinates $x_1 - x_2$, we obtain
\begin{eqnarray}
 & & \int d ^4 (x_1 - x_2) \int d^4 x' \Pi ^{\mu \nu} ( x_1, x') b_{\nu} ( x' , x_2 ) e^{-ip(x_1 - x_2)} \nonumber \\
 & & = \int d ^4 (x_1 - x_2) \int d^4 x' \Pi ^{\mu \nu} ( x_1 - x' : X + (x' - x_2 )/2 ) b_{\nu} ( x' - x_2 : X + (x' - X_1 )/2 ) e^{-ip (x_1 - x_2)} \nonumber \\
 & & = \tilde{\Pi}^{\mu \nu} ( p, X) \exp \left( - \frac{1}{2} \partial _p ^{\Pi} \partial _X ^b \right) \exp \left( \frac{1}{2} \partial _p ^b \partial _X ^{\Pi} \right) \tilde{b}_{\nu} ( p, X) \nonumber \\
 & & \sim \tilde{\Pi}^{\mu \nu} (p,X) \tilde{b}_{\nu} (p, X) ,
\end{eqnarray}
in which $\partial_{p}^{\Pi}$ is a partial derivative with respective to $p$ acting on $\tilde{\Pi}$; other $\partial$'s should be interpreted in similar ways; the juxtapositions of two $\partial$'s stand for four-dimensional contractions; the last expression is the approximation to the lowest order with respective to the gradient in $X$, which is justified by our assumption. In deriving the third line of the above equations, we employ the following relations:
\begin{eqnarray}
 \Pi ^{\mu \nu} ( x_1 - x' : X + (x' - x_2)/2) &=& \exp \left( \frac{x' - x_2}{2} \partial _{X} ^{\Pi} \right) \Pi ^{\mu \nu} ( x_1 - x' : X), \\
 b ^{\mu} ( x' - x_2 : X + (x' - x_1)/2) &=& \exp \left( \frac{x' - x_1}{2} \partial _{X} ^{b} \right) b ^{\mu} ( x' - x_2 : X) .
\end{eqnarray}
Fourier-transforming the left hand side of Eq.~(\ref{maxwell_equation_discussion_Wigner_representation}) also with respective to the relative coordinates $x_1-x_2$, we obtain finally the "local" Maxwell equation as follows:
\begin{eqnarray}
 - p^2 \tilde{b}^{\mu} (p, X) + p^{\nu} p^{\mu} \tilde{b}_{\nu} (p,X) = \tilde{\Pi}^{\mu \nu} (p, X) \tilde{b}_{\nu} (p,X) . \label{maxwell_equation_wigner_representation_fourier}
\end{eqnarray}
Note that we also ignore the derivative with respect to $X$ in the kinetic part of the Maxwell equation, which is again valid under the current assumption.
We then consider the dispersion relation for the probe photon in the point-wise fashion, plugging the polarization tensor obtained locally this way.
This is nothing but the WKB approximation for the propagation of probe photon.
Note that the momentum of the probe photon is hence not the one in the asymptotic states~\cite{2014PhRvD..89l5003D} but the local one defined at each point in the background electromagnetic wave. It should be also stressed that the derived refractive index is a local quantity. Although such a quantity may not be easy to detect in experiments, this is regardless the main accomplishment in this paper.

\begin{figure}[htbp]
 \begin{center}
  \includegraphics[width=8.0cm]{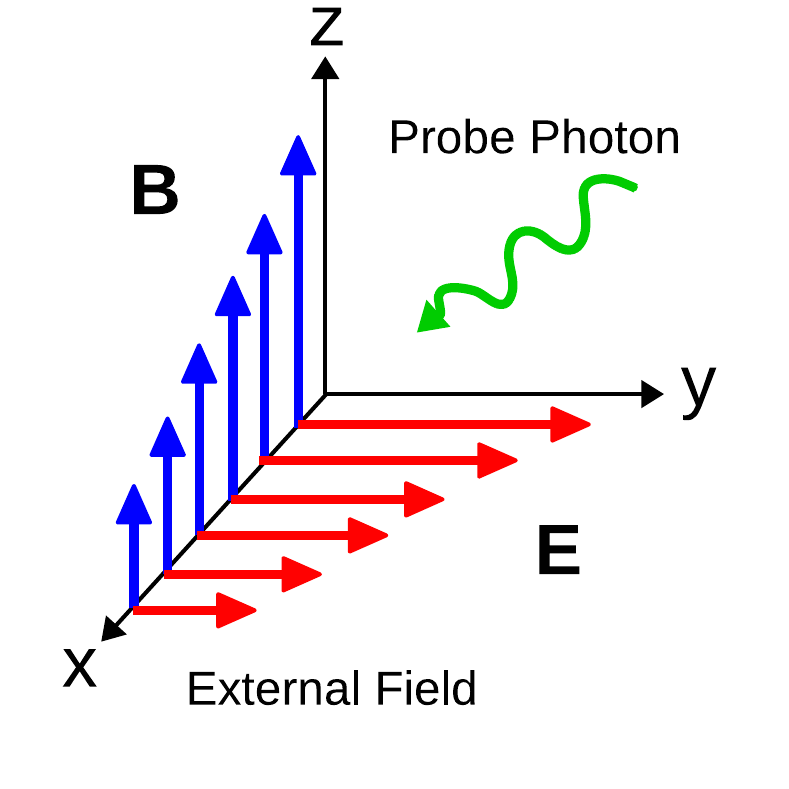}
  \caption{Schematic picture of the system considered in this paper. The external field consists of a non-uniform electric and magnetic fields denoted by E and B, respectively, and the probe photon.}
  \label{system}
 \end{center}
\end{figure}
We now proceed to the actual calculations. The induced electromagnetic current is written as
\begin{eqnarray}
 & & \langle  j^{\mu} \rangle = \frac{e}{2} \int ^{\infty} _{0} ds e^{-im^2 s} \nonumber \\
 & & \times \mathrm{tr} \left[ \langle x(s)| \hat{\Pi}^{\mu} (s) U(s) + U(s) \hat{\Pi}^{\mu} (0) | x(0) \rangle - i \sigma ^{\mu \nu} \langle x(s) |\hat{\Pi}_{\nu} (s) U(s) - U(s) \hat{\Pi}_{\nu} (0) |x(0) \rangle  \right] . \hspace{0.8cm} \label{current}
\end{eqnarray}
In this expression, we drop for brevity the superscript~${}^{(0)}$,
which means the unperturbed states,
and the subscript~${}_I$, which stands for the operators in the interaction picture.
We use these notations in the following.

The proper-time evolution operator $U(s)$ is given as
\begin{eqnarray}
U(s) &=& 1 - i \int ^{s} _0 du \delta H (u) \nonumber \\
&=& 1- i \int ^{s} _0 du \left\{ e \hat{\Pi}^{\alpha} (u) b_{\alpha} \exp \left[ -i k_{\delta} \hat{x}^{\delta} (u) \right] \right. \nonumber \\
& & + e b_{\alpha} \exp \left[ - i k_{\delta} \hat{x}^{\delta} (u) \right] \hat{\Pi}^{\alpha} (u) \left.  + \frac{1}{2} e \sigma ^{\alpha \beta} (u) g_{\alpha \beta} \exp \left[ -i k_{\delta} \hat{x} ^{\delta} (u) \right] \right\} \label{operatorU} 
\end{eqnarray}
for the present case.
Note that $\sigma ^{\alpha \beta} (u)$ is a proper-time-dependent operator in the interaction picture, which is defined as
\begin{eqnarray}
\sigma ^{\alpha \beta} (u) = e^{i H u} \sigma ^{\alpha \beta} e^{-i H u}.
\end{eqnarray}
The explicit expression of $\sigma ^{\alpha \beta} (u)$ can be easily obtained
to the first order of perturbation for the current Hamiltonian
evaluated at $u=0$ \footnote{\mbox{The Hamiltonian is proper-time-independent and can be evaluated at any time.}}
as
$H = - \hat{\Pi}^2 (0) + \frac{1}{2} e \sigma ^{\mu \nu}  F_{\mu \nu} (0)$:
\begin{eqnarray}
\sigma^{\alpha \beta} (u) &\simeq& \left[ 1 + \frac{i}{2} e u ( \sigma f ) \left( 1 + \Omega \xi (0) \right) \right] \sigma ^{\alpha \beta} \left[ 1- \frac{i}{2} e u ( \sigma f) \left( 1 + \Omega \xi (0) \right) \right] , \label{sigmat}
\end{eqnarray}
where we employ the abbreviation $( \sigma f) \equiv \sigma ^{\mu \nu} f_{\mu \nu}$.
Note that although $\xi (0)$ does not have a spinor structure and commutes with $(\sigma g) \equiv \sigma ^{\alpha \beta} g_{\alpha \beta}$,
$\xi (s)$ may have a nontrivial spinor structure induced by the proper-time evolution.

The amplitudes of $\langle x(s)| \hat{\Pi} ^{\mu} (s) U (s) |x (0) \rangle$ and $\langle x(s)| U(s) \hat{\Pi}^{\mu} (0) |x (0) \rangle$
can now be expressed with the unperturbed operators and states.
The calculations are involved, though, and given in Appendix~\ref{method}.
The final expressions are given as
\begin{eqnarray}
& & \langle x(s)| \hat{\Pi} ^{\mu} (s) U (s) |x (0) \rangle \nonumber \\ 
&=& \langle x(s)| \hat{\Pi}^{\mu} (s) |x(0)  \rangle  \nonumber \\
 & &  -i \int ^s _0 du \langle x(s)| 2 e b_{\alpha}  \hat{\Pi}^{\mu} (s)  \hat{\Pi}^{\alpha} (u) \exp \left[ -i k_{\delta} \hat{x}^{\delta} (u) \right]  |x(0)  \rangle \nonumber \\
 & & -i \int ^s _0 du \langle x(s)| - e b_{\alpha} k^{\alpha} \hat{\Pi}^{\mu} (s)   \exp \left[ -i k_{\delta} \hat{x}^{\delta} (u)  \right] |x(0)  \rangle \nonumber \\
 & & + \int ^s _0 du \langle x(s)| \hat{\Pi}^{\mu} (s) \exp \left[ -i k_{\delta} \hat{x}^{\delta} (u) \right] |x (0) \rangle \nonumber \\
 & & \times  \left( - \frac{ie}{2} \right) \left[ ( \sigma g ) + \frac{ieu}{2} \left\{ ( \sigma f ) ( \sigma g ) - ( \sigma g) ( \sigma f) \right\} + \frac{e^2 u^2}{4} ( \sigma f) ( \sigma g )( \sigma f) \right] \nonumber \\
 & & + \int ^s _0 du \langle x(s)| \hat{\Pi} ^{\mu} (s) \exp \left[ -i k_{\delta} \hat{x}^{\delta} (u) \right] |x (0) \rangle \nonumber \\
 & & \times \left( - \frac{ie}{2} \right) \left[ \frac{ieu}{2} \left\{ ( \sigma f ) ( \sigma g ) - ( \sigma g ) ( \sigma f) \right\} ( \Omega \xi ) + \frac{e^2 u^2}{2} ( \sigma f) ( \sigma g)( \sigma f ) ( \Omega \xi )   \right] , \label{pisu}
\end{eqnarray}
\begin{eqnarray}
& & \langle x(s)| U(s) \hat{\Pi}^{\mu} (0) |x (0)  \rangle \nonumber \\
&=& \langle x (s)| \hat{\Pi}^{\mu} (0) |x (0) \rangle \nonumber \\
 & & -i \int ^s _0 du \langle  x(s)| 2e b_{\alpha} \hat{\Pi}^{\alpha} (u) \exp \left[ -i k_{\delta} \hat{x}^{\delta} (u) \right] \hat{\Pi}^{\mu} (0) |x (0) \rangle \nonumber \\
 & & -i \int ^s _0 du \langle x (s)| -e b_{\alpha} k^{\alpha} \exp \left[ -i k_{\delta} \hat{x}^{\delta} (u) \right] \hat{\Pi}^{\mu} (0) |x (0) \rangle \nonumber \\
 & & + \int ^s _0 du \langle x (s)| \exp \left[ -i k_{\delta} \hat{x}^{\delta} (u) \right] \hat{\Pi}^{\mu} (0) |x (0) \rangle  \nonumber \\
 & & \times \left( - \frac{ie}{2} \right) \left[ ( \sigma g) + \frac{ieu}{2} \left\{ ( \sigma f )( \sigma g) - ( \sigma g) ( \sigma f)  \right\} + \frac{e^2 u^2}{4} ( \sigma f) ( \sigma g) ( \sigma f)   \right] \nonumber \\
 & & + \int ^s _0 du \langle x(s)| \exp \left[ -i k_{\delta} \hat{x}^{\delta} (u) \right] \hat{\Pi}^{\mu} (0) |x (0) \rangle \nonumber \\
 & & \times \left( - \frac{ie}{2} \right) \left[ \frac{ieu}{2} \left\{ ( \sigma f) ( \sigma g) - ( \sigma g) ( \sigma f) \right\}  ( \Omega \xi ) + \frac{e^2 u^2}{2} ( \sigma f )( \sigma g) ( \sigma f) ( \Omega \xi )  \right] \nonumber \\
 & & + \int ^s _0 du \langle x (s)| \exp \left[ -i k_{\delta} \hat{x}^{\delta} (u) \right] |x (0)  \rangle \nonumber \\
 & & \times \left( - \frac{ie}{2} \right) ( -i n^{\mu} ) \left[ \frac{ieu}{2} \left\{ ( \sigma f )( \sigma g) - ( \sigma g)( \sigma f) \right\} \Omega + \frac{e^2 u^2}{2} ( \sigma f) ( \sigma g) ( \sigma f)  \Omega  \right] . \label{upi0}
\end{eqnarray}

Then the induced electromagnetic current can be calculated by inserting these amplitudes in Eq.~(\ref{current}).
The operators
$\hat{\Pi} ^{\mu} (s)$, $\hat{\Pi} ^{\mu} (0)$, $\hat{\Pi} ^{\mu} (u)$ and $\exp [ -ik _{\alpha} \hat{x}^{\alpha} (u)]$
that appear in these expressions can be written in terms of the operators $\hat{x}^{\mu} (s)$ and $\hat{x}^{\mu} (0)$
as given in Eqs.~(\ref{pis}),~(\ref{pi0}),~(\ref{pit}) and~(\ref{expxt}).
Since the operators $\hat{x}^{\mu} (s)$ and $\hat{x}^{\mu} (0)$ do not commute with each other, we need to permute them
with the help of the commutation relations for these operators so that all
$\hat{x}^{\mu} (s)$ should sit to the left of all $\hat{x}^{\mu} (0)$. 
The details are given in Appendix~\ref{permutation}.
Note that the commutators such as $[ \hat{x}^{\mu} (s) , \hat{x}^{\nu} (0)]$ are operators
and hence we need to calculate commutation relations like $[ \hat{x}^{\mu} (s) , [\hat{x}^{\nu}  (s) ,  \hat{x}^{\lambda} (0)] ]$.
After these permutations, various amplitudes can be easily obtained from the following relations:
\begin{eqnarray}
 & &  \langle x(s)| \hat{x}^{\mu} (s) = x^{\mu} \langle x(s)| , \\
 & & \langle x(s)| \xi (s) = \xi  \langle x(s)| , \\
& & \hat{x}^{\mu} (0) |x(0) \rangle  = x^{\mu} |x(0) \rangle , \\
& & \xi (0) |x (0) \rangle = \xi |x(0) \rangle . 
\end{eqnarray}

In deriving Eqs.~(\ref{sigmat})-(\ref{upi0}), we consider only the neighborhood
of the coordinate origin,
the linear size of which is much shorter than the wavelength
of the background plane-wave but larger than the wavelength of the probe photon.
As mentioned earlier, however, the origin is arbitrary and one can shift the coordinates so that the point of interest should coincide with the origin.
Hence the results are actually applicable to any point.
More discussions on this point will be found in Appendix~\ref{renormalization}.
Note also that Furry's theorem dictates that the number of external fields that appear
in the expression of the induced electromagnetic current should be even, the details of which can be
found in Appendix~\ref{furry}.

After all these considerations and calculations, the induced electromagnetic current is given to
the lowest order of the perturbation and $\Omega$.
The details are presented in Appendix~\ref{current_expression}.
Since the induced electromagnetic current $\langle j_{\mu} \rangle$ is vanishing in the absence of the probe photon,
it is generated by its presence and is should be proportional to it:
\begin{eqnarray}
 \langle j_{\mu} (x) \rangle = {\Pi_{\mu}} ^{\nu} (k, x)  b_{\nu} (k ,x )  \exp ( -i k_{\alpha} x^{\alpha} ) . \label{polarization_tensor}
\end{eqnarray} 
Note that we employ the local approximation here again.

In this expression, the probe photon is given as $b_{\nu} \exp ( -i k_{\alpha} x^{\alpha})$
and the proportionality coefficient ${\Pi _{\mu}}^{\nu}$ is nothing but the polarization tensor at each point.
Following Ritus~\cite{ritus72}, we decompose the polarization tensor so obtained as
\begin{eqnarray}
{\Pi_{\mu}}^{\nu} &=&  \int ^{\infty} _0 ds \int ^s _0 du \: \left[ \Pi _1 (fk)_{\mu} (fk)^{\nu} +  \Pi _2 (\tilde{f}k)_{\mu} (\tilde{f} k)^{\nu} + \Pi _3 G_{\mu} G^{\nu} \right] , \label{polarization_tensor}
\end{eqnarray}
with three mutually orthogonal vectors
\begin{eqnarray}
& & (fk)_{\mu} = {f_{\mu}}^{\nu} k_{\nu}, \ (\tilde{f} k)_{\mu} = \tilde{f}_{\mu} {}^{\nu} k_{\nu}, \ G_{\mu} = \frac{k_{\alpha} k^{\alpha}}{k_{\beta} {f^{\beta}}_{\gamma} {f^{\gamma}}_{\delta} k^{\delta}} {f_{\mu}}^{\nu}  {f_{\nu}}^{\lambda} k_{\lambda} ,
\end{eqnarray}
where $\tilde{f}^{\mu \nu} = \varepsilon ^{\mu \nu \rho \sigma} f_{\rho \sigma} /2$ is
the dual tensor of $f^{\mu \nu}$.
Here $\varepsilon ^{\mu \nu \rho \sigma}$ is
the Levi-Civita antisymmetric symbol, which satisfies $\varepsilon ^{0123} = 1$.
The following abbreviations $(kk) = k_{\mu} k^{\mu}$ and  
$(kffk) = k_{\mu} {f^{\mu}}_{\nu} {f^{\nu}}_{\lambda} k^{\lambda}$ are also used~\cite{ritus72}.
Then the coefficients are given as follows:
\begin{eqnarray}
& & \Pi_1 = \frac{e^2 e^{-im^2 s}}{72 \pi ^2 s^3 (kffk)} \exp \left[ i \left( u - \frac{u^2}{s} \right) (kk) \right] \left( -1 + \exp \left[ \frac{i(s-u)^2 u^2 e^2 (kffk) }{3s} \right] \right) \left( -18 i - 9s (kk) \right) \nonumber \\
  & & + \frac{e^2 e^{-im^2 s}}{72 \pi ^2 s^3} \exp \left[ i \left( u - \frac{u^2}{s} \right) (kk) + \frac{i(s-u)^2 u^2 e^2 (kffk) }{3s} \right] e^2 \left( - \frac{18}{s} \right) u ( s^3 - 3 s^2 u +4 su^2 -2 u^3) \nonumber \\
 & & + \frac{e^2 e^{-im^2 s}}{72 \pi ^2 s^3} \exp \left[ i \left( u - \frac{u^2}{s} \right) (kk) + \frac{i(s-u)^2 u^2 e^2 (kffk) }{3s} \right] \nonumber \\
& & \times \left[ \left( \frac{2}{s^2} e^2 ( \Omega kn) (s-u)u ( 6s^4 + 22 s^3 u - 79 s^2 u^2 + 78 su^3 - 36 u^4) \right)\right. \nonumber \\
& & + \left( -2 i \right) e^2 (\Omega kn) (kk) (s-u)^2 u (3s^2 - su -3 u^2) \nonumber \\
 & & \left. + \left( - \frac{4 i}{s^2} \right) e^4 ( kffk) ( \Omega kn) (s-u)^3 u^2 ( 3s^4 - 7s^3 u + 5 s^2 u^2 + 4 s u^3 - 6u^4 ) \right], \label{pi1}
\end{eqnarray}
\begin{eqnarray}
 & & \Pi _2 =  \frac{e^2 e^{-im^2 s}}{72 \pi ^2 s^3 (kffk)} \exp \left[ i \left( u - \frac{u^2}{s} \right) (kk) \right] \left( -1 + \exp \left[ \frac{i(s-u)^2 u^2 e^2 (kffk) }{3s} \right] \right) \left( -18 i - 9s (kk) \right)  \nonumber \\
& & + \frac{e^2 e^{-im^2 s}}{72 \pi ^2 s^3} \exp \left[ i \left( u - \frac{u^2}{s} \right) (kk) + \frac{i(s-u)^2 u^2 e^2 (kffk) }{3s} \right] e^2 ( - 18 ) su (s-u) \nonumber \\
 & & + \frac{e^2 e^{-im^2 s}}{72 \pi ^2 s^3} \exp \left[ i \left( u - \frac{u^2}{s} \right) (kk) + \frac{i(s-u)^2 u^2 e^2 (kffk) }{3s} \right] \nonumber \\
& & \times \left[ \left( \frac{2}{s} e^2 ( \Omega kn) (s-u) u ( 6s^3 + 4s^2 u - 7s u^2 + 6 u^3) \right)\right. \nonumber \\
& & + \left( -2 i \right) e^2 (\Omega kn) (kk) (s-u)^2 u (3s^2 - su -3 u^2) \nonumber \\
 & & \left. + \left( - 4i \right) e^4 ( kffk) ( \Omega kn) (s-u)^3 u^2 ( 3s^2 - su - 3u^2 ) \right] , \label{pi2}
    \end{eqnarray}
\begin{eqnarray}
 & &  \Pi_3 = \frac{e^2 e^{-im^2 s}}{72 \pi ^2 s^4 (kk)} \exp \left[ i \left( u - \frac{u^2}{s} \right)  (kk) \right] \left( -1 + \exp \left[ \frac{i (s-u)^2 u^2 e^2 (kffk)}{3s} \right] \right) \nonumber \\
 & & \times ( -6 is + (-3s^2 + 16su - 16 u^2) (kk))  \nonumber \\
 & & + \frac{e^4 (kffk) e^{-im^2 s}}{324 \pi ^2 s^4 (kk)^2} \exp \left[ i \left( u - \frac{u^2}{s} \right) (kk) + \frac{i(s-u)^2 u^2 e^2 (kffk) }{3s} \right] \nonumber \\
& & \times \left[ 3i (s^3 - 6s^2 u + 6su^2 ) - 2u^2 (s^2 -3su + 2u^2 )^2 e^2 (kffk) -12 (s-2u)^2 (s-u) u (kk) \right] \nonumber \\
 & & - \frac{i e^2 e^{-im^2 s}}{27 (kk)^2 \pi ^2 s^4} \exp \left[ i \left( u - \frac{u^2}{s} \right) (kk) + \frac{i(s-u)^2 u^2 e^2 (kffk) }{3s} \right] \nonumber \\
 & & \times (\Omega kn) e^2 (kffk) (s-u) u (s^2 - 5su + 5u^2) \nonumber \\
& & - \frac{e^2 e^{-im^2s}}{972 (kk)^2 \pi ^2 s^5} \exp \left[ i \left( u - \frac{u^2}{s} \right) (kk) + \frac{i(s-u)^2 u^2 e^2 (kffk) }{3s} \right] \nonumber \\
 & & \times ( \Omega kn) e^2 (kffk) (s-u)^2 u \left[ 2 e^2 (kffk) (3s^5 -19 s^4 u + 6s^3 u^2 + 87 s^2 u^3 - 138 s u^4 + 60 u^5 ) \right. \nonumber \\
 & & \hspace{2cm} \left. -9 ( 6s^3 + 9 s^2 u -46 s u^2 + 40 u^3 ) (kk) \right] \nonumber \\
& & - \frac{i e^2 e^{-im^2 s}}{2916 (kk)^2 \pi ^2 s^5} \exp \left[ i \left( u - \frac{u^2}{s} \right) (kk) + \frac{i(s-u)^2 u^2 e^2 (kffk) }{3s} \right] \nonumber \\
 & & \times  ( \Omega kn) e^2 (kffk) (s-u)^2 u ( 3s^2 - su - 3u^2)  \left[ 4 e^4 (kffk)^2 u^2 ( s^2 - 3su + 2 u^2) ^2  \right. \nonumber \\
 & & \hspace{2cm} \left. + 24 e^2 (kffk) (kk) (s-2u)^2 (s-u) u + 9 ( 3s^2 - 16 su + 16 u^2) (kk)^2 \right] , \hspace{0.8cm} \label{pi3}
\end{eqnarray}
where $( \Omega kn) = \Omega k_{\mu} n^{\mu}$ is the inner product of the momentum vectors of
the external plane-wave and the probe photon.
The refractive indices for physical modes are related to $\Pi _1$ and $\Pi _2$.

The proper-time integration in Eq.~(\ref{polarization_tensor}) or its pre-decomposition form,
Eq.~(\ref{currentlong}), has to be done numerically.
The original form is not convenient for this purpose and we rotate the integral
path by $-\pi /3$ in the complex plane so that the integral could converge
exponentially as $s$ goes to infinity.
Note that the rotation angle is arbitrary as long as it is in the range of $(0, - \pi /3]$.
The refractive index is then obtained by solving the Maxwell equation reduced in the following form:
\begin{eqnarray}
{A_{\mu}}^{\nu} ( k) b_{\nu}  = 0, \label{maxwell_equation}
\end{eqnarray}
with ${A_{\mu}}^{\nu} = - (kk) {\delta_{\mu}}^{\nu} + k_{\mu} k^{\nu} + {\Pi _{\mu}}^{\nu}$.
The probe photon is hence described as a non-trivial solution of this homogeneous equation
and its dispersion relation is obtained from the relation
$\mathrm{det} A = 0$. Note that not all of them are physical.
Unphysical modes are easily eliminated, however,
by calculating the electric and magnetic field strengths, which are gauge-invariant.
It is then found that only two of them associated with $\Pi_1$ and $\Pi_2$ are physical as expected.
Note that the four momenta of the probe photon thus obtained are no longer null
in accordance with the refractive indices different from unity.
The polarization vectors are also obtained simultaneously.

\begin{table}[htbp]
\caption{Eigenmodes of the probe photons with different 4-momenta} \label{eigenmode}
\begin{center}
 \begin{tabular}{lll} \hline
probe momentum $k_{\mu}$ & eigenmode & mode name \\
\hline
\raisebox{-0.3cm}[0cm][-0.1cm]{$(k_0, k_1 ,0,0)$} & $(0,0,1,0) _{\mu}$ & x2 mode \\
 & $(0,0,0,1) _{\mu}$ & x3 mode \\
\raisebox{-0.3cm}[0cm][-0.1cm]{$(k_0 , 0, k_2 , 0 )$} & $ ( k_2 , - k_2 , k_0 ,0) _{\mu}$ & y1 mode \\
 & $ ( 0,0,0,1) _{\mu}$ & y3 mode \\
\raisebox{-0.3cm}[0cm][-0.1cm]{$(k_0 , 0, 0 , k_3 )$} & $  ( k_3 , - k_3,0, k_0) _{\mu}$ & z1 mode \\
 & $  ( 0 , 0 , 1 ,0) _{\mu}$ & z2 mode \\
\raisebox{-0.3cm}[0cm][-0.1cm]{$(k_0 , \frac{k_i}{\sqrt{3}},  \frac{k_i}{\sqrt{3}},  \frac{k_i}{\sqrt{3}} )$} & $ ( A , B , 1 ,0) _{\mu}$  & s2 mode \footnote{} \\
  & $ ( A , B , 0 ,1) _{\mu}$ & s3 mode \\ \hline
  \\[-12pt]
 \multicolumn{3}{l}{${}^5$ $A$, $B$ are constants written with $k_0$ and $k_i$.}
 \end{tabular} 
\end{center}
\end{table}

In the next section, we show the results of some numerical evaluations. As representative cases,
we consider four propagating directions of the probe photon as summarized in Table~\ref{eigenmode}.
Since the background plane-wave is assumed to have a definite propagation direction ($x$-direction) and linear polarization ($y$-direction),
these four directions are not equivalent. For each propagation direction, there are two physical eigenmodes,
as mentioned above, which are in general different from each other, having distinctive dispersion relations,
i.e., the background is birefringent.

\section{Results}
In this section, we numerically evaluate the refractive index $N$, which is defined as
$N= | \vector{k}| / k_0$.
Firstly, the crossed fields are considered and then the first order correction
$\delta N$ in the gradient expansion is calculated for the plane-wave field.
The eigenmodes of the probe photon depend on the propagation direction as already mentioned.
The refractive index is complex in general with the real part representing 
the phase velocity of the probe photon divided by the light speed and the imaginary part
indicating the decay, possibly via electron-positron pair creations.
Since the deviation of the refractive index from unity is usually much smaller than 
unity, only the deviations are shown in the following: $\mathrm{Re}[N-1]$ and
$\mathrm{Im}[N]$.

Note that for all cases considered in this paper,
the refractive indices, both real and imaginary parts, of the y1 and z2 modes
are identical and so are those of the y3 and z1 modes.
Although the exact reason for this phenomenon is not known to us for the moment,
the following should be mentioned:
the polarization tensor $\Pi ^{\mu \nu}$ is expressed as the sum of
three contributions proportional to
$(fk)^{\mu} (fk)^{\nu}$, $( \tilde{f} k) ^{\mu} ( \tilde{f} k) ^{\nu}$ and $G^{\mu} G^{\nu}$
given as Eq.~(\ref{polarization_tensor});
each pair of the modes that have the identical refractive index
are actually eigenmodes of either $(fk)^{\mu} (fk)^{\nu}$
or $( \tilde{f} k)^{\mu} ( \tilde{f} k)^{\nu}$.
We will show 
these degenerate modes with the same color in figures hereafter.

\subsection{Crossed Fields}
\begin{figure}[htbp]
 \begin{center}
  \includegraphics[width=8.6cm]{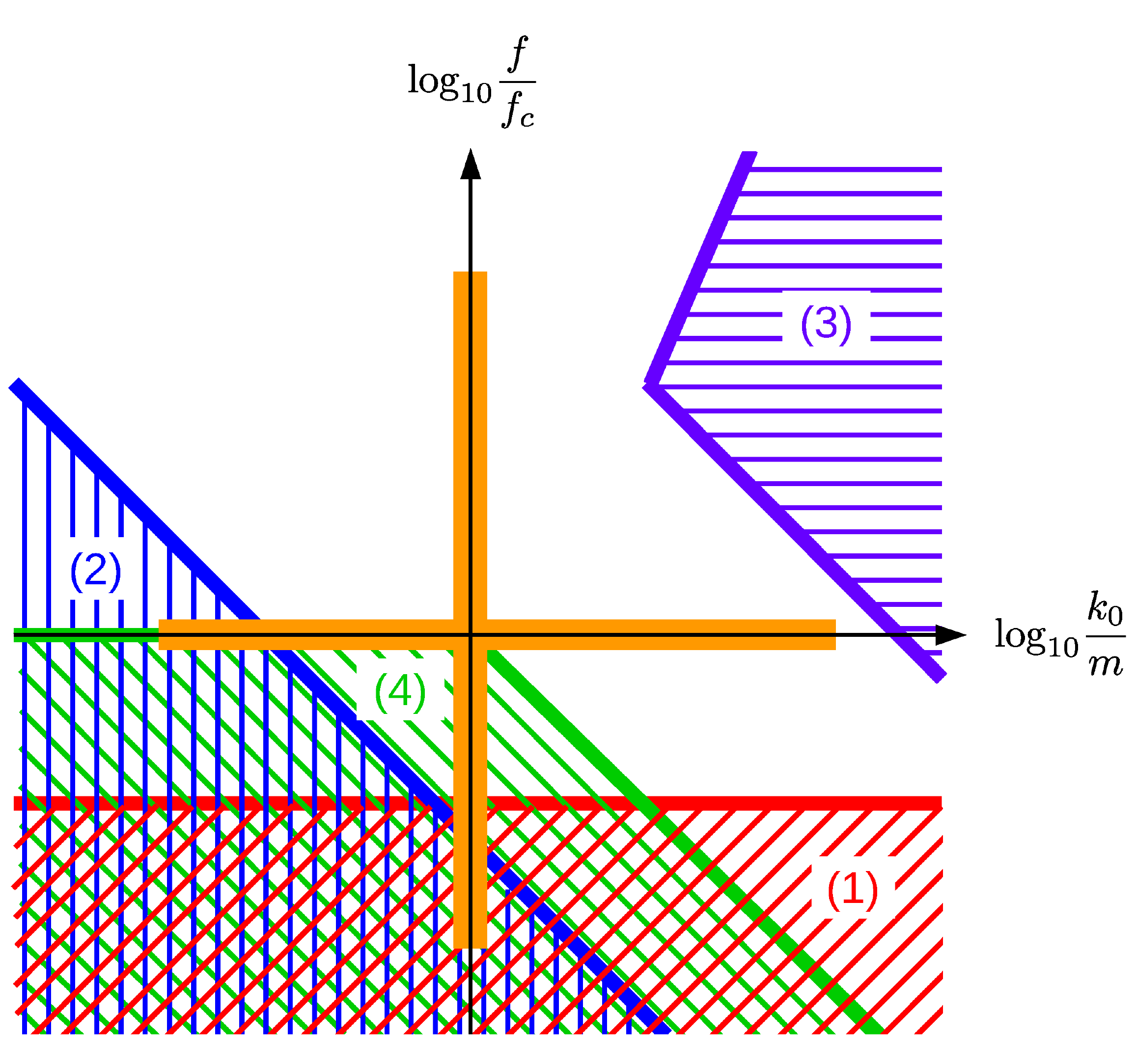}
  \caption{Regions in the plane of the strength of external crossed fields and the probe-photon energy that have been already explored. Each region is labeled as follows: region~(1) is the weak-field limit $f/f_c \ll 1$~\cite{1952PhDT........21T}; region~(2) is for the weak-field or low-energy limit $e^2 k_{\mu} {f^{\mu}}_{\nu} f^{\nu \lambda} k_{\lambda} / m^6 \ll 1$~\cite{Baier1967b,narozhnyi69,ritus72}; region~(3) corresponds to the strong-field and high-energy limit $1 \ll e^2 k_{\mu} {f^{\mu}}_{\nu} f^{\nu \lambda} k_{\lambda} / m^6 \ll (k_0 / m)^6 \times \alpha ^{-3}$ studied in~\cite{narozhnyi69}, where $\alpha = e^2/ 4 \pi$ is the fine-structure constant; region~(4) is the region that satisfies $(f/f_c) \lesssim 1$ and $(f/f_c) \times ( k_0 / m) \lesssim 1$ explored in~\cite{heinzl}. The orange lines indicate the regions, in which the refractive indices are computed numerically in this paper. Note that our method can treat the whole region in this figure in principle.}
  \label{region}
 \end{center}
\end{figure}

As mentioned in Introduction, the vacuum polarization in the crossed fields
was already obtained by many authors.
The refractive index was also evaluated both analytically and
numerically~\cite{1952PhDT........21T,Baier1967b,narozhnyi69,ritus72,heinzl}.
The regions in the plane of the field strength $f$ and the probe-photon energy $k_0$
that have been investigated in these papers are summarized in Fig.~\ref{region}.
It is apparent from the figure that there is still an unexplored region, which is unshaded.
And that is the target of this paper.
The parameter ranges we adopted in this paper are displayed in orange in the same figure: we
first calculate the refractive index for the external field
of the critical value to validate our formulation by comparing our results
with those in the previous studies; then we vary the strength of the external field.

\begin{figure}[htbp]
 \begin{center}
  \includegraphics[width=8.6cm]{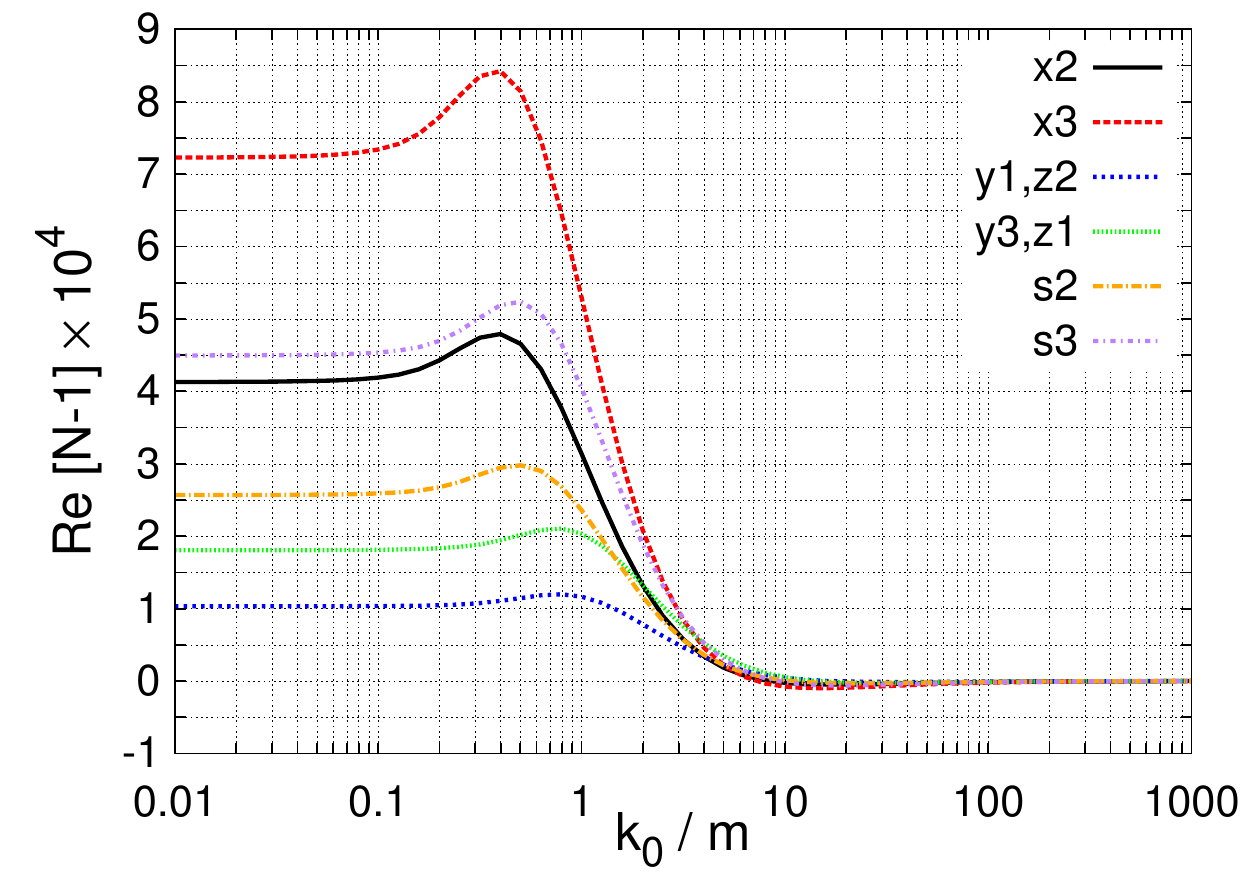}
  \caption{Plot of $\mathrm{Re} [N-1]$ as a function of the probe-photon energy in the crossed field. Here $N$ is a refractive index. We set $f/f_c =1$. Colors specify different modes.}
  \label{graph_crossed_field_a1_b_re_all}
 \end{center}
\end{figure}

The polarization tensor ${\Pi _{\mu}}^{\nu}$ in the crossed field
is obtained by simply taking the limit of $\Omega \rightarrow 0$ in Eq.~(\ref{polarization_tensor}).
Setting the strength of the external field to the critical value $f /f_c = 1$,
we compute the refractive indices for the range of $0.01 \leq k_0 /m  \leq 1000$\footnote{Shore studied the refractive index of super-critical magnetic fields for a wider range of the photon energy~\cite{2007NuPhB.778..219S}. The results are similar to ours for the crossed field.}.
Note that the low-energy regime $(k_0/m \lesssim 1)$ has been investigated already as shown in Fig.~\ref{region}.
The real part $\mathrm{Re} [N-1]$ is shown in Fig.~\ref{graph_crossed_field_a1_b_re_all}
with colors indicating different modes of the probe photon.

\begin{figure}[htbp]
 \begin{center}
  \includegraphics[width=8.6cm]{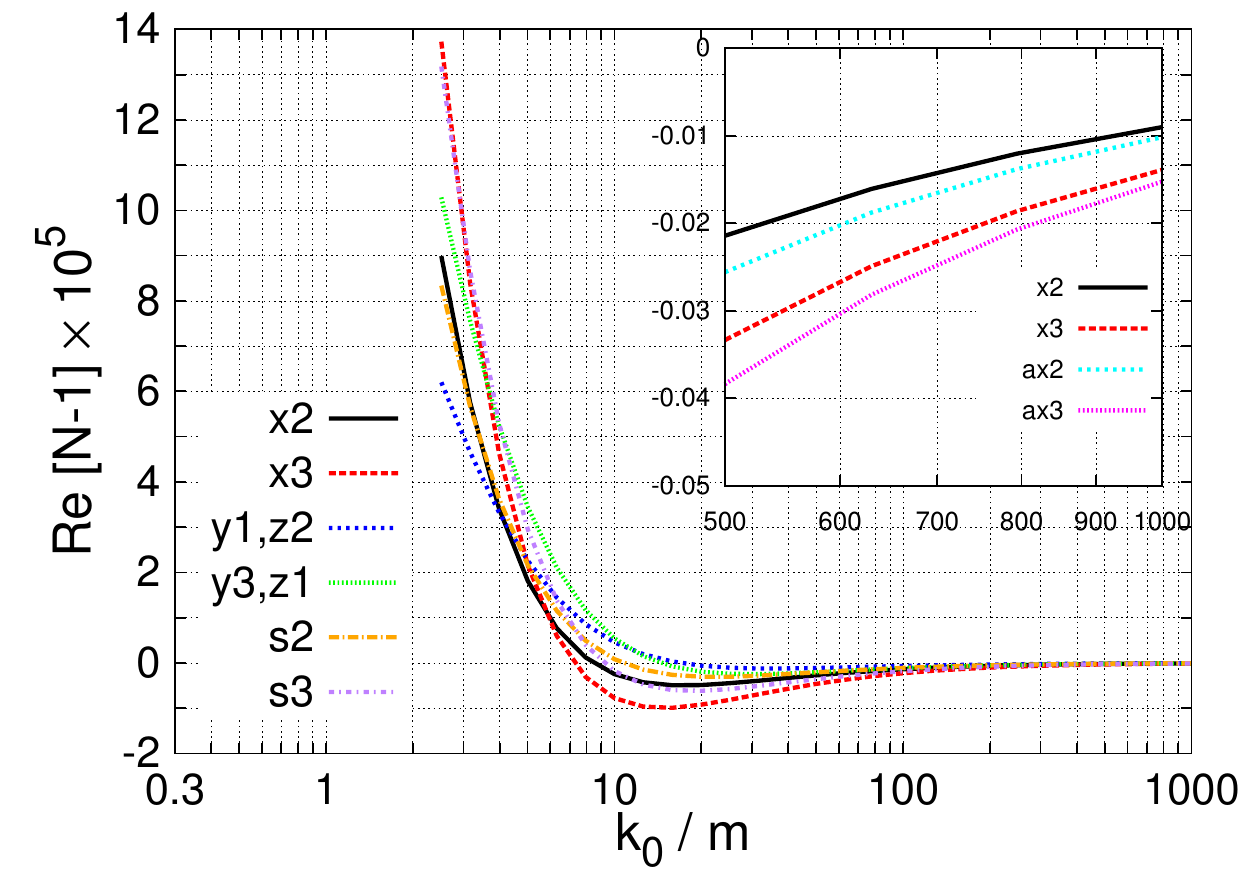}
  \caption{Same as Fig.~\ref{graph_crossed_field_a1_b_re_all} but for high energies alone on a different vertical scale. The inset zooms into the high-energy range of $500 \leq k_0/m \leq 1000$ and asymptotic formulae are also shown as ax2 and ax3 for the x2 and x3 modes, respectively.}
  \label{graph_crossed_field_a1_b_re_high_energy}
 \end{center}
\end{figure}

It is found that the deviation of the refractive index from unity is of the order of $10^{-4}$. 
As $k_0/m$ gets smaller, the refractive index approaches the values in the weak-field or
low-energy limit (region (2) in Fig.~\ref{region}),
which are written as
\begin{eqnarray}
& & N_{\mathrm{x2}} \simeq 1 + \frac{2 \alpha}{45 \pi} \frac{\kappa ^2 m^2}{k_0 ^2}, \label{weak_re_x2} \\
& &  N_{\mathrm{x3}} \simeq 1 + \frac{7 \alpha}{90 \pi} \frac{\kappa ^2 m^2}{k_0 ^2}, \label{weak_re_x3}
\end{eqnarray}
for the x2 and x3 modes, respectively,
where $\kappa ^2 = e^2 k_{\mu} {f^{\mu}}_{\nu} f^{\nu \lambda} k_{\lambda} /m^6 = e^2 f^2 (k_0 + k_1 )^2 /m^6$
is the product of the probe photon energy and the field strength normalized
by the critical value.
Then the typical value of $N_{\mathrm{x2}} -1$ can be estimated as
\begin{eqnarray}
 N_{\mathrm{x2}} - 1 \simeq \frac{8 \alpha}{45 \pi} \left( \frac{f}{f_c} \right) ^2 \sim 4.1 \times 10^{-4} \left( \frac{I}{4.6 \times 10^{29} \mathrm{W/cm^2}} \right) ,
\end{eqnarray}
where $I = f^2 / 4 \pi$ is the intensity of the plane wave.
The results are hence in agreement with what was already published in~\cite{Baier1967b,narozhnyi69,ritus72,heinzl}.
The refractive indices depend on the propagation direction of the 
probe photon: the modulus $|\mathrm{Re} [N-1]|$ is larger for the photon propagating in the opposite direction
to the background plane-wave (the x mode) than 
those going perpendicularly (the y/z modes); the s mode that propagates obliquely
lies normally in between although the modulus is greater for the s3 mode than for the x2 mode.
The photons polarized in the $z$-direction have larger moduli in general except
the z mode, which propagates in this direction, has a greater modulus when
it is polarized in the x-direction.
These trends are also true for other results obtained below in this paper.

As $k_0/m$ becomes larger than $\sim 10$, all the refractive indices for different
propagation directions appear to converge to unity,
which is consistent with~\cite{1952PhDT........21T,heinzl}.
This is more apparent in Fig.~\ref{graph_crossed_field_a1_b_re_high_energy},
which zooms into the region of $3 \lesssim k_0 /m \leq 1000$.
It is also seen in the same figure that $\mathrm{Re} [N-1]$ is negative
and the modulus $| \mathrm{Re} [N-1] |$ decreases for $k_0/m \gtrsim 10$.
This trend is consistent with the high-energy limits given in~\cite{narozhnyi69}
(region (3) in Fig.~\ref{region}),
which are written as
\begin{eqnarray}
& & N_{\mathrm{x2}} \simeq 1 - \frac{\sqrt{3} \alpha m^2}{14 \pi ^2 k_0 ^2} (3 \kappa )^{2/3} \Gamma ^4 \left( \frac{2}{3} \right) \left( 1 - i \sqrt{3} \right), \label{high_energy_limit_x2} \\
& & N_{\mathrm{x3}} \simeq 1 - \frac{3 \sqrt{3} \alpha m^2}{28 \pi ^2 k_0 ^2} ( 3 \kappa )^{2/3} \Gamma ^4 \left( \frac{2}{3} \right) \left( 1 - i \sqrt{3} \right) , \ \label{high_energy_limit_x3}
\end{eqnarray}
for the x2 and x3 modes, respectively. In our formulation, these results
are reproduced by putting $e^{-im^2 s}$ to unity and setting $(kk) = k_0 ^2 - k_1 ^2$ 
equal to zero in Eqs.~(\ref{pi1}), (\ref{pi2}) and~(\ref{pi3})
for the polarization tensor ${\Pi_{\mu}}^{\nu}$
or Eqs.~(\ref{termK}) and ~(\ref{current_crossed}) for the induced electromagnetic
current $\langle j_{\mu} \rangle$.
Note, however, that our numerical results for $\mathrm{Re} [N-1]$ are not yet settled
to the asymptotic limits with deviations of $\sim 10\%$ still remaining at $k_0/m \sim 1000$.
In this figure, the high energy limits for the x2 and x3 modes are displayed as the lines
labeled as ax2 and ax3, respectively.
The imaginary parts, on the other hand, have already reached the asymptotic limits
at $k_0/m \sim 1000$ (see below).

\begin{figure}[htbp]
 \begin{center}
  \includegraphics[width=8.6cm]{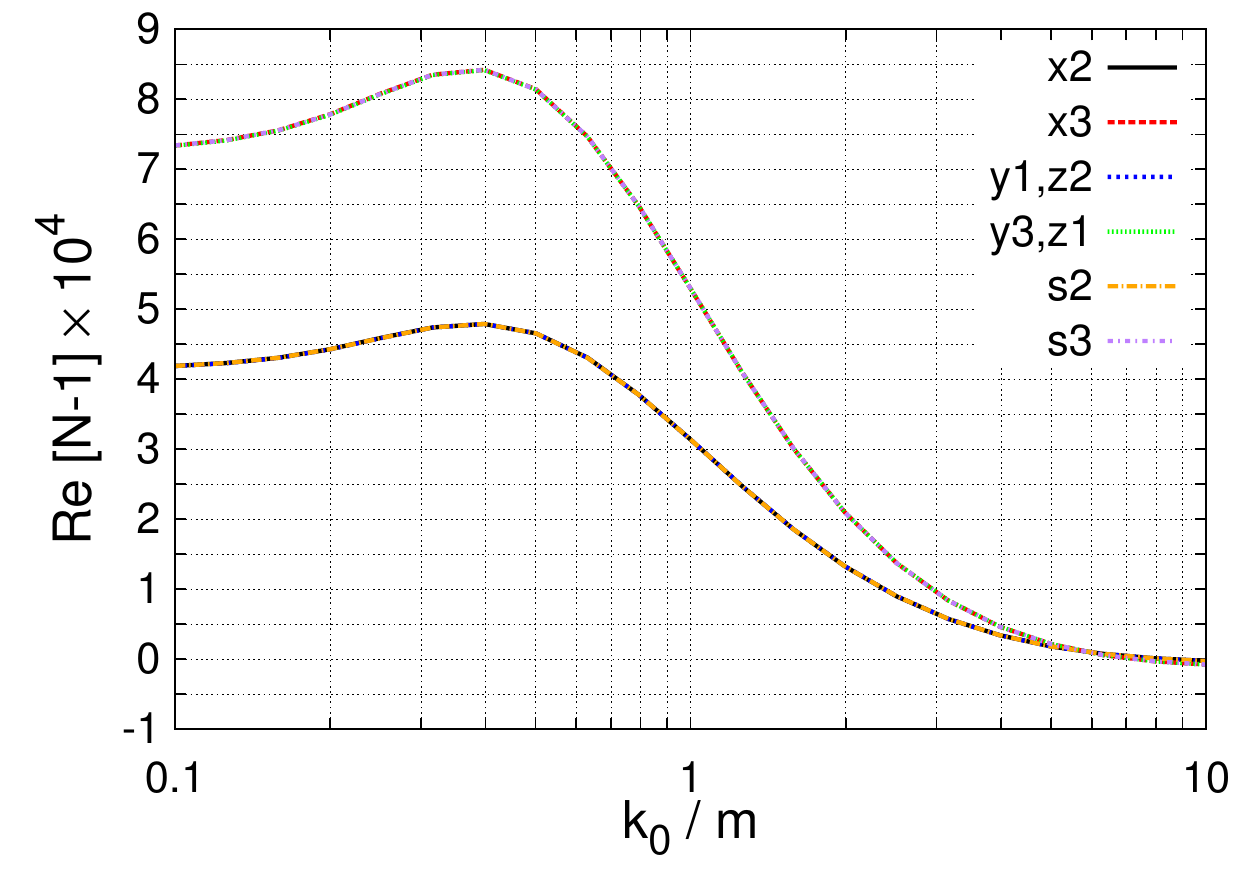}
  \caption{Same figure as Fig.~\ref{graph_crossed_field_a1_b_re_all} but for $f_{\mathrm{rd}} = f_c$ in the energy range of $0.1 \leq k_0 /m \leq 10$ for all the modes.}
  \label{graph_crossed_field_a1_b_rd}
 \end{center}
\end{figure}

Toll~\cite{1952PhDT........21T} pointed out that unless the Poynting vectors
of the probe photon and the external field are parallel to each other,
an appropriate Lorentz transformation makes them anti-parallel
and, as a result, the refractive index depends only on the reduced field
strength $f_{\mathrm{rd}}$
\begin{eqnarray}
 f_{\mathrm{rd}} = f \sin ^2 \left( \frac{\theta}{2} \right) 
\end{eqnarray}
as long as the field strength is not much larger than the critical value.
Here $\theta$ is the angle between the Poynting vectors of the probe photon
and the external field.
We hence redraw Fig.~\ref{graph_crossed_field_a1_b_re_all} as
Fig.~\ref{graph_crossed_field_a1_b_rd} in the range of $0.1 \leq k_0 /m \leq 10$
after adjusting the external-field strength so that $f_{\mathrm{rd}} = f_c$
for all the modes.
As expected, the x3, s3, y3 and z1 modes become identical, which is also true
for the x2, s2, y1 and z2 modes. The relation also holds for the imaginary part.
It is important that these relations are obtained as a result of separate calculations
for different propagation directions in our formulation,
the fact that guarantees the correctness of our calculations.

\begin{figure}[htbp]
 \begin{center}
  \includegraphics[width=8.6cm]{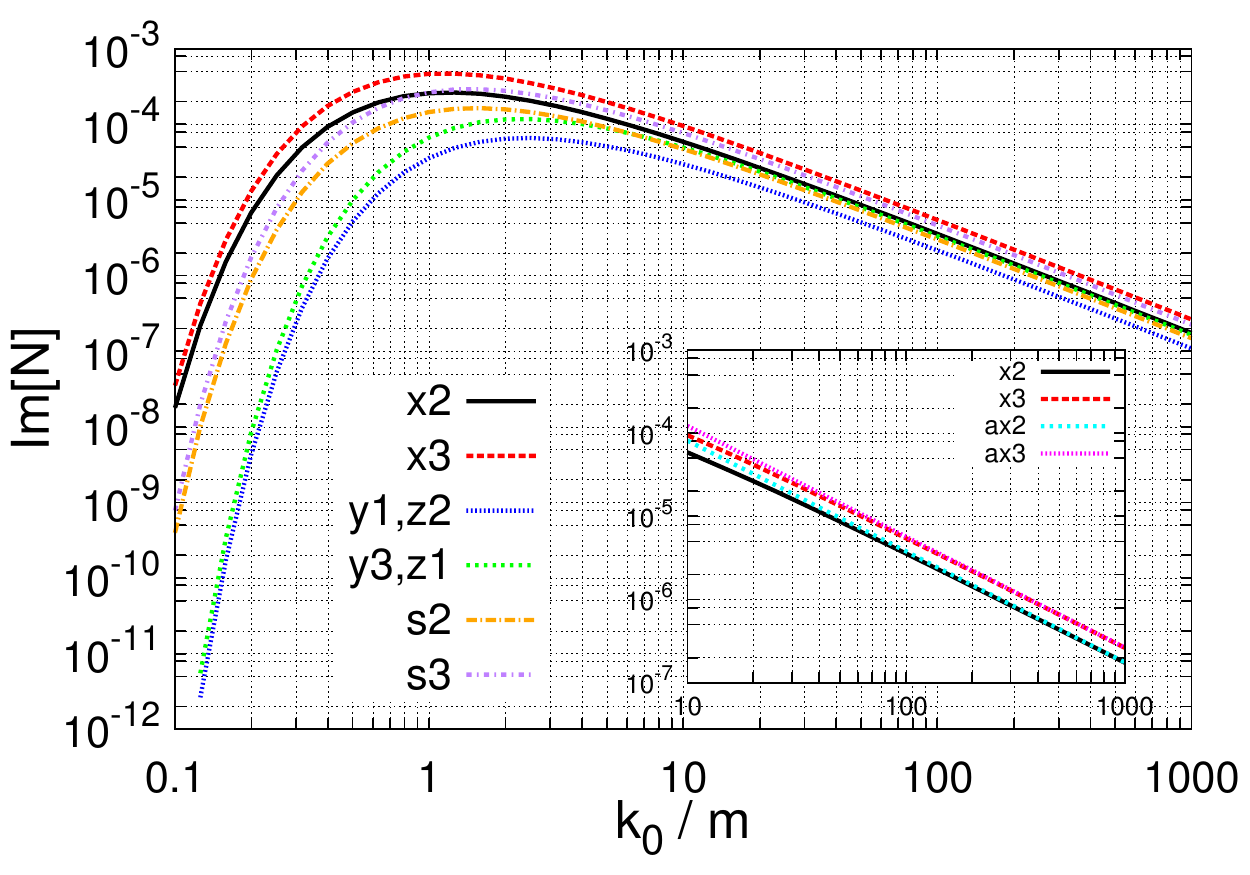}
  \caption{ Same as Fig.~\ref{graph_crossed_field_a1_b_re_all} but for the imaginary part of refractive index $\mathrm{Im} [N]$. The inset shows the behavior in the high-energy regime as in Fig.~\ref{graph_crossed_field_a1_b_re_high_energy}. The lines labeled as ax2 and ax3 show the high-energy limit expressed as Eqs.~(\ref{high_energy_limit_x2}) and (\ref{high_energy_limit_x3}), respectively.}
  \label{graph_crossed_field_a1_b_im}
 \end{center}
\end{figure}

\begin{figure}[htbp]
 \begin{center}
  \includegraphics[width=8.6cm]{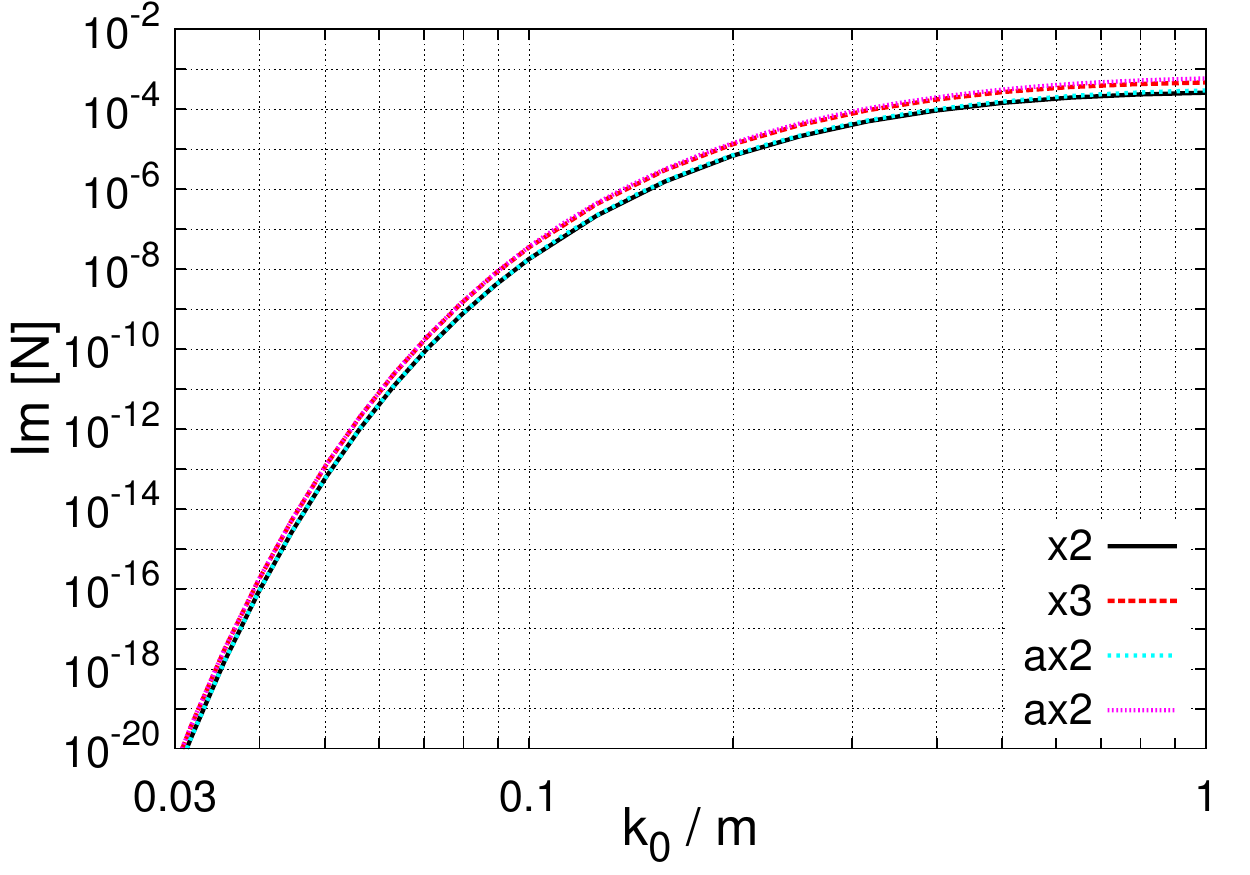}
  \caption{Same as Fig.~\ref{graph_crossed_field_a1_b_im} but for the low-energy range. The lines labeled as ax2 and ax3 show the weak-field or low-energy limits expressed as Eqs.~(\ref{weak_imaginary_x2}) and~(\ref{weak_imaginary_x3}), respectively.}
  \label{graph_approx_a1_b_im}
 \end{center}
\end{figure}

The imaginary part of the refractive index $\mathrm{Im} [N]$ is
shown in Fig.~\ref{graph_crossed_field_a1_b_im} for the same case.
It is found that the imaginary part is non-vanishing down to $k_0=0$ although
it diminishes very rapidly for $k_0/m \lesssim 0.1$.
It is also seen that $\mathrm{Im} [N]$ for each photon mode reaches its maximum
at $k_0 /m \sim 1$ and it decreases monotonically for higher energies.
These behaviors are also consistent with the known limits~\cite{narozhnyi69}.
In fact, as mentioned above, they are already settled to the asymptotic values at $k_0/m \sim 1000$
as shown in the inset of the figure. 
The imaginary parts $\mathrm{Im}[N]$ for different modes follow the general trend mentioned
earlier for $|\mathrm{Re} [N-1]|$ with the x3 mode being the largest
and the y1/z2 being the smallest except around $k_0/m \sim 1$, where some crossings occur. 

The imaginary part of the refractive index in the weak-field or low-energy
(region~(2) in Fig.~\ref{region}) was considered in \cite{ritus72,heinzl}.
Although we cannot obtain the analytic expression,
we try to compute the imaginary part numerically in this regime.
The results are displayed in Fig.~\ref{graph_approx_a1_b_im} for the x2 and x3 modes
in the range of $0.03 \lesssim k_0 /m \leq 1$.
The lines labeled as ax2 and ax3 are the results obtained in~\cite{ritus72},
which are expressed as
\begin{eqnarray}
 \mathrm{Im} [N_{\mathrm{x2}}] \simeq \frac{1}{8} \sqrt{\frac{3}{2}} \frac{\alpha \epsilon}{\nu} e^{- \frac{4}{3 \epsilon \nu}}, \label{weak_imaginary_x2} \\
 \mathrm{Im} [N_{\mathrm{x3}}] \simeq \frac{1}{4} \sqrt{\frac{3}{2}} \frac{\alpha \epsilon}{\nu} e^{- \frac{4}{3 \epsilon \nu}} , \label{weak_imaginary_x3}
\end{eqnarray}
where $\epsilon = f /f_c$ and $\nu = k_0 /m$.
It is found that the imaginary parts $\mathrm{Im} [N]$ are better approximated in this regime
by Eqs.~(\ref{weak_imaginary_x2}) and (\ref{weak_imaginary_x3}) rather than by
\begin{eqnarray}
 \mathrm{Im} [N_{\mathrm{x2}}] \simeq \frac{4 \alpha \epsilon ^2}{45} \frac{4}{3 \epsilon \nu} e^{- \frac{4}{3 \epsilon \nu}} ,\\
 \mathrm{Im} [N_{\mathrm{x3}}] \simeq \frac{7 \alpha \epsilon ^2}{45} \frac{4}{3 \epsilon \nu} e^{- \frac{4}{3 \epsilon \nu}},
\end{eqnarray}
obtained in~\cite{heinzl}.

\begin{figure}[htbp]
 \begin{center}
  \includegraphics[width=8.6cm]{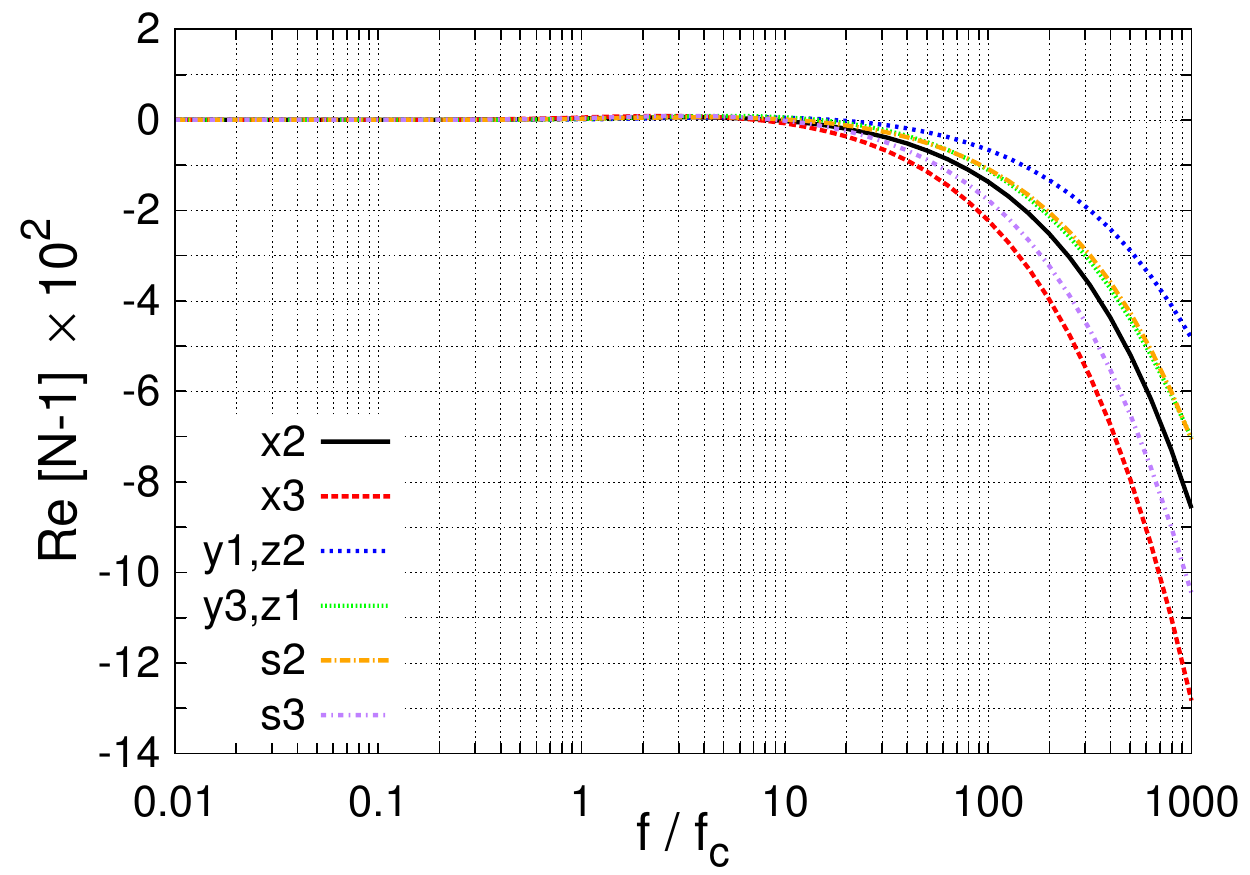}
  \caption{Plot of $\mathrm{Re} [N-1]$ as a function of the field strength. We assume $k_0 /m = 1$ this time.}
  \label{graph_crossed_field_b1_a_re_all}
 \end{center}
\end{figure}

\begin{figure}[htbp]
 \begin{center}
  \includegraphics[width=8.6cm]{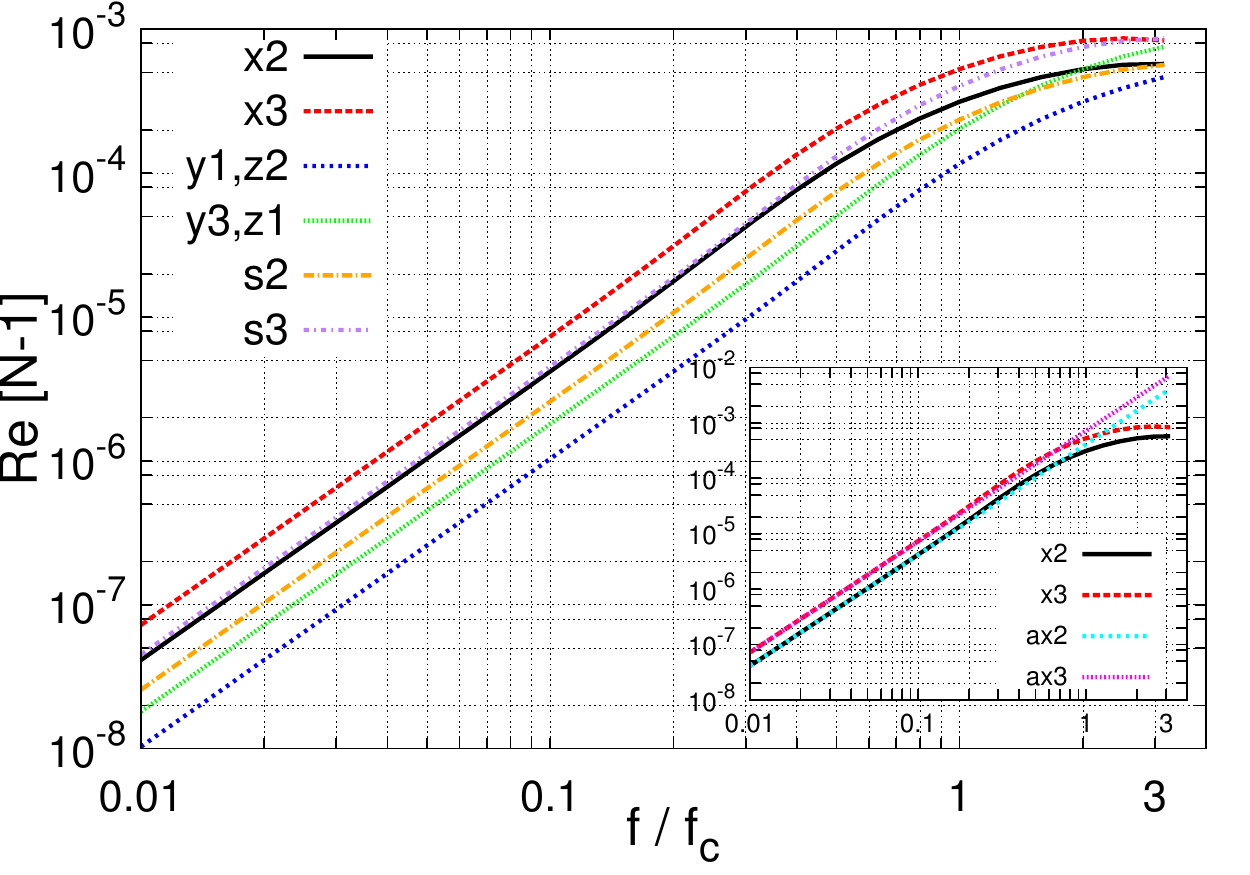}
  \caption{Same as Fig.~\ref{graph_crossed_field_b1_a_re_all} but for weak fields. The inset shows the comparison between our numerical results and asymptotic expressions, Eqs.~(\ref{weak_re_x2}) and (\ref{weak_re_x3}), labeled as ax2 and ax3 for the x2 and x3 modes, respectively, in the weak-field or low-energy limits.}
  \label{graph_crossed_field_b1_a_re_weak}
 \end{center}
\end{figure}

Next we show the dependence of the refractive index on the external-field strength,
setting $k_0/m=1$. This has never been published in the literature before.
In Fig.~\ref{graph_crossed_field_b1_a_re_all}, $\mathrm{Re} [N-1]$
is shown as a function of $f/f_c$ in the range of $0.01 \leq f/f_c \leq 1000$. Figure 
\ref{graph_crossed_field_b1_a_re_weak}
zooms in to the range of $0.01 \leq f/f_c \leq 3$,
setting the vertical axis in the logarithmic scale. 
The quadratic behavior observed for $0.01 \leq f/f_c \lesssim 0.5 $
is in accord with the weak-field or low-energy limits~\cite{narozhnyi69},
which are given as ax2 and ax3 for the x2 and x3 modes in the inset of this figure, respectively.
$\mathrm{Re} [N-1]$ is negative at $f/f_c \gtrsim 10$,
which is consistent with the earlier findings.
The modulus $| \mathrm{Re} [N-1] |$ is an increasing function of $f$
at $f/f_c \gtrsim 10$.

\begin{figure}[htbp]
 \begin{center}
  \includegraphics[width=8.6cm]{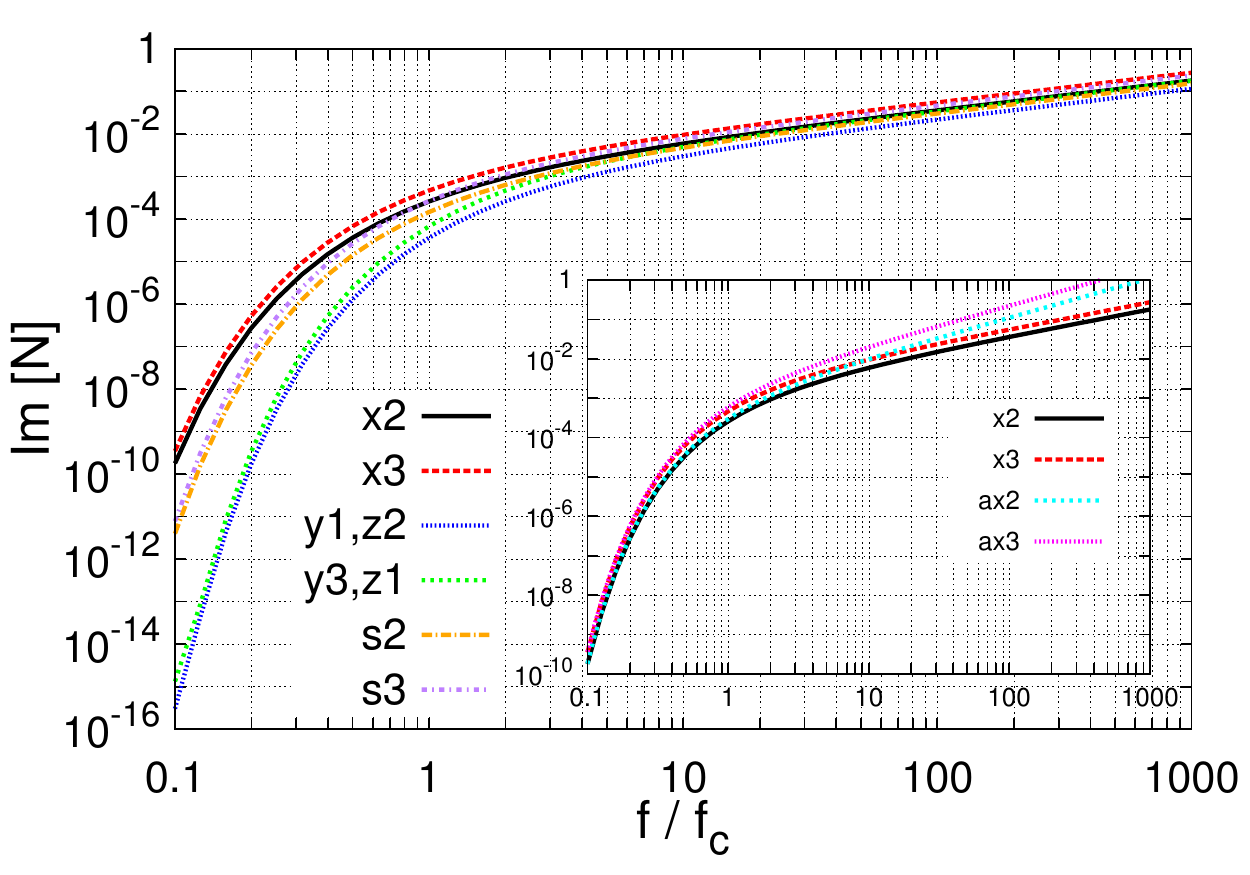}
  \caption{Same figure as Fig.~\ref{graph_crossed_field_b1_a_re_all} but for the imaginary part of refractive index $\mathrm{Im} [N]$. The inset shows the comparison between our numerical results and the asymptotic expressions in the weak-field or low-energy limits, Eqs.~(\ref{weak_imaginary_x2}) and (\ref{weak_imaginary_x3}), labeled as ax2 and ax3 for the x2 and x3 modes, respectively.}
  \label{graph_crossed_field_b1_a_im}
 \end{center}
\end{figure}

The imaginary part $\mathrm{Im} [N]$ is shown in Fig.~\ref{graph_crossed_field_b1_a_im}. 
It increases monotonically with the external-field strength.
The slopes are steeper at $f/f_c \lesssim 0.5$, which is consistent with the analytic expression
in the weak-field or low-energy limit of $\mathrm{Im} [N]$~\cite{ritus72,heinzl}. The inset of this
figure shows the comparison of our numerical results with the asymptotic limits,
Eqs.~(\ref{weak_imaginary_x2}) and (\ref{weak_imaginary_x3}), labeled as ax2 and ax3
for the x2 and x3 modes, respectively. They almost coincide with each other at $f/f_c \lesssim 0.5$.
Note, on the other hand, that the behavior of the imaginary part at high field-strengths
has not been reported in the literature.

\subsection{Plane-Wave}
We next consider the ``local'' refractive index for the plane wave field, which is also
original in this paper.
We evaluate numerically the polarization tensor is given in Eqs.~(\ref{polarization_tensor}), (\ref{pi1})-(\ref{pi3}) and solve the Maxwell equation, Eq.~(\ref{maxwell_equation}), obtained in the gradient expansion.
Since our formulation is based on the perturbation theory, it is natural to express
the refractive index in the plane wave as $N + \delta N$,
where $N$ is the refractive index for the crossed field and $\delta N$ is the
correction from the temporal and spatial non-uniformities.
As mentioned for the crossed field, the refractive indices for the y1 and z2 modes
are identical to each other.
In fact, the relevant components of the Maxwell equations,
Eq.~(\ref{maxwell_equation}), are the same for these modes.
This is also true for the y3 and z1 modes.

It is found that the correction $\delta N$ starts indeed with the linear order of $\Omega/m$ for both
the real and imaginary part. It is then written as 
\begin{eqnarray}
 \delta N = ( C_{\mathrm{Re}} + i C_{\mathrm{Im}} ) \times \Omega /m + O( (\Omega /m) ^2)
\end{eqnarray}
and the numerical values of the coefficients $C_{\mathrm{Re}}$ and $C_{\mathrm{Im}}$
are given for $k_0/m=1$ and $f/f_c=1$ in Table~\ref{fitting}.
The temporal and spatial variations are found to mainly affect the imaginary part:
$| \mathrm{Im} [ \delta N] | > | \mathrm{Re} [ \delta N]|$ from these results.
It is also seen that $\mathrm{Im} [\delta N]$ is larger for the
photons propagating in the opposite direction to the external plane-wave ($x$-direction)
as in the crossed field limit.
The real parts $\mathrm{Re} [\delta N]$ are negative for photons other than those
propagating perpendicularly to the external plane-wave.
The modulus $| \mathrm{Re} [N + \delta N ]|$
is hence reduced for these modes by the field variation.

\begin{table}
 \caption{Proportionality coefficients in the correction $\delta N$ from temporal and spatial non-uniformities ${}^7$}
\begin{center}
\begin{tabular}{ccc} \hline
mode & $C_{\mathrm{Re}}$ & $C_{\mathrm{Im}}$ \\
\hline
x2 & $-1.30 \times 10^{-3}$  & $3.16 \times 10^{-3}$  \\
x3 & $-3.08 \times 10^{-3}$ & $5.17 \times 10^{-3}$ \\
y1,z2  & $1.42 \times 10^{-4}$ & $4.35 \times 10^{-4}$ \\
y3,z1 & $1.83 \times 10^{-4}$  & $8.28 \times 10^{-4}$ \\
s2 & $-3.10 \times 10^{-4}$ & $1.79 \times 10^{-3}$ \\
s3 & $-9.69 \times 10^{-4}$  & $3.11 \times 10^{-3}$ \\ \hline
   \\[-12pt]
 \multicolumn{3}{l}{${}^7$ $k_0/m = 1$ and $f/f_c = 1$.}
\end{tabular}
\end{center}
\label{fitting}
\end{table}

\begin{figure}[htbp]
 \begin{center}
  \includegraphics[width=8.6cm]{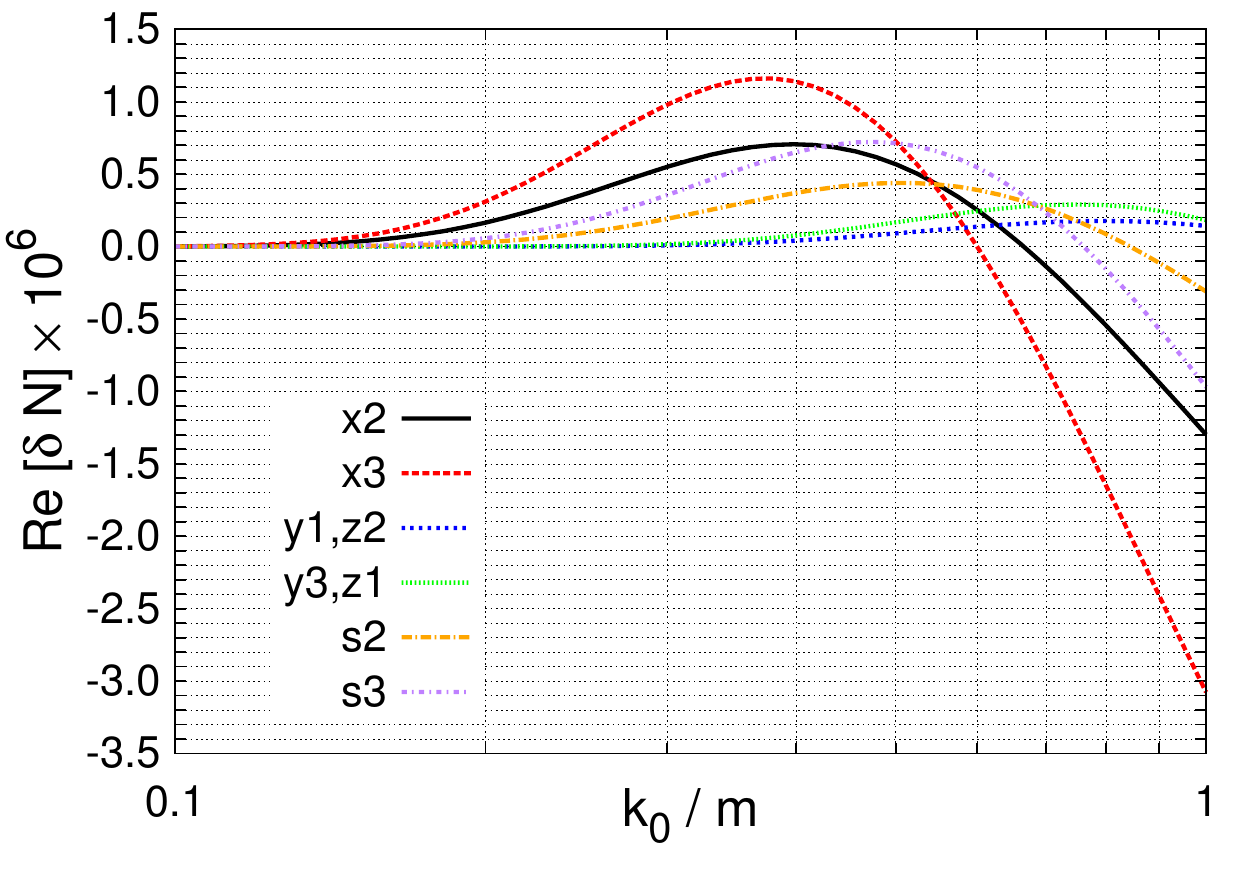}
  \caption{Plot of the correction to the real part of the refractive index for the crossed field from the temporal and spatial variations in the plane-wave field. The field strength is set to the critical value, i.e., $f/f_c=1$ and the frequency of the external wave field is chosen as $\Omega /m =10^{-3}$.}
  \label{threshold_a1_first_re}
 \end{center}
\end{figure}

\begin{figure}[htbp]
 \begin{center}
  \includegraphics[width=8.6cm]{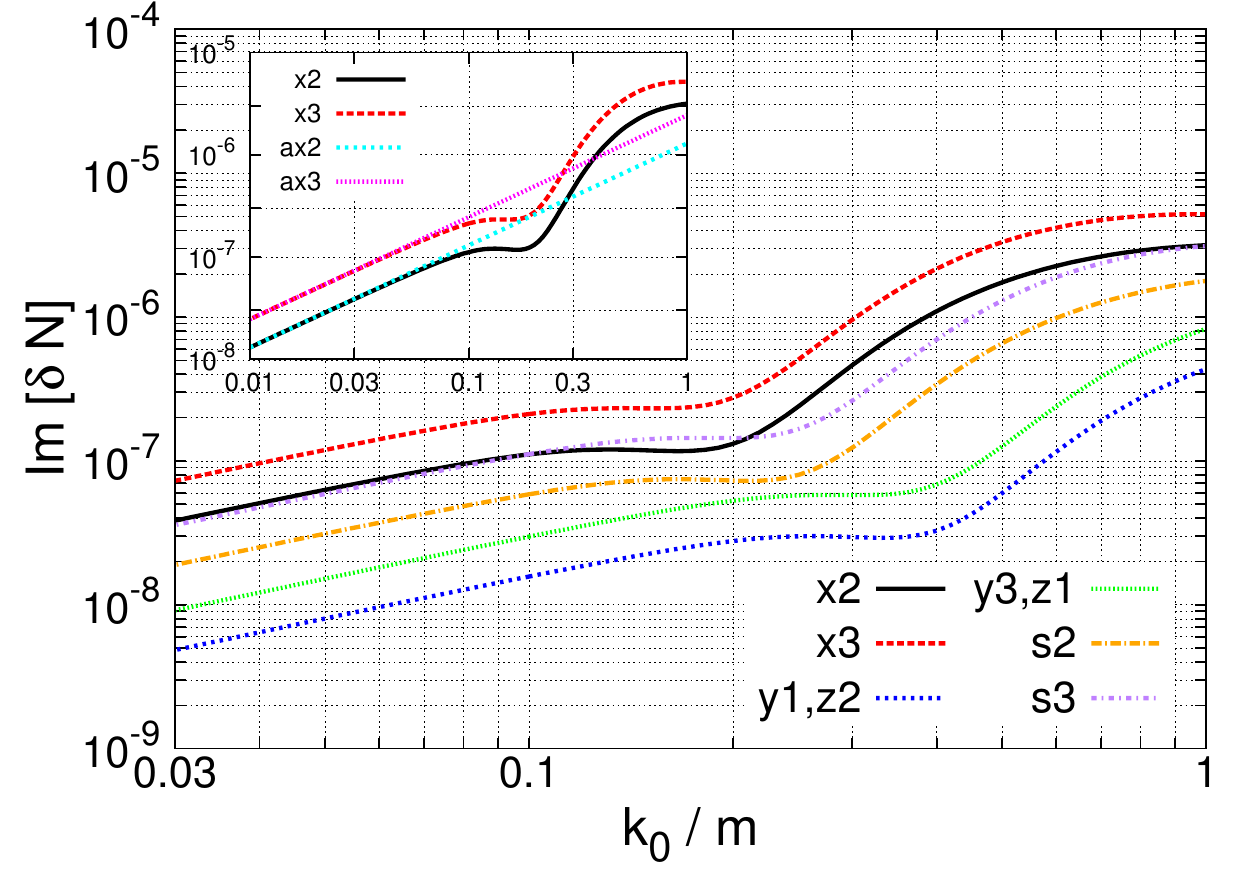}
  \caption{Same as Fig.~\ref{threshold_a1_first_re} but for $\mathrm{Im} [\delta N]$. The inset shows the comparison for the x2 and x3 modes between the asymptotic limits (Eqs.~(\ref{Im_delta_N_x2}) and (\ref{Im_delta_N_x3}) labeled as ax2 and ax3) and the numerically computed results.}
  \label{threshold_a1_first_im}
 \end{center}
\end{figure}

We next present the dependence on $k_0 /m$ of $\delta N$ for $f /f_c= 1$, $\Omega /m = 10^{-3}$
in Figs.~\ref{threshold_a1_first_re} and~\ref{threshold_a1_first_im}.
The real part $\mathrm{Re} [\delta N]$ is exhibited in Fig.~\ref{threshold_a1_first_re}.
It is seen that the real part can be both positive and negative:
it tends to be negative at higher values
of $k_0/m$ although the range depends on the mode; in fact, the
values of the photon energy, above which $\delta N$ gets positive,
are smaller for the photons propagating oppositely
to the external plane-wave.
$\mathrm{Re} [\delta N]$ is much smaller than $\mathrm{Re} [N-1]$
for the crossed field at $0.1 \leq k_0/m \leq 1$ and decreases very rapidly like
$\mathrm{Im} [N]$ for the crossed field.

\begin{figure}[htbp]
 \begin{center}
  \includegraphics[width=8.6cm]{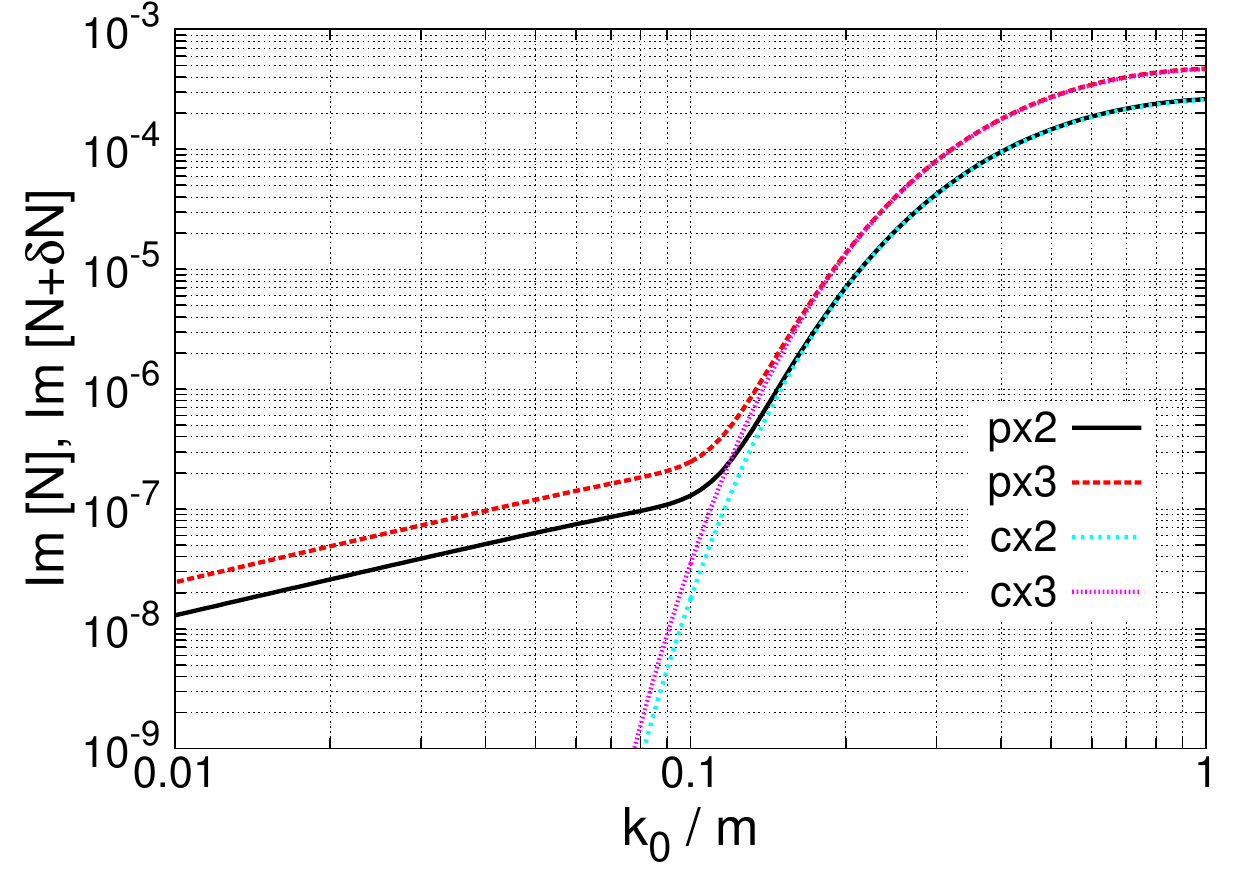}
  \caption{Comparison for the x2 and x3 modes of $\mathrm{Im} [N + \delta N]$ labeled as px2 and px3, respectively, and $\mathrm{Im} [N]$ labeled as cx2 and cx3, respectively, as a function of the probe-photon energy, $k_0/m$. Here we set $f/f_c = 1$ as in the previous two figures.}
  \label{threshold_a1_first_im_comparison}
 \end{center}
\end{figure}

The imaginary part $\mathrm{Im} [\delta N]$ is shown in Fig.~\ref{threshold_a1_first_im}
for the probe-photon energies of $0.03 \leq k_0 /m \leq 1$.
The inset indicates the comparison for the x2 and x3 modes between the numerically computed results of
$\mathrm{Im} [ \delta N]$ and the asymptotic values in the weak-field or low-energy limits.
The expressions of $\mathrm{Im} [ \delta N]$ in this regime are given from
Eqs.~(\ref{pi1}) and (\ref{pi2}) as
\begin{eqnarray}
& &  \mathrm{Im} [ \delta N_{\mathrm{x2}} ] \simeq \frac{263 \alpha}{3780 \pi} \left( \frac{\Omega k_{\mu} n^{\mu}}{m^2} \right) \frac{\kappa ^2 m^2}{k_0 ^2} , \label{Im_delta_N_x2} \\
 & & \mathrm{Im} [ \delta N_{\mathrm{x3}} ] \simeq \frac{71 \alpha}{540 \pi} \left( \frac{\Omega k_{\mu} n^{\mu}}{m^2} \right) \frac{\kappa ^2 m^2}{k_0 ^2}, \label{Im_delta_N_x3}
\end{eqnarray} 
where $\Omega k_{\mu} n^{\mu} / m^2 = \Omega ( k_0 + k_1 ) / m^2$ is the product of the momentum of the external plane-wave and that of the probe photon normalized by the electron mass and is a representative term in the gradient expansion $F_{\mu \nu} \sim f_{\mu \nu} (1+ \Omega \xi)$, being proportional to $\Omega$ with the proportional factor $k_{\mu} n^{\mu}$ originating from the commutation relation of $\xi$ that accompanies $\Omega$; $\kappa ^2 = e^2 k_{\mu} {f^{\mu}}_{\nu} f^{\nu \lambda} k_{\lambda} /m^6 = e^2 f^2 (k_0 + k_1 )^2 /m^6$ as previously defined in Eq.~(\ref{weak_re_x2}).
Equations~(\ref{Im_delta_N_x2}) and (\ref{Im_delta_N_x3}) are convenient for the evaluation of the typical value of $\mathrm{Im} [\delta N]$:
\begin{eqnarray}
 \mathrm{Im} [ \delta N]  &\sim& \alpha \left( \frac{\Omega}{m} \right) \left( \frac{k_0}{m} \right) \left( \frac{f}{f_c} \right)^2 \nonumber \\
 &\sim& 7 \times 10^{-6} \left( \frac{\Omega}{0.5 \mathrm{keV}} \right) \left( \frac{k_0}{510 \mathrm{keV}} \right) \left( \frac{I}{4.6 \times 10^{29} \mathrm{W/cm^2}}\right) ,
\end{eqnarray}
where $I$ is the intensity of the external electromagnetic wave.
It is found from the inset that $\mathrm{Im} [\delta N]$ is well approximated
for $\Omega / m = 10^{-3}$ by the asymptotic expressions at $k_0 / m \leq 0.03$
for $f/f_c = 1$.
There occurs a dent at $k_0 / m \simeq 0.2$ and $\mathrm{Im} [\delta N]$
rises more rapidly with $k_0/m$ at larger energies, where $\mathrm{Im} [N]$
of the crossed field also becomes substantial.
The location of the dent depends on the propagation direction of the probe photon,
with the x (y/z) mode having the smallest (largest) value of $k_0/m$ at the dent, respectively.

Since the imaginary part of the refractive index declines rapidly below these energies
for the crossed field, it is dominated by the first-order correction $\mathrm{Im} [\delta N]$
from the temporal and spatial variations in the plane-wave at these low energies.
In fact, the latter is commonly more than 10 times larger than the former
$\mathrm{Im} [\delta N] \gtrsim 10 \times \mathrm{Im} [N]$ at $k_0 /m \lesssim 0.1$.
See also Fig.~\ref{threshold_a1_first_im_comparison}, where we plot $\mathrm{Im} [N+ \delta N]$
and $\mathrm{Im} [N]$ as a function of $k_0/m$.
This is especially the case of the probe photons propagating
transversally to the background plane-wave.
In accordance with the trend for the crossed field,
the x (y/z) modes have largest (smallest) moduli $|\mathrm{Im} [ \delta N]|$
and s modes come in between in general.

\begin{figure}[htbp]
 \begin{center}
  \includegraphics[width=8.6cm]{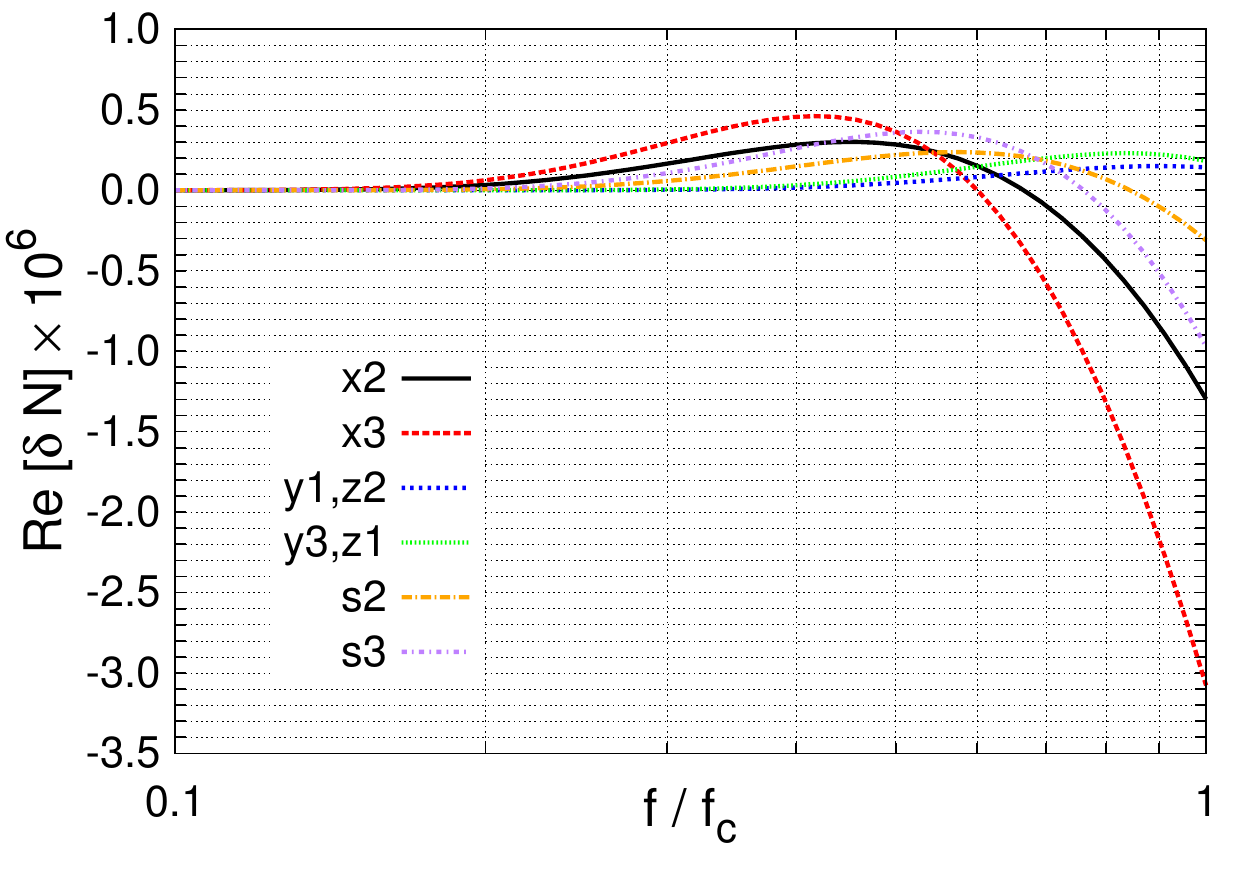}
  \caption{$\mathrm{Re} [\delta N]$ for the external plane-wave field of the frequency of $\Omega /m = 10^{-3}$. The probe-photon energy is chosen as $k_0/m = 1$.}
  \label{threshold_b1_first_re}
 \end{center}
\end{figure}

\begin{figure}[htbp]
 \begin{center}
  \includegraphics[width=8.6cm]{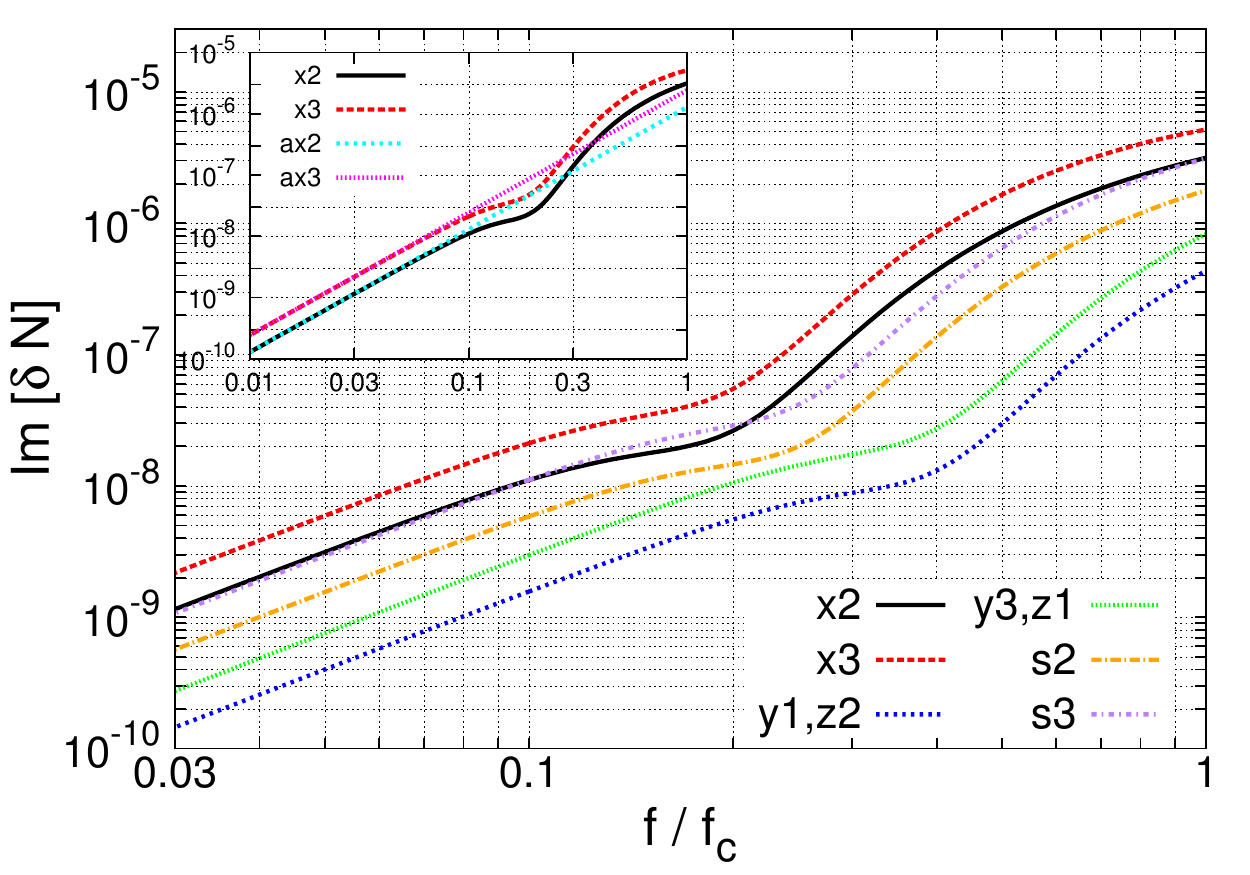}
  \caption{Same figure as Fig.~\ref{threshold_b1_first_re} but for $\mathrm{Im} [\delta N]$. The inset shows the comparison for the x2 and x3 modes between the asymptotic limits (Eqs.~(\ref{Im_delta_N_x2}) and (\ref{Im_delta_N_x3})) labeled as ax2 and ax3, respectively, and the numerically computed results for the x2 and x3 modes labeled as x2 and x3, respectively.}
  \label{threshold_b1_first_im}
 \end{center}
\end{figure}

\begin{figure}[htbp]
 \begin{center}
  \includegraphics[width=8.6cm]{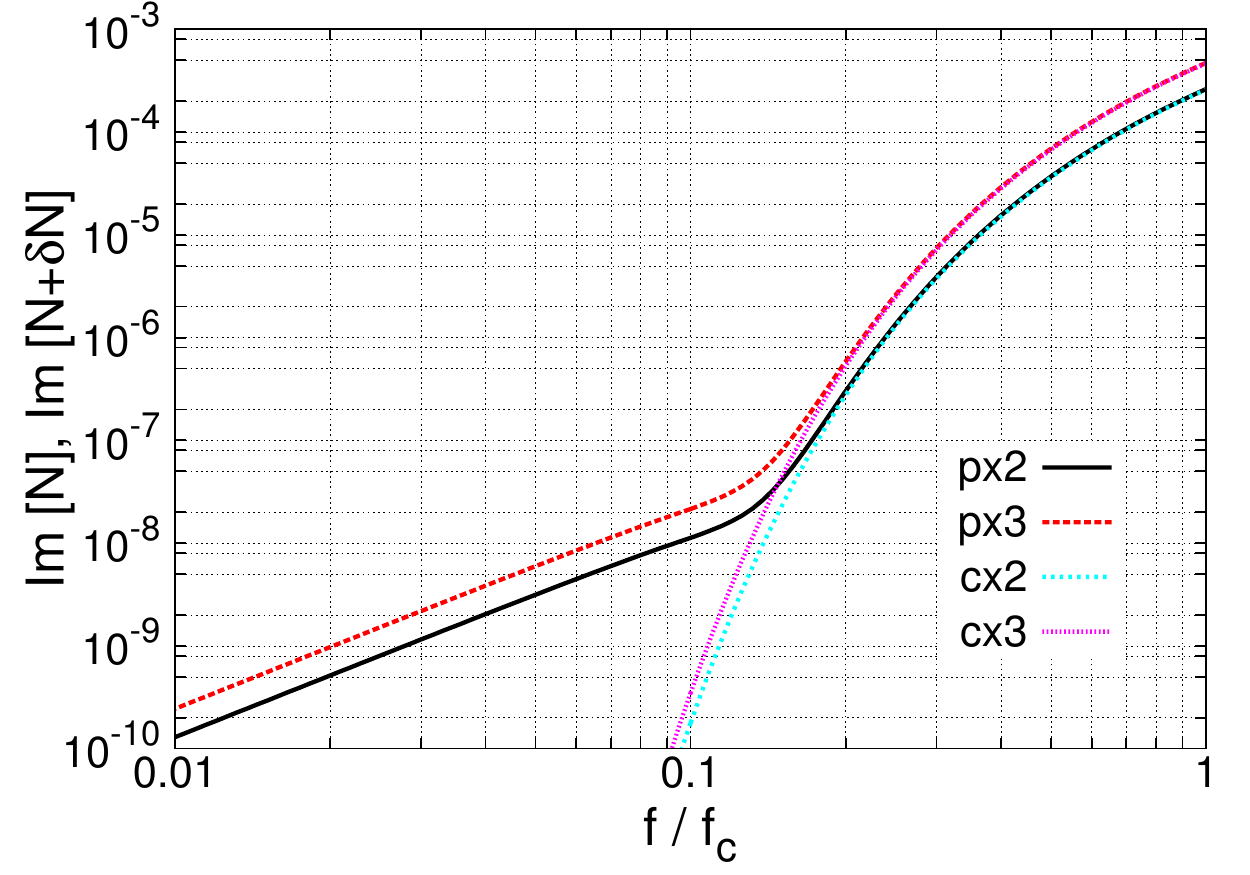}
  \caption{Same as Fig.~\ref{threshold_a1_first_im_comparison} but as a function of field strength, $f/f_c$. Here $k_0/ m$ is set to 1 as in the previous two figures.}
  \label{threshold_b1_first_im_comparison}
 \end{center}
\end{figure}

Finally, we look into the dependence of $\delta N$ on the field strength in the range of
$f/ f_c \leq 1$. The real and imaginary parts of $\delta N$
are shown in Figs.~\ref{threshold_b1_first_re} and \ref{threshold_b1_first_im}, respectively.
The probe-photon energy is set to $k_0 /m = 1$ and the frequency of the external field
assumed to be $\Omega /m = 10^{-3}$ again, though the results scale with the latter linearly.
It is evident that the results are quite similar to those shown in
Figs.~\ref{threshold_a1_first_re} and~\ref{threshold_a1_first_im}:
the real part, $\mathrm{Re} [\delta N]$, has a hump at $f/f_c \sim 0.5$
whereas the imaginary part, $\mathrm{Im} [\delta N]$,
is quadratic in $f/f_c$ at weak fields and becomes dominant over the crossed-field contribution,
$\mathrm{Im} [N]$, at $f/f_c \lesssim 0.1$;
the order in the magnitudes of $\mathrm{Im} [ \delta N]$ for different modes is the same as
that in Fig.~\ref{threshold_a1_first_re};
$\mathrm{Re} [\delta N]$ is negative at a certain range of $f/f_c$, which depends on the mode,
occurring for stronger fields for the mode propagating transversally to the background plane-field.
The reason for these behaviors is the following: although $\delta N$ depends
not only on the product of $k_0/m$ and $(f/f_c)^2$ but also on $k^{\mu} k_{\mu} /m^2$,
the latter dependence is minuscule in the regime we consider here.
As a result, the dependence of the refractive index on $k_0/m$ can be translated
into that of $f/f_c$.
In fact, the numerical results for $\mathrm{Im} [\delta N]$ are well-approximated
by the same asymptotic formulae, Eqs.~(\ref{Im_delta_N_x2}) and (\ref{Im_delta_N_x3}),
in the weak-field regime $f/f_c \lesssim 0.1$,
which can be seen in the inset of Fig.~\ref{threshold_b1_first_im};
$\mathrm{Im} [\delta N]$ has a dent at $f/f_c \sim 0.2$ and changes its behavior
at larger field-strengths, where the crossed-field contribution, $\mathrm{Im} [N]$,
becomes large, overwhelming $\mathrm{Im} [\delta N]$. See also Fig.~\ref{threshold_b1_first_im_comparison}, where we plot $\mathrm{Im} [N + \delta N]$ and $\mathrm{Im} [N]$ as a function of $f/f_c$.

\section{Summary and Discussion}

In this paper we have developed a perturbation theory adapted to Schwinger's proper-time method
to calculate the induced electromagnetic current, which should be plugged into the Maxwell equations
to obtain the refractive indices, for the external, linearly polarized plane-waves,
considering them as the unperturbed states and regarding a probe photon as the perturbation to them.
Although this is nothing new and indeed was already employed previously~\cite{Adler71},
our formulation is based on the interaction picture, a familiar tool in quantum mechanics
and referred to also as the Furry picture in strong-field QED,
rather than utilizing the properties of particular electromagnetic fields from the beginning. 
Moreover, assuming that the wavelength of the external plane-wave is much longer than
the Compton wavelength of electron and employing the gradient expansion,
we have evaluated locally the polarization tensor
via the induced electromagnetic current to the lowest order of the spatial
and temporal variations of the external fields, which is the main achievement in this paper.
It has been shown that the vacuum polarization is given locally by the field strength
and its gradient of the external plane-waves at each point.
We have then considered the dispersion relations for the probe photons propagating
in various directions and derived the local refractive indices.

We have first evaluated them for the crossed fields,
which are the long-wavelength limit of the plane-waves.
In so doing, the field strength and the energy of the probe photon are not limited
but are allowed to take any values.
Note that even for the crossed field not all the parameter regime has been investigated
and we have explored those portions unconsidered so far. 
We have shown that the
refractive index is larger for the photons propagating oppositely
to the external field than for those propagating perpendicularly.
We have also confirmed some limiting cases that were already known
in the literature analytically or numerically~\cite{Baier1967b,narozhnyi69,ritus72,heinzl},
particularly
the behavior in the weak external fields demonstrated in~\cite{1952PhDT........21T}.
Note, however, that the assumption of a fixed classical background field becomes rather questionable at field strengths near or, in particular, above the critical field strength, since the back reactions to the background field from pair creations should be then taken into account. This issue is certainly much beyond the scope of this paper and in spite of this conceptual problem we think that the results in such very strong fields are still useful to understand the scale and qualitative behavior of the corrections from the field gradient. 

We have then proceeded to the evaluation of the refractive index for the plane-wave to
the lowest order of the temporal and spatial variations
of the background field.
The local correction $\delta N$ to the refractive index
for the crossed field $N$ has been numerically evaluated for the first time.
We have demonstrated that the modulus of its imaginary part is larger than that of the
real part, i.e., the field variations mainly affect
the imaginary part of the refractive index.
Note that the refractive index we have obtained in this study is local, depending on the local field-strength and its gradient, and is meaningful in the sense of the WKB approximation.
This is in contrast to the refractive index averaged over the photon path in~\cite{2014PhRvD..89l5003D}.

In the optical laser experiments ($\Omega /m \sim 10^{-6}$), the refractive index may be
approximated very well by that for the crossed field.
The correction from the field variations is typically $|\delta N| \sim 10^{-5} \times \mathrm{Re} [N -1]$.
The weak-field limit may be also justified, since the current maximum
laser-intensity $\sim 2.0 \times 10^{22} \mathrm{W/cm^2}$ is still much lower than
the critical value, $4.6 \times 10^{29} \mathrm{W/cm^2}$.
Then the numerical results given in
Fig.~\ref{graph_crossed_field_b1_a_re_weak} are applicable:
$\mathrm{Re} [N-1] \sim 10^{-4} \times (f/f_c)^2$ for the probe photon with $k_0/m = 1$,
which corresponds to $\sim 10^{-8}$ at $\sim 10^{25} \mathrm{W/cm^2}$,
the power expected for future laser facilities such as ELI.
Note that how to observe the local refractive index in the electromagnetic wave is
a different issue and the averaged one will be better
suited for experiments~\cite{2014PhRvD..89l5003D}.

Unlike for the optical laser, the field variations may not be ignored for x-ray lasers
with $\Omega /m \simeq 10^{-2}$.
We find from Fig.~\ref{graph_crossed_field_b1_a_re_weak} and Table~\ref{fitting}
that the refractive index for the crossed field and the first-order correction to it are
$|N-1| \sim 10^{-4}$ and $|\delta N| \sim 10^{-5}$, respectively,
for the probe photon with $k_0 /m =1$ propagating oppositely to the external fields
with the critical field strength.
It may be more interesting that the imaginary part of the first-order correction,
$\mathrm{Im} [\delta N]$, becomes larger than that for the crossed
field $\mathrm{Im} [N]$ at $f/f_c \lesssim 0.1$ for $k_0/m= 1$ or at $k_0/m \lesssim 0.1$
for $f/f_c=1$.
It should be noted, however, that the suppression is much relaxed by the presence of the temporal and spatial variations in the background plane-field.
This is because the imaginary part of the refractive index is
exponentially suppressed for the crossed-field while it is suppressed only
by powers for the plane-wave.

Very strong electromagnetic fields and their temporal and/or spatial variations may be
also important for some astronomical phenomena.
For example, burst activities called giant flares and short bursts have been observed in
magnetars, i.e., strongly magnetized neutron stars~\cite{mcgill}.
Although the energy source of these activities is thought to be the magnetic fields of magnetars,
the mechanism of bursts is not understood yet.
In the analysis of the properties of the emissions from these bursts, the results obtained in this paper may be useful.

As for the burst mechanism, one interesting model related with the strong field variation was proposed by some authors~\cite{HH98,HH99,HH05},
in which they considered shock formations in electromagnetic waves propagating in strong magnetic fields around the magnetar.
The shock dissipation may produce a fireball of electrons and positrons via pair creations. 
Their discussion is based on the Rankine-Hugoniot-type jump condition
and the Euler-Heisenberg Lagrangian,
which is certainly not able to treat the close vicinity of the shock wave, since the
shock is essentially a discontinuity.
Note, however, that our result in this paper is not very helpful for this problem,
either, since the field variation is very rapid and has quite short wavelengths and,
moreover, finite amplitudes of waves are essential for shock formation while our method is limited to the linear level.
It is hence needed to extend the formulation to accommodate these nonlinear effects somehow, which will be a future task.

\section*{Acknowledgement}
This work was supported by the Grants-in-Aid for the
Scientific Research from the Ministry of Education, Culture,
Sports,  Science,  and  Technology  (MEXT)  of  Japan
(No. 24103006, No. 24244036, and No. 16H03986), the
HPCI Strategic Program of MEXT, MEXT Grant-in-Aid
for  Scientific  Research  on  Innovative  Areas
''New
Developments in Astrophysics Through Multi-Messenger
Observations  of  Gravitational  Wave  Sources''
(Grant
Number A05 24103006).

\providecommand{\noopsort}[1]{}\providecommand{\singleletter}[1]{#1}

\appendix

\section{Detailed Derivations} \label{method}

We begin with the following transformation amplitudes:
$\langle x^{(0)} (s) | \hat{\Pi} ^{\mu} _{I} (s) U (s) |x (0) \rangle$, $\langle x^{(0)} (s) | U (s) \hat{\Pi} ^{\mu} _{I} (0) | x(0) \rangle$.
They are written as
\begin{eqnarray}
& &  \langle x(s)| \hat{\Pi} ^{\mu} (s) U (s) |x (0) \rangle \nonumber \\ 
 &=&  \langle x(s)| \hat{\Pi} ^{\mu} (s) \left[ 1- i  \int ^s _0 du \bigg\{ e \hat{\Pi}^{\alpha} (u) b_{\alpha} \exp \left[ -i k_{\delta} \hat{x}^{\delta} (u) \right]  \right. \nonumber \\
 & & \left. \left. + e b_{\alpha} \exp \left[ - i k_{\delta} \hat{x}^{\delta} (u) \right] \hat{\Pi}^{\alpha} (u) + \frac{1}{2} e \sigma ^{\alpha \beta} (u) g_{\alpha \beta} (u)   \right\} \right] |x(0) \rangle , \\
& & \langle x(s)| U(s) \hat{\Pi}^{\mu} (0) |x (0) \rangle \nonumber \\
 &=& \langle x(s)| \left[ 1- i \int ^s _0 du \bigg\{ e \hat{\Pi}^{\alpha} (u) b_{\alpha} \exp \left[ -i k_{\delta} \hat{x}^{\delta} (u) \right] \right. \nonumber \\
 & & \left. \left. + e b_{\alpha} \exp \left[ - i k_{\delta} \hat{x}^{\delta} (u) \right] \hat{\Pi}^{\alpha} (u) + \frac{1}{2} e \sigma ^{\alpha \beta} (u) g_{\alpha \beta} (u)  \right\}  \right] \hat{\Pi}^{\mu} (0) |x (0) \rangle
\end{eqnarray}
with the proper-time evolution operator given in Eq.~(\ref{operatorU}).
In this expression, $g_{\alpha \beta} (u) = g_{\alpha \beta} \exp \left[ -i k_{\delta} \hat{x}^{\delta} (u) \right]$.
We rearrange the first two terms in the integrand as
\begin{eqnarray}
& & \hat{\Pi}^{\mu} (s) \left( e \hat{\Pi}^{\alpha} (u) b_{\alpha}\exp \left[ -i k_{\delta} \hat{x}^{\delta} (u) \right] +  e b_{\alpha}\exp \left[ -i k_{\delta} \hat{x}^{\delta} (u) \right] \hat{\Pi}^{\alpha} (u)  \right) \nonumber \\
& & = 2e b_{\alpha} \hat{\Pi}^{\mu} (s) \hat{\Pi}^{\alpha} (u) \exp \left[ -i k_{\delta} \hat{x}^{\delta} (u) \right] - e b_{\alpha} k^{\alpha} \hat{\Pi}^{\mu} (s) \exp \left[ -i k_{\delta} \hat{x}^{\delta} (u) \right] , \label{pipibexp} \\
 & & \left( e \hat{\Pi}^{\alpha} (u) b_{\alpha}\exp \left[ -i k_{\delta} \hat{x}^{\delta} (u) \right] +  e b_{\alpha}\exp \left[ -i k_{\delta} \hat{x}^{\delta} (u) \right] \hat{\Pi}^{\alpha} (u)  \right) \hat{\Pi}^{\mu} (0)  \nonumber \\
& & = 2e b_{\alpha} \hat{\Pi}^{\alpha} (u) \exp \left[ -i k_{\delta} \hat{x}^{\delta} (u) \right] \hat{\Pi}^{\mu} (0) - e b_{\alpha} k^{\alpha} \exp \left[ -i k_{\delta} \hat{x}^{\delta} (u) \right] \hat{\Pi}^{\mu} (0) , \label{pibexppi}
\end{eqnarray}
using by the following relation
\begin{eqnarray}
 \exp \left[ -ik _{\delta} \hat{x}^{\delta} (u) \right] \hat{\Pi}_{\alpha} (u) &=& \left\{ \hat{\Pi}_{\alpha} (u) + \left[ -i k_{\delta} \hat{x}^{\delta} (u) , \hat{\Pi} _{\alpha} (u) \right] \right\}  \exp \left[ -ik_{\delta} \hat{x}^{\delta} (u) \right] \nonumber \\
&=& \hat{\Pi} _{\alpha} (u) \exp \left[ -i k_{\delta} \hat{x}^{\delta} (u) \right] - k_{\alpha} \exp \left[ -i k_{\delta} \hat{x}^{\delta} (u) \right] ,
\end{eqnarray}
which is obtained from Eqs.~(\ref{expab}) and (\ref{canonical}). The calculations of the remaining terms in the integrand, $\langle x(s) | \hat{\Pi}^{\mu} (s) \left( -i \int ^s _0 du \frac{1}{2} e \sigma ^{\alpha \beta} (u) g_{\alpha \beta} (u) \right) | x(0) \rangle$ and $\langle x(s) | \left( -i \int ^s _0 du \frac{1}{2} e \sigma ^{\alpha \beta} (u) g_{\alpha \beta} (u) \right) \hat{\Pi}^{\mu} (0) | x(0) \rangle$, proceed as follows:
\begin{eqnarray}
& & \langle x(s)| \hat{\Pi}^{\mu} (s) \int ^s _0 du \left( - \frac{ie}{2} \right) \sigma ^{\alpha \beta} (u) g_{\alpha \beta} (u) |x (0) \rangle \nonumber \\
 &\simeq& \langle x(s)| \int ^s _0 du \hat{\Pi}^{\mu} (s) \exp \left[ -i k_{\delta} \hat{x}^{\delta} (u) \right] \nonumber \\
 & & \times \left( - \frac{ie}{2} \right) \left[ ( \sigma g) + \frac{ieu}{2} \left\{ ( \sigma f ) ( \sigma g) - ( \sigma g ) ( \sigma f) \right\} + \frac{e^2 u^2}{4} ( \sigma f ) ( \sigma g)( \sigma f) \right] |x (0) \rangle \nonumber \\
& & + \langle x(s)| \int ^s _0 du \hat{\Pi}^{\mu} (s) \exp \left[ -i k_{\delta} \hat{x}^{\delta} (u) \right] \left( - \frac{ie}{2} \right) \nonumber \\
 & &  \times  \left[ \frac{ieu}{2} \left\{ (\sigma f ) ( \sigma g) - ( \sigma g) (\sigma f ) \right\} ( \Omega \xi (0)) + \frac{e^2 u^2}{2} ( \sigma f)( \sigma g)( \sigma f) ( \Omega \xi (0)) \right] |x(0) \rangle ,
\end{eqnarray}
\begin{eqnarray}
& &  \langle x(s)| \int ^s _0 du \left( - \frac{ie}{2} \right) g_{\alpha \beta} (u) \sigma ^{\alpha \beta} (u)  \hat{\Pi}^{\mu} (0) |x(0) \rangle \nonumber \\
&\simeq& \langle x(s)| \int ^s _0 du \left( - \frac{ie}{2} \right)  \exp \left[ -i k_{\delta} \hat{x}^{\delta} (u) \right] \hat{\Pi}^{\mu} (0) \nonumber \\
 & & \times \left[ (\sigma g ) + \frac{ieu}{2} \left\{ ( \sigma f ) ( \sigma g) - ( \sigma g )( \sigma f) \right\} + \frac{e^2 u^2}{4} ( \sigma f )( \sigma g)( \sigma f)  \right] |x (0) \rangle \nonumber \\
& & + \langle x(s)| \int ^s _0 du \left( - \frac{ie}{2} \right) \exp \left[ -i k_{\delta} \hat{x}^{\delta} (u) \right] \hat{\Pi}^{\mu} (0) \nonumber \\
& & \times \left[ \frac{ieu}{2} \left\{ ( \sigma f ) ( \sigma g) - ( \sigma g ) ( \sigma f ) \right\} ( \Omega \xi (0)) + \frac{e^2 u^2}{2} ( \sigma f ) ( \sigma g)( \sigma f) ( \Omega \xi (0)) \right] |x (0) \rangle \nonumber \\
& & + \langle x(s)| \int ^s _0 du \left( - \frac{ie}{2}\right)  \exp \left[ -i k_{\delta} \hat{x}^{\delta} (u) \right] ( -i n^{\mu} ) \nonumber \\
 & & \times \left[ \frac{ieu}{2} \left\{ ( \sigma f ) ( \sigma g) - ( \sigma g ) ( \sigma f ) \right\} \Omega + \frac{e^2 u^2}{2} ( \sigma f )( \sigma g) ( \sigma f ) \Omega \right] |x (0) \rangle .
\end{eqnarray}
On the second lines in the above equations, we employed the expansion of
$\sigma^{\alpha\beta}(u)$ given in Eq.~(\ref{sigmat}).
The resultant expressions with Eqs.~(\ref{pipibexp}), (\ref{pibexppi}) give Eqs.~(\ref{pisu}) and (\ref{upi0}).
Note that all operators in these expressions, i.e.,
$\hat{\Pi}^{\mu} (s)$, $\hat{\Pi}^{\mu} (0)$, $\hat{\Pi}^{\mu} (u)$ and $\hat{x}^{\mu} (u)$,
are defined in the interaction picture.

Remaining are the evaluations of the transformation amplitudes such as
\begin{eqnarray}
 \langle x(s) | \hat{\Pi}^{\mu} (s) | x (0) \rangle , \label{amplitudepis} \\
 \langle x(s) | \hat{\Pi}^{\mu} (0) | x (0) \rangle , \label{amplitudepi0} \\
 \langle x(s) | \exp \left[ -i k_{\delta} \hat{x}^{\delta} (u) \right] | x (0) \rangle , \label{amplitudeexp} \\
 \langle x(s) | \hat{\Pi} ^{\alpha} (u) \exp \left[ -i k_{\delta} \hat{x}^{\delta} (u) \right] | x (0) \rangle , \label{amplitudepitexp} \\ 
 \langle x(s) | \hat{\Pi} ^{\mu} (s) \exp \left[ -i k_{\delta} \hat{x}^{\delta} (u) \right] | x (0) \rangle , \label{amplitudepisexp} \\
 \langle x(s) | \exp \left[ -i k_{\delta} \hat{x}^{\delta} (u) \right] \hat{\Pi} ^{\mu} (0) | x (0) \rangle , \label{amplitudeexppi0} \\
 \langle x(s) | \hat{\Pi} ^{\mu} (s) \hat{\Pi} ^{\alpha} (u) \exp \left[ -i k_{\delta} \hat{x}^{\delta} (u) \right] | x (0) \rangle , \label{amplitudepispitexp} \\
 \langle x(s) | \hat{\Pi} ^{\alpha} (u) \exp \left[ -i k_{\delta} \hat{x}^{\delta} (u) \right] \hat{\Pi}^{\mu} (0) | x (0) \rangle .  \label{amplitudepitexppi0}
\end{eqnarray} 
Each operator in these amplitudes can be represented with
$\hat{x}^{\mu} (s)$ and $\hat{x}^{\mu} (0)$.
For example, $\hat{\Pi}^{\mu} (s)$ and $\hat{\Pi}^{\mu} (0)$ are
derived from Eqs.~(\ref{Pis}) and (\ref{Pi0}) to the lowest order of $\Omega$ as
\begin{eqnarray}
\hat{\Pi} ^{\mu} (s) &=& \frac{\hat{x}^{\mu} (s) - \hat{x}^{\mu} (0)}{2s} + \frac{e}{2} {f^{\mu}}_{\nu} ( \hat{x}^{\nu} (s) - \hat{x}^{\nu} (0) )  + \Omega \frac{e}{2} {f^{\mu}}_{\nu} ( \hat{x}^{\nu} (s) - \hat{x}^{\nu} (0) ) \left( \frac{2}{3} \xi (s) + \frac{1}{3} \xi (0) \right) \nonumber \\
& & + n^{\mu} e^2 f^2 s \left( \frac{1}{6} \xi (s) - \frac{1}{6} \xi (0) \right) + \Omega  n^{\mu} e^2 f^2 s \left( \frac{1}{4} \xi ^2 (s) - \frac{1}{6} \xi (s) \xi (0) - \frac{1}{12} \xi ^2 (0) \right) \nonumber \\
 & & + \frac{1}{4}  \Omega es n^{\mu} ( \sigma f) , \label{pis}
\end{eqnarray} 
\begin{eqnarray}
 \hat{\Pi}^{\mu} (0) &=& \frac{\hat{x}^{\mu} (s) - \hat{x}^{\mu} (0)}{2s} - \frac{e}{2} {f^{\mu}}_{\nu} ( \hat{x}^{\nu} (s) - \hat{x}^{\nu} (0) ) + \Omega  \frac{e}{2} {f^{\mu}}_{\nu} ( \hat{x}^{\nu} (s) - \hat{x}^{\nu} (0) ) \left( - \frac{1}{3} \xi (s) - \frac{2}{3} \xi (0) \right) \nonumber \\
& & + n^{\mu} e^2 f^2 s \left( \frac{1}{6} \xi (s) - \frac{1}{6} \xi (0) \right) + \Omega n^{\mu} e^2 f^2 s \left( \frac{1}{12} \xi ^2 (s) + \frac{1}{6} \xi (s) \xi (0) - \frac{1}{4} \xi ^2 (0) \right) \nonumber \\
& & - \frac{1}{4} \Omega es n^{\mu} ( \sigma f) . \label{pi0}
\end{eqnarray}
Using the fact that the left hand side (and hence the right hand side also)
of Eq.~(\ref{pi0}) is independent of $s$, we obtain the operator $\hat{x}^{\mu} (u)$
in terms of $\hat{x} ^{\mu}(s)$ and $\hat{x}^{\mu} (0)$ as
\begin{eqnarray}
& & \hat{x}^{\mu} (u) \nonumber \\
& & = \hat{x}^{\mu} (0) + \frac{u}{s} ( \hat{x}^{\mu} (s) - \hat{x}^{\mu} (0) ) \nonumber \\
& & + e {f^{\mu}}_{\nu} ( \hat{x}^{\nu} (s) - \hat{x}^{\nu} (0)) \left[ - u + \frac{u^2}{s} + \Omega \left\{ \left( - \frac{u}{3} + \frac{1}{3} \frac{u^3}{s^2} \right) \xi (s) + \left( - \frac{2}{3} u +  \frac{u^2}{s} - \frac{1}{3} \frac{u^3}{s^2} \right) \xi (0) \right\} \right] \nonumber \\
& & + n^{\mu} e^2 f^2  \left\{ \left( \frac{su}{3} - u^2 + \frac{2}{3} \frac{u^3}{s} \right) \xi (s) + \left( - \frac{su}{3} + u^2 - \frac{2}{3} \frac{u^3}{s} \right) \xi (0) \right\} \nonumber \\
& & + \Omega n^{\mu} e^2 f^2 \left\{ \left( \frac{su}{6} - \frac{1}{3} u^2 - \frac{1}{3} \frac{u^3}{s} + \frac{1}{2} \frac{u^4}{s^2} \right) \xi ^2 (s) \right. \nonumber \\
& & \ \ \ \ \ \left. + \left( \frac{su}{3} - \frac{4}{3} u^2 + 2 \frac{u^3}{s} - \frac{u^4}{s^2} \right) \xi (s) \xi (0) + \left( - \frac{su}{2} + \frac{5}{3} u^2 - \frac{5}{3} \frac{u^3}{s} + \frac{1}{2} \frac{u^4}{s^2} \right) \xi ^2 (0) \right\} \nonumber \\
& & + \frac{1}{2} \Omega e \sigma ^{\nu \lambda} f_{\nu \lambda} n^{\mu} \left( u^2 - su \right) . \label{xt}
\end{eqnarray}
Replacing $s$ with $u$ in Eq.~(\ref{pis}) and plugging Eq.~(\ref{xt})
into Eq.~(\ref{pis}), we can express $\hat{\Pi}^{\mu} (u)$ as
\begin{eqnarray}
& &  \hat{\Pi} ^{\mu} (u) \nonumber \\
&=& \frac{\hat{x}^{\mu} (s) - \hat{x}^{\mu} (0)}{2s} \nonumber \\
& & + e {f^{\mu}}_{\nu} ( \hat{x}^{\nu} (s) - \hat{x}^{\nu} (0) ) \left[ - \frac{1}{2} + \frac{u}{s} +  \Omega \left\{ \left( \frac{1}{2} \left( \frac{u}{s} \right) ^2 - \frac{1}{6} \right) \xi (s) + \left( - \frac{1}{2} \left( \frac{u}{s} \right)^2 + \frac{u}{s} - \frac{1}{3} \right) \xi (0) \right\} \right] \nonumber \\
& & + n^{\mu} e^2 f^2 s \left[ \left\{ \frac{1}{6} - \frac{u}{s} + \left( \frac{u}{s} \right) ^2 \right\} \xi (s) + \left\{ - \frac{1}{6} + \frac{u}{s} - \left( \frac{u}{s} \right) ^2 \right\} \xi (0) \right] \nonumber \\
& & + \Omega n^{\mu} e^2 f^2 s \left[ \left\{ \frac{1}{12} - \frac{1}{3} \left( \frac{u}{s} \right) - \frac{1}{2} \left( \frac{u}{s} \right) ^2 + \left( \frac{u}{s} \right)^3 \right\} \xi ^2 (s) \right. \nonumber \\
 & & \left. + \left\{ \frac{1}{6} - \frac{4}{3} \frac{u}{s} + 3 \left( \frac{u}{s} \right) ^2 - 2 \left( \frac{u}{s} \right)^ 3 \right\} \xi (s) \xi (0) + \left\{ - \frac{1}{4} + \frac{5}{3} \frac{u}{s} - \frac{5}{2} \left( \frac{u}{s} \right) ^2 + \left( \frac{u}{s} \right) ^3 \right\} \xi ^2 (0) \right] \nonumber \\
 & & + \Omega e \sigma ^{\nu \lambda} f_{\nu \lambda} n^{\mu} s \left( - \frac{1}{4} + \frac{1}{2} \frac{u}{s} \right) . \label{pit}
\end{eqnarray}

It is now easy to evaluate the amplitudes in Eqs.~(\ref{amplitudepis}) and~(\ref{amplitudepi0}),
which appear in the induced electromagnetic current as
$\mathrm{tr} \left( \langle x(s) | \hat{\Pi}^{\mu} (s) + \hat{\Pi}^{\mu} (0) | x(0) \rangle \right)$
and $\mathrm{tr} \left( \sigma ^{\mu \nu}  \langle x(s) | \hat{\Pi}^{\mu} (s) - \hat{\Pi}^{\mu} (0) | x(0) \rangle \right)$.
They are given as
\begin{eqnarray}
 \mathrm{tr} \left( \langle x(s) | \hat{\Pi}^{\mu} (s) + \hat{\Pi}^{\mu} (0) | x(0) \rangle \right)&\simeq& \mathrm{tr} \left( \langle x(s) | 0 | x(0) \rangle \right) = 0 ,
\end{eqnarray}
\begin{eqnarray}
\mathrm{tr} \left( \sigma ^{\mu \nu} \langle x(s) | \hat{\Pi}_{\nu} (s) - \hat{\Pi}_{\nu} (0) | x(0) \rangle \right) \simeq \mathrm{tr} \left[ \sigma ^{\mu \nu}  n_{\nu} ( \sigma f) \right] \frac{1}{i ( 4 \pi )^2 s} \frac{e}{2} \Omega = 0 ,
\end{eqnarray}
where we used the following relation
\begin{eqnarray}
 \langle x(s) | x(0) \rangle = \frac{1}{i ( 4 \pi )^2 s^2} \left( {\bf 1} - \frac{ies}{2} ( \sigma f) ( 1 + \Omega \xi) \right) ,
\end{eqnarray}
which is derived from Eq.~(\ref{plac}).
There is hence no contribution to the induced electromagnetic current from
$\langle x(s) | \hat{\Pi}^{\mu} (s) | x(0) \rangle$ and
$\langle x(s) | \hat{\Pi}^{\mu} (0) | x(0) \rangle$.

The amplitude given in Eq.~(\ref{amplitudeexp}) is calculated to the linear order of $\Omega$
by using the Zassenhaus formula:
\begin{eqnarray}
e^{X+\Omega Y} &\simeq& e^{X} e^{\Omega Y} e^{- \frac{1}{2} [X, \Omega Y]} e^{\frac{1}{6} ( 2 [ \Omega Y, [X, \Omega Y]] + [X, [X, \Omega Y]] )} \nonumber \\
&\simeq& e^X + e^X \Omega Y + e^X \left( - \frac{1}{2} \left[ X , \Omega Y \right] \right)  + e^X \frac{1}{6} \left[ X , \left[ X , \Omega Y \right] \right] . 
\end{eqnarray}
In this expression, $X$ stands collectively for the terms that do not include
$\Omega$ in the argument of the exponential function in Eq.~(\ref{amplitudeexp})
whereas $\Omega Y$ represents those terms that depend on $\Omega$. 
The commutation relations in this equation are evaluated as follows:
\begin{eqnarray}
& & \left[ X , \Omega Y \right] \nonumber \\
& & = i \Omega ( k \cdot n )^2 e^2 f^2 \left( - \frac{4}{3} s u^2 + \frac{10}{3} u^3 - 2 \frac{u^4}{s} \right) \xi (s) + i \Omega ( k \cdot n ) ^2 e^2 f^2 \left( - \frac{2}{3} s^2 u + \frac{10}{3} s u^2 - \frac{14}{3} u^3 + 2 \frac{u^4}{s} \right) \xi (0 ) \nonumber \\
& & \hspace{0.5cm} + i \Omega ( k \cdot n ) e k_{\beta} {f^{\beta}}_{\nu} \left[ \hat{x}^{\nu} (s) - \hat{x}^{\nu} (0) \right]  \left( - \frac{2}{3} s u + 2 u^2 - \frac{4}{3} \frac{u^3}{s} \right) , \\
& & \left[ X , \left[ X, \Omega Y \right] \right] = i \Omega ( k \cdot n ) ^3 e ^2 f^2 \left( \frac{4}{3} s^3 u - \frac{16}{3} s^2 u^2 + 8 s u^3 - 4 u^4 \right) . 
\end{eqnarray}

Putting these results together, we obtain the explicit expression
of the exponential operator suited for the calculation of the amplitude as
\begin{eqnarray}
 & & e^{-i k_{\alpha} \hat{x}^{\alpha} (u) } \nonumber \\
 && =  \exp \left\{ -i k_{\alpha} \left[ \frac{u}{s} \hat{x}^{\alpha} (s) + e {f^{\alpha}}_{\beta} \hat{x}^{\beta} (s) \left( - u + \frac{u^2}{s} \right) + n^{\alpha} e^2 f^2 \left( \frac{su}{3} - u^2 + \frac{2}{3} \frac{u^3}{s} \right) \xi (s) \right] \right\} \nonumber \\
& & \hspace{0.0cm} \times  \left( 1  + \Omega \left\{ ie k_{\alpha} {f^{\alpha}}_{\beta} \left[ \hat{x}^{\beta} (s) - \hat{x}^{\beta} (0) \right] \left[ \left( \frac{u}{3} - \frac{1}{3} \frac{u^3}{s^2} \right) \xi (s) + \left( \frac{2}{3} u - \frac{u^2}{s} + \frac{1}{3} \frac{u^3}{s^2} \right) \xi (0) \right] \right.  \right. \nonumber \\
& & \hspace{0.0cm} + i  ( k \cdot n ) e^2 f^2 \left[ \left( - \frac{su}{6} + \frac{1}{3} u^2 + \frac{1}{3} \frac{u^3}{s} - \frac{1}{2} \frac{u^4}{s^2} \right) \xi ^2 (s) \right. \nonumber \\
& & \hspace{0.5cm} + \left. \left( - \frac{su}{3} + \frac{4}{3} u^2 - 2 \frac{u^3}{s} + \frac{u^4}{s^2} \right) \xi (s) \xi (0) + \left( \frac{su}{2} - \frac{5}{3} u^2 + \frac{5}{3} \frac{u^3}{s} - \frac{1}{2} \frac{u^4}{s^2} \right) \xi ^2 (0) \right]  \nonumber \\
& & \hspace{0.0cm} + ie ( k \cdot n ) k_{\alpha} {f^{\alpha}}_{\beta} \left[ \hat{x}^{\beta} (s) - \hat{x}^{\beta} (0) \right] \left( - \frac{2}{3} u^2 + \frac{4}{3} \frac{u^3}{s} - \frac{2}{3} \frac{u^4}{s^2} \right)  \nonumber \\
& & \hspace{0.0cm} + i ( k \cdot n )^2 e^2 f^2 \xi (s) \left( \frac{1}{3} s u^2 - 2 u^3 + 3 \frac{u^4}{s} - \frac{4}{3} \frac{u^5}{s^2} \right)  + i ( k \cdot n )^2 e^2 f^2 \xi (0) \left( - s u^2 + \frac{10}{3} u^3 - \frac{11}{3} \frac{u^4}{s} + \frac{4}{3} \frac{u^5}{s^2} \right)  \nonumber \\
& & \hspace{0.0cm} + i ( k \cdot n) ^3 e^2 f^2 \left( - \frac{2}{3} s^3 u + \frac{14}{9} s^2 u^2 - \frac{4}{9} s u^3 - \frac{10}{9} u^4 + \frac{2}{3} \frac{u^5}{s} \right) \left. \left. + ie ( \sigma f ) ( k \cdot n) \left( \frac{1}{2} su - \frac{1}{2} u^2 \right)   \right\} \right) \nonumber \\
& &  \hspace{0.0cm} \times \exp \left\{ -i k_{\mu} \left[ \left( 1 - \frac{u}{s} \right) \hat{x}^{\mu} (0) + e {f^{\mu}}_{\nu} \hat{x}^{\nu} (0) \left( u - \frac{u^2}{s} \right) + n^{\mu} e^2 f^2 \left( - \frac{su}{3} + u^2 - \frac{2}{3} \frac{u^3}{s} \right) \xi (0) \right] \right\} \nonumber \\
& &  \hspace{0.0cm} \times \exp \left[ i ( k)^2 \left( u- \frac{u^2}{s} \right) + i ( k \cdot n ) ^2 e^2 f^2 \left( - \frac{1}{3} su^2 + \frac{2}{3} u^3 - \frac{1}{3} \frac{u^4}{s} \right) \right] . \label{expxt}
\end{eqnarray}
The transformation amplitude is then given as
\begin{eqnarray}
& & \langle x(s)| \exp \left[ -i k_{\mu} \hat{x}^{\mu} (u) \right] |x (0) \rangle \nonumber \\
& & = \langle x (s)| x (0) \rangle \times \exp \left( -i k_{\mu} x^{\mu} \right)  \exp \left[ i (k )^2 \left( u- \frac{u^2}{s} \right) +  i ( k \cdot n)^2 e^2 f^2 \left( - \frac{1}{3} su^2 + \frac{2}{3} u^3 - \frac{1}{3} \frac{u^4}{s} \right) \right] \nonumber \\
& & \times \left\{ 1 +  \Omega \left[ i  (k \cdot n) ^2 e^2 f^2 \xi \left( - \frac{2}{3} su^2 + \frac{4}{3} u^3 - \frac{2}{3} \frac{u^4}{s} \right) \right. \right. \nonumber \\
& & \hspace{1cm} \left. \left.   + i ( k \cdot n )^3 e^2 f^2 \left( - \frac{2}{3} s^3 u + \frac{14}{9} s^2 u^2 - \frac{4}{9} s u^3 - \frac{10}{9} u^4 + \frac{2}{3} \frac{u^5}{s} \right)  + ie ( \sigma f ) ( k \cdot n) \left( \frac{1}{2} su - \frac{1}{2} u^2 \right) \right] \right\} . \hspace{1cm}
\end{eqnarray}

We next calculate the amplitudes in Eqs. (\ref{amplitudepitexp}) - (\ref{amplitudeexppi0}).
The operators $\hat{\Pi} ^{\mu} (u), \hat{\Pi} ^{\mu} (s), \hat{\Pi} ^{\mu} (0)$ are written
in terms of $\hat{x}^{\mu} (s)$ and $\hat{x}^{\mu} (0)$ and
the amplitudes can be calculated after re-arranging the order of operators.
We first consider the rearrangement of
$\hat{x}^{\alpha} (0) \exp \left[ -i k_{\mu} \hat{x}^{\mu} (u) \right]$.
Using the relations
\begin{eqnarray}
 B e^{-A} &=& e^{-A} B + e^{-A} \left[ A , B \right] + \frac{1}{2}  e^{-A} \left[ A, \left[A , B \right] \right] , \label{bexp-a} \\
e^A B &=& B e^A + \left[ A , B \right] e^A + \frac{1}{2} \left[ A, \left[ A , B \right] \right] e^A , \label{expab}
\end{eqnarray}
which are derived from Hadamard's lemma
\begin{eqnarray}
e^A B e^{-A} = B + \left[ A, B \right] + \frac{1}{2} \left[ A , \left[ A , B \right] \right] ,
\end{eqnarray}
one can obtain 
\begin{eqnarray}
& & \hat{x}^{\alpha} (0) \exp \left[ -i k_{\mu} \hat{x}^{\mu} (u) \right] \nonumber \\
&=& \exp \left[ -i k_{\mu} \hat{x}^{\mu} (u) \right] \nonumber \\
 & & \times \left[ \hat{x}^{\alpha} (0) - 2u k^{\alpha} + 2e ( f^{\alpha \mu} k_{\mu} ) u^2 - \frac{4}{3} u^3 n^{\alpha} ( k \cdot n ) e^2 f^2 \right. \nonumber \\
 & & \hspace{0.5cm} + \Omega e ( f^{\alpha \mu} k_{\mu} ) \left\{ \frac{2}{3} \frac{u^3}{s} \xi (s) + \left( 2 u^2 - \frac{2}{3} \frac{u^3}{s} \right) \xi (0 ) \right\} + \Omega e n^{\alpha} k_{\mu} {f^{\mu}}_{\nu} \left( \hat{x}^{\nu} (s) - \hat{x}^{\nu} (0) \right) \left( - \frac{2}{3} \frac{u^3}{s} \right) \nonumber \\
 & & \hspace{0.5cm} + \Omega n^{\alpha} ( k \cdot n ) e^2 f^2 \left\{ \left( \frac{2}{3} u^3 - 2 \frac{u^4}{s} \right) \xi (s) + \left( - \frac{10}{3} u^3 + 2 \frac{u^4}{s} \right) \xi (0) \right\} \nonumber \\
& & \hspace{0.5cm} + \Omega e ( f^{\alpha \mu} k_{\mu} ) ( k \cdot n ) \left( - \frac{4}{3} u^3 \right)  +  \Omega n^{\alpha} ( k  \cdot n )^2 e^2 f^2 (2 u^4 ) \biggr] ,
\end{eqnarray}
which is still inappropriate for the calculation of the amplitudes because some $\hat{x} ^{\mu} (s)$ are sitting to the right of $\exp \left[ -i k_{\delta} \hat{x}^{\delta} (u) \right]$, which contains $\hat{x} ^{\mu} (0)$. We hence have to rearrange further the terms that contain $\hat{x}^{\mu} (s)$ to obtain
\begin{eqnarray}
& & \hat{x}^{\alpha} (0) \exp \left[ -i k_{\mu} \hat{x}^{\mu} (u) \right] \nonumber \\
 &=& \exp \left[ -i k_{\mu} \hat{x}^{\mu} (u) \right] \left[ \hat{x}^{\alpha} (0) + \Omega e ( f^{\alpha \mu} k_{\mu} )  \left( 2 u^2 - \frac{2}{3} \frac{u^3}{s} \right) \xi (0 ) \right. \nonumber \\
 & & \hspace{0.5cm} + \Omega e n^{\alpha} k_{\mu} {f^{\mu}}_{\nu} \hat{x}^{\nu} (0)  \frac{2}{3} \frac{u^3}{s}  \left. + \Omega n^{\alpha} ( k \cdot n ) e^2 f^2 \left( - \frac{10}{3} u^3 + 2 \frac{u^4}{s} \right) \xi (0) \right] \nonumber \\
 & & + \left[ \Omega e ( f^{\alpha \mu} k_{\mu} ) \frac{2}{3} \frac{u^3}{s} \xi (s)  \right. + \Omega e n^{\alpha} k_{\mu} {f^{\mu}}_{\nu} \hat{x}^{\nu} (s) \left( - \frac{2}{3} \frac{u^3}{s} \right)   \nonumber \\
 & & \hspace{0.5cm} \left. + \Omega n^{\alpha} ( k \cdot n ) e^2 f^2 \left( \frac{2}{3} u^3 - 2 \frac{u^4}{s} \right) \xi (s) \right] \exp \left[ -i k_{\mu} \hat{x}^{\mu} (u) \right] \nonumber \\
& & + \exp \left[ -i k_{\mu} \hat{x}^{\mu} (u) \right]  \left[  - 2u k^{\alpha} + 2e ( f^{\alpha \mu} k_{\mu} ) u^2 - \frac{4}{3} u^3 n^{\alpha} ( k \cdot n ) e^2 f^2 \right.  \nonumber \\ 
& & \hspace{0.5cm} + \Omega e ( f^{\alpha \mu} k_{\mu} ) ( k \cdot n ) \left( \frac{4}{3} \frac{u^4}{s} - \frac{8}{3} u^3 \right)  + \Omega n^{\alpha} ( k  \cdot n )^2 e^2 f^2 \left( - \frac{8}{3} \frac{u^5}{s} + \frac{14}{3} u^4 \right) \biggr] .  \nonumber \\
\end{eqnarray}
This is the expression suitable for the calculation of the transformation amplitudes.

The re-arrangement of $ \exp \left[ -i k_{\mu} \hat{x}^{\mu} (u) \right] \hat{x}^{\alpha} (s)$ goes similarly.
The amplitudes of these operators are then written as follows:
\begin{eqnarray}
& & \langle x(s)| \hat{x}^{\mu} (0 ) \exp \left[ -i k_{\delta} \hat{x}^{\delta} (u) \right] | x(0) \rangle \nonumber \\
 & & = \langle x(s)| \exp \left[ -i k_{\delta} \hat{x}^{\delta} (u) \right] | x(0) \rangle \nonumber \\
 & & \times \left[ x^{\mu} + \Omega e f^{\mu \nu} k_{\nu} \xi 2 u^2 + \Omega n^{\mu} (k \cdot n) e^2 f^2 \xi \left( - \frac{8}{3} u^3 \right) \right.  - 2u k^{\alpha} + 2 e f^{\mu \nu} k_{\nu} u^2  \nonumber \\
& & \ \ \ \left. - \frac{4}{3} u^3 n^{\mu} ( k \cdot n ) e^2 f^2 + \Omega e f^{\mu \nu} k_{\nu} ( k \cdot n ) \left( \frac{4}{3} \frac{u^4}{s} - \frac{8}{3} u^3 \right) + \Omega n^{\mu} ( k \cdot n )^2 e^2 f^2 \left( - \frac{8}{3} \frac{u^5}{s} + \frac{14}{3} u^4 \right) \right] ,  \label{amplitudex0exp} \hspace{1.0cm}\\
& & \langle x(s)| \exp \left[ -i k_{\delta} \hat{x}^{\delta} (u) \right] \hat{x}^{\mu} (s) | x(0) \rangle \nonumber \\
& & = \langle x(s)| \exp \left[ -i k_{\delta} \hat{x}^{\delta} (u) \right] |x (0) \rangle \nonumber \\
& & \times \left[ x^{\mu} + \Omega e f^{\mu \nu} k_{\nu} \xi ( -2 s^2 + 4 su -2 u^2) + \Omega n^{\mu} (k \cdot n) e^2 f^2 \left( - \frac{8}{3} s^3 + 8 s^2 u - 8 s u^2 + \frac{8}{3} u^3 \right) \xi \right. \nonumber \\
& & \ \ \ + 2 ( u-s) k^{\mu} + 2 e f^{\mu \nu} k_{\nu} ( -s^2 + 2su - u^2 ) + n^{\mu} ( k \cdot n ) e^2 f^2 \left( - \frac{4}{3} s^3 + 4 s^2 u - 4 s u^2 + \frac{4}{3} u^3 \right) \nonumber \\
& & \ \ \ + \Omega e (k \cdot n ) f^{\mu \nu} k_{\nu} \left( \frac{4}{3} s^3 - \frac{8}{3} s^2 u + \frac{8}{3} u^3 - \frac{4}{3} \frac{u^4}{s} \right) \nonumber \\
& & \ \ \ \left. + \Omega n^{\mu} ( k \cdot n )^2 e^2 f^2 \left( 2s^4 - \frac{16}{3} s^3 u + \frac{4}{3} s^2 u^2 + 8 s u^3 - \frac{26}{3} u^4 + \frac{8}{3} \frac{u^5}{s} \right) \right] ,
\end{eqnarray}
The quadratic terms in $x$, e.g., $\langle x (s) | {f^{\mu}}_{\nu} \hat{x}^{\nu} (0) \xi (0) \exp \left[ -i k_{\delta} \hat{x}^{\delta} (u) \right] |x (0) \rangle$, can be calculated by successive commutations.
All results combined, the amplitude of $\hat{\Pi}^{\mu} (u) \exp \left[ -i k_{\delta} \hat{x}^{\delta} (u) \right]$ is given as 
\begin{eqnarray}
& & \langle x(s)| \hat{\Pi}^{\alpha} (u) \exp \left[ -i k_{\mu} \hat{x}^{\mu} (u) \right] |x (0)\rangle  \nonumber \\
& & =  \langle  x(s) | \exp \left[ -i k_{\mu} \hat{x}^{\mu} (u) \right] | x (0) \rangle  \nonumber \\
& & \ \ \ \times \left[ \frac{u}{s} k^{\alpha} + e f^{\alpha \beta} k_{\beta} \left( \frac{u^2}{s} - u \right) + n^{\alpha} ( k \cdot n ) e^2 f^2 \left( \frac{1}{3} su - u^2 + \frac{2}{3} \frac{u^3}{s} \right) \right.  + \Omega e f^{\alpha \beta} k_{\beta} \xi \left( \frac{u^2}{s} - u \right) \nonumber \\
& & + \Omega n^{\alpha} ( k \cdot n) e^2 f^2 \xi \left( \frac{2}{3} su - 2 u^2 + \frac{4}{3} \frac{u^3}{s} \right) + \Omega e f^{\alpha \beta} k_{\beta} ( k \cdot n ) \left( \frac{4}{3} u^2 - \frac{8}{3} \frac{u^3}{s} + \frac{4}{3} \frac{u^4}{s^2} \right) \nonumber \\
& &  + \Omega n^{\alpha} ( k \cdot n)^2 e^2 f^2 \left( - su^2 + 4 u^3 - 5 \frac{u^4}{s} + 2 \frac{u^5}{s^2} \right) \left. + \Omega e ( \sigma f ) n^{\alpha} \left( - \frac{1}{4} s + \frac{1}{2} u \right) \right] .
\end{eqnarray}
Similar expressions are obtained for the amplitudes of $\hat{\Pi}^{\mu} (s) \exp \left[ -i k_{\delta} \hat{x}^{\delta} (u) \right]$ and $\exp \left[ -i k_{\delta} \hat{x}^{\delta} (u) \right] \hat{\Pi}^{\mu} (0)$, which are shown, respectively, as follows:
\begin{eqnarray}
& & \langle x(s)| \hat{\Pi}^{\mu} (s) \exp \left[ -i k_{\alpha} \hat{x}^{\alpha} (u) \right] | x(0)  \rangle \nonumber \\
& & = \langle  x(s)| \exp \left[ -i k_{\alpha} \hat{x}^{\alpha} (u) \right] | x(0) \rangle \nonumber \\
& & \ \ \ \times \left[ \Omega e f^{\mu \nu} k_{\nu} \xi \left( u - \frac{u^2}{s} \right) + \Omega n^{\mu} ( k \cdot n ) e^2 f^2 \xi \left( \frac{2}{3} su - 2 u^2 + \frac{4}{3} \frac{u^3}{s} \right) + \left( u - \frac{u^2}{s} \right) e f^{\mu \nu} k_{\nu} \right.  \nonumber \\ 
& & \ \ \ \ \  + \left( \frac{1}{3} su - u^2 + \frac{2}{3} \frac{u^3}{s} \right) n^{\mu} ( k \cdot n) e^2 f^2 + \frac{u}{s} k^{\mu} + \left( - \frac{2}{3} u^2 + \frac{4}{3} \frac{u^3}{s} - \frac{2}{3} \frac{u^4}{s^2} \right) \Omega e f^{\mu \nu} k_{\nu} ( k \cdot n) \nonumber \\
 & & \ \ \ \ \ + \left( - \frac{1}{3} su^2 + 2 u^3 - 3 \frac{u^4}{s} + \frac{4}{3} \frac{u^5}{s^2} \right) \Omega n^{\mu} ( k \cdot n) ^2 e^2 f^2  \left. + \frac{1}{4}  \Omega e s n^{\mu} ( \sigma f ) \right] ,
\end{eqnarray}
\begin{eqnarray}
& & \langle x(s)| \exp \left[ -i k_{\alpha} \hat{x}^{\alpha} (u) \right] \hat{\Pi}^{\mu} (0) | x(0) \rangle \nonumber \\
& & = \langle x(s)| \exp \left[ -i k_{\alpha} \hat{x}^{\alpha} (u) \right] |x (0) \rangle \nonumber \\
& & \times \left[ \Omega e f^{\mu \nu} k_{\nu} \xi \left( u - \frac{u^2}{s} \right) + \Omega n^{\mu} ( k \cdot n) e^2 f^2 \xi \left( \frac{2}{3} su - 2 u^2 + \frac{4}{3} \frac{u^3}{s} \right) + \left( \frac{u}{s} -1 \right) k^{\mu} \right. \nonumber \\
 & & \ \ \  + \left( u - \frac{u^2}{s} \right) e f^{\mu \nu} k_{\nu} + \left( \frac{1}{3} su -  u^2 + \frac{2}{3} \frac{u^3}{s} \right) n^{\mu} ( k \cdot n) e^2 f^2 \nonumber \\
 & & \ \ \ + \Omega \left( - \frac{2}{3} u^2 + \frac{4}{3} \frac{u^3}{s} - \frac{2}{3} \frac{u^4}{s^2} \right) e f^{\mu \nu} k_{\nu} ( k \cdot n ) \nonumber \\
& & \ \ \ + \Omega \left( - su^2 + \frac{10}{3} u^3 - \frac{11}{3} \frac{u^4}{s} + \frac{4}{3} \frac{u^5}{s^2} \right) n^{\mu} ( k \cdot n )^2 e^2 f^2  \left. + \left( - \frac{1}{4} \right) \Omega es n^{\mu} ( \sigma f ) \right] .
\end{eqnarray}

Finally, Eqs.~(\ref{amplitudepispitexp}) and~(\ref{amplitudepitexppi0}) are calculated. We rewrite them in terms of $\langle x(s)| \exp \left[ -i k_{\delta} \hat{x}^{\delta} (u) \right] | x(0) \rangle$ and $\langle x(s)| \hat{\Pi} ^{\alpha} (u) \exp \left[ -i k_{\delta} \hat{x}^{\delta} (u) \right] | x(0) \rangle$, which have been already evaluated. In so doing, the products of the operators such as $\hat{x}^{\mu} (0) \hat{\Pi} ^{\alpha} (u) \exp \left[ -i k_{\delta} \hat{x}^{\delta} (u) \right]$ in $\hat{\Pi}^{\mu} (s) \hat{\Pi} ^{\alpha} (u) \exp \left[ -i k_{\delta} \hat{x}^{\delta} (u) \right]$ and $\hat{\Pi} ^{\alpha} (u) \exp \left[ -i k_{\delta} \hat{x}^{\delta} (u) \right] \hat{x}^{\mu} (s)$ in $\hat{\Pi} ^{\alpha} (u) \exp \left[ -i k_{\delta} \hat{x}^{\delta} (u) \right] \hat{\Pi} ^{\mu} (0)$ have to be rearranged. To accomplish it, we need the following commutation relations for $\hat{\Pi} ^{\mu} (u)$, which are obtained from the results given in Appendix B:
\begin{eqnarray}
& &  \left[ \hat{x}^{\mu} (0) , \hat{\Pi} ^{\alpha} (u) \right] = - i \eta ^{\mu \alpha} + 2u ie f^{\mu \alpha} - 2 u^2 i n^{\mu} n^{\alpha} e^2 f^2 + \Omega i e f^{\mu \alpha} \xi (s) \left( \frac{u^2}{s} \right) \nonumber \\
 & & \ \ \ + \Omega ie f^{\mu \alpha} \xi (0) \left( 2u - \frac{u^2}{s} \right) + \Omega i n^{\mu} n^{\alpha} e^2 f^2 \xi (s) \left( u^2 - 4 \frac{u^3}{s} \right) \nonumber \\
 & & \ \ \ + \Omega  i n^{\mu} n^{\alpha} e^2 f^2 \xi (0) \left( -5 u^2 + 4 \frac{u^3}{s} \right) + \Omega ie n^{\mu} {f^{\alpha}}_{\beta} \left( \hat{x}^{\beta} (s) - \hat{x}^{\beta} (0) \right) \left( - \frac{u^2}{s} \right) ,
\end{eqnarray}
\begin{eqnarray}
& & \left[ \hat{\Pi}^{\alpha} (u) , \hat{x}^{\mu} (s) \right]  = i \eta ^{\alpha \mu} + ( -2s +2u ) ie f^{\alpha \mu} + ( 2s^2 - 4su + 2 u^2 ) i n^{\alpha} n^{\mu} e^2 f^2 \nonumber \\
 & & \ \ \ + \Omega ie f^{\alpha \mu} \xi (s) \left( \frac{u^2}{s} -s \right) + \Omega ie f^{\alpha \mu} \xi (0) \left( - \frac{u^2}{s} + 2u - s \right) \nonumber \\
 & & \ \ \ + \Omega i n^{\alpha} n^{\mu} e^2 f^2 \xi (s) \left( s^2 + 2su - 7 u^2 + 4 \frac{u^3}{s} \right) \nonumber \\
 & & \ \ \ + \Omega i n^{\alpha} n^{\mu} e^2 f^2 \xi (0) \left( 3s^2 - 10 su + 11 u^2 - 4 \frac{u^3}{s} \right) \nonumber \\
 & & \ \ \ + \Omega ie n^{\mu} {f^{\alpha}}_{\beta} \left( \hat{x}^{\beta} (s) - \hat{x}^{\beta} (0) \right) \left( \frac{u^2}{s} - 2u + s \right) ,
\end{eqnarray}
The employment of these relations produces the following results:
\begin{eqnarray}
& & \langle x(s)| \hat{x}^{\mu} (0) \hat{\Pi}^{\alpha} (u) \exp \left[ -i k_{\delta} \hat{x}^{\delta} (u) \right] |x(0) \rangle \nonumber \\
 & & = \langle x(s)| \exp \left[ - i k_{\delta} \hat{x}^{\delta} (u) \right] |x(0) \rangle \nonumber \\
 & & \times \left[ -i \eta ^{\mu \alpha} + 2u ie f^{\mu \alpha} - 2 u^2 i n^{\mu} n^{\alpha} e^2 f^2 + i \Omega e n^{\alpha} f^{\mu \nu} k_{\nu} \frac{2}{3} \frac{u^3}{s} \right. + i \Omega e n^{\mu} f^{\alpha \beta} k_{\beta} \left( - \frac{4}{3} \frac{u^3}{s} \right) \nonumber \\
 & & + i \Omega e (k \cdot n) f^{\mu \alpha} \left( -4 u^2 + 2 \frac{u^3}{s} \right) + i \Omega  n^{\mu} n^{\alpha} ( k \cdot n) e^2 f^2 \left( \frac{28}{3} u^3 - \frac{20}{3} \frac{u^4}{s} \right) \nonumber \\
 & & +  i \Omega  n^{\mu} n^{\alpha} e^2 f^2 \xi ( -4 u^2 ) + i \Omega e f^{\mu \alpha} \xi 2 u \biggr] \nonumber \\
 & & + \langle x(s)| \hat{\Pi}^{\alpha} (u) \exp \left[ -i k_{\delta} \hat{x}^{\delta} (u) \right] |x(0) \rangle \nonumber \\
 & & \times \biggl[ x^{\mu} -2u k^{\mu} + 2e f^{\mu \nu} k_{\nu} u^2 - \frac{4}{3} u^3 n^{\mu} ( k \cdot n) e^2 f^2 + \Omega e f^{\mu \nu} k_{\nu} ( k \cdot n) \left( \frac{4}{3} \frac{u^4}{s} - \frac{8}{3} u^3 \right) \nonumber \\
 & & \ \ \ + \Omega n^{\mu} ( k \cdot n) ^2 e^2 f^2 \left( - \frac{8}{3} \frac{u^5}{s} + \frac{14}{3} u^4 \right) \nonumber \\
 & & \left. + \Omega e f^{\mu \nu} k_{\nu} ( 2 u^2 ) \xi + \Omega n^{\mu} ( k \cdot n) e^2 f^2 \left( - \frac{8}{3} u^3 \right) \xi \right] ,
\end{eqnarray}
\begin{eqnarray}
& & \langle x(s)| \hat{\Pi}^{\alpha} (u) \exp \left[ -i k_{\delta} \hat{x}^{\delta} (u) \right] \hat{x}^{\mu} (s) | x(0)  \rangle \nonumber \\
& & = \langle x(s) | \exp \left[ -i k_{\delta} \hat{x}^{\delta} (u) \right] |x (0)  \rangle \nonumber \\
& & \ \ \ \ \times \biggl[  i \Omega e f^{\alpha \mu} \xi ( 2u - 2s ) + i \Omega  n^{\alpha} n^{\mu} e^2 f^2 \xi ( 4 s^2 - 8su + 4 u^2 ) +  i \eta ^{\alpha \mu} + ( 2u -2s ) ie f^{\alpha \mu}  \nonumber \\
 & & \ \ \ + ( 2s^2 - 4su + 2 u^2 ) i n^{\alpha} n^{\mu} e^2 f^2 + i \Omega  e n^{\alpha} f^{\mu \nu} k_{\nu} \left( - \frac{4}{3} s^2 + 2su - \frac{2}{3} \frac{u^3}{s} \right) \nonumber  \\
 & & \ \ \ + i \Omega e n^{\mu} f^{\alpha \beta} k_{\beta} \left( \frac{2}{3} s^2 - 2 u^2 + \frac{4}{3} \frac{u^3}{s} \right) + i \Omega e ( k \cdot n) f^{\alpha \mu} \left( 2 \frac{u^3}{s} - 4 u^2 + 2su \right) \nonumber \\
 & & \ \ \ \left. + i \Omega  n^{\mu} n^{\alpha} ( k \cdot n ) e^2 f^2 \left( - \frac{8}{3} s^3 + \frac{4}{3} s^2 u + 12 su^2 - \frac{52}{3} u^3 + \frac{20}{3} \frac{u^4}{s} \right) \right] \nonumber \\
& & + \langle x(s)| \hat{\Pi}^{\alpha} (u) \exp \left[ -i k_{\delta} \hat{x}^{\delta} (u) \right] |x (0) \rangle \nonumber \\
& & \ \ \ \times \biggl[ x ^{\mu} + \Omega e f^{\mu \nu} k_{\nu} \xi ( - 2s^2 + 4su - 2 u^2 )  + \Omega n^{\mu} ( k \cdot n) e^2 f^2 \xi \left( - \frac{8}{3} s^3 + 8 s^2 u - 8 s u^2 + \frac{8}{3} u^3 \right) \nonumber \\
& & \ \ \ +  2 (u-s ) k^{\mu} + 2e f^{\mu \nu} k_{\nu} ( -s ^2 + 2su - u^2 ) + n^{\mu} ( k \cdot n ) e^2 f^2 \left( - \frac{4}{3} s^3 + 4s^2 u - 4 s u^2 + \frac{4}{3} u^3 \right) \nonumber \\
 & &  \ \ \ + \Omega e ( k \cdot n) f^{\mu \nu} k_{\nu} \left( \frac{4}{3} s^3 - \frac{8}{3} s^2 u + \frac{8}{3} u^3 - \frac{4}{3} \frac{u^4}{s} \right) \nonumber \\
 & & \ \ \ \left.  + \Omega n^{\mu} ( k \cdot n) ^2 e^2 f^2 \left( 2 s^4 - \frac{16}{3} s^3 u + \frac{4}{3} s^2 u^2 + 8 s u^3 - \frac{26}{3} u^4 + \frac{8}{3} \frac{u^5}{s} \right) \right] ,
\end{eqnarray}

We are now ready to write down the amplitudes of the addition $\langle x(s)| \hat{\Pi}^{\mu} (s) \hat{\Pi}^{\alpha} (u) \exp \left[ -i k_{\delta} \hat{x}^{\delta} (u) \right] + \hat{\Pi}^{\alpha} (u) \exp \left[ -i k_{\delta} \hat{x}^{\delta} (u) \right] \hat{\Pi}^{\mu} (0) | x(0) \rangle$ and the subtraction $\langle x(s)| \hat{\Pi}^{\mu} (s) \hat{\Pi}^{\alpha} (u) \exp \left[ -i k_{\delta} \hat{x}^{\delta} (u) \right] - \hat{\Pi}^{\alpha} (u) \exp \left[ -i k_{\delta} \hat{x}^{\delta} (u) \right] \hat{\Pi}^{\mu} (0) | x(0) \rangle$, which appear in the induced electromagnetic current. The results are as follows:
\begin{eqnarray}
& & \langle x(s)| \hat{\Pi}^{\mu} (s) \hat{\Pi}^{\alpha} (u) \exp \left[ -i k_{\delta} \hat{x}^{\delta} (u) \right] + \hat{\Pi}^{\alpha} (u) \exp \left[ -i k_{\delta} \hat{x}^{\delta} (u) \right] \hat{\Pi}^{\mu} (0) | x(0) \rangle \nonumber \\
&=& \langle x(s)| \exp \left[ -i k_{\delta} \hat{x}^{\delta} (u) \right] |x (0) \rangle \nonumber \\
& & \times \left[ i e f^{\mu \alpha} \left( 1 - 2 \frac{u}{s} \right) + k^{\mu} k^{\alpha} \left( - \frac{u}{s} + 2 \frac{u^2}{s^2} \right) + e k^{\mu} f^{\alpha \beta} k_{\beta} \left( u - 3 \frac{u^2}{s} + 2 \frac{u^3}{s^2} \right) \right. \nonumber \\
& & \ \ \ + e f^{\mu \nu} k_{\nu} k^{\alpha} \left( 2 \frac{u^2}{s} - 2 \frac{u^3}{s^2} \right)  + e^2 f^{\mu \nu} k_{\nu} f^{\alpha \beta} k_{\beta} \left( -2 u^2 + 4 \frac{u^3}{s} - 2 \frac{u^4}{s^2} \right) \nonumber \\
& & \ \ \  + k^{\mu} n^{\alpha} ( k \cdot n) e^2 f^2 \left( - \frac{1}{3} su + \frac{5}{3} u^2 - \frac{8}{3} \frac{u^3}{s} + \frac{4}{3} \frac{u^4}{s^2} \right)  \nonumber \\
& & \ \ \ + f^{\mu \nu} k_{\nu} n^{\alpha} (k \cdot n) e^3 f^2 \left( \frac{2}{3} su^2 - \frac{8}{3} u^3 + \frac{10}{3} \frac{u^4}{s} - \frac{4}{3} \frac{u^5}{s^2} \right) + n^{\mu} k^{\alpha} (k \cdot n) e^2 f^2 \left( \frac{2}{3} u^2 - 2 \frac{u^3}{s} + \frac{4}{3} \frac{u^4}{s^2} \right)  \nonumber \\
& & \ \ \ + n^{\mu} f^{\alpha \beta} k_{\beta} ( k \cdot n) e^3 f^2 \left( - \frac{2}{3} s u^2 + \frac{8}{3} u^3 - \frac{10}{3} \frac{u^4}{s} + \frac{4}{3} \frac{u^5}{s^2} \right) + i n^{\mu} n^{\alpha} e^2 f^2 \left( \frac{1}{3} s - 2u + 2 \frac{u^2}{s} \right) \nonumber \\
 & & \ \ \ \left.  + n^{\mu} n^{\alpha} ( k \cdot n )^2 e^4 f^4 \left( \frac{2}{9} s^2 u^2 - \frac{4}{3} s u^3 + \frac{26}{9} u^4 - \frac{8}{3} \frac{u^5}{s} + \frac{8}{9} \frac{u^6}{s^2} \right) + i \eta ^{\mu \alpha} \frac{1}{s} \right] \nonumber \\
 & & (\mathrm{The \ expression \ continues \ to \ the \ next \ page.}) \nonumber
\end{eqnarray}  
\begin{eqnarray}
& & (\mathrm{The \ expression \ is \ continued.}) \nonumber \\
&+& i \Omega \langle x(s) | \exp \left[ -i k_{\delta} \hat{x}^{\delta} (u) \right] |x (0)  \rangle \nonumber \\
 & & \times \left[ e f^{\mu \alpha} (k \cdot n) \left( \frac{1}{3} s - \frac{5}{3} u + 4 \frac{u^2}{s} - 2 \frac{u^3}{s^2} \right) + e f^{\mu \alpha} \xi \left( 1 - 2 \frac{u}{s} \right) \right. \nonumber \\
 & & \ \ \ + ie (k \cdot n ) k^{\mu} f^{\alpha \beta} k_{\beta} \left( \frac{4}{3} u^2 - \frac{16}{3} \frac{u^3}{s} + \frac{20}{3} \frac{u^4}{s^2} - \frac{8}{3} \frac{u^5}{s^3} \right) + i e \xi k^{\mu} f^{\alpha \beta} k_{\beta} \left( -u + 3 \frac{u^2}{s} - 2 \frac{u^3}{s^2} \right) \nonumber \\
 & & \ \ \ + ie (k \cdot n) f^{\mu \nu} k_{\nu} k^{\alpha} \left( \frac{4}{3} \frac{u^3}{s} - \frac{8}{3} \frac{u^4}{s^2} + \frac{4}{3} \frac{u^5}{s^2} \right) + ie \xi f^{\mu \nu} k_{\nu} k^{\alpha} \left( - 2 \frac{u^2}{s} + 2 \frac{u^3}{s^2} \right) \nonumber \\
 & & \ \ \ + i e^2 ( k \cdot n) f^{\mu \nu} k_{\nu} f^{\alpha \beta} k_{\beta} \left( -4 u^3 + 12 \frac{u^4}{s} - 12 \frac{u^5}{s^2} + 4 \frac{u^6}{s^3} \right) + ie^2 \xi f^{\mu \nu} k_{\nu} f^{\alpha \beta} k_{\beta} \left( 4 u^2 - 8 \frac{u^3}{s} + 4 \frac{u^4}{s^2} \right) \nonumber \\
 & & \ \ \ + i k^{\mu} n^{\alpha} ( k \cdot n )^2 e^2 f^2 \left( -s u^2 + 6 u^3  - 13 \frac{u^4}{s} + 12 \frac{u^5}{s^2} - 4 \frac{u^6}{s^3} \right) \nonumber \\
 & & \ \ \ + i k^{\mu} n^{\alpha} \xi ( k \cdot n) e^2 f^2 \left( \frac{2}{3} su - \frac{10}{3} u^2 + \frac{16}{3} \frac{u^3}{s} - \frac{8}{3} \frac{u^4}{s^2} \right) \nonumber \\
& & \ \ \ + ie k^{\mu} n^{\alpha} ( \sigma f ) \left( - \frac{1}{4} s + u - \frac{u^2}{s} \right) + e f^{\mu \nu} k_{\nu} n^{\alpha} \left( - \frac{1}{3} s + \frac{1}{3} u - \frac{2}{3} \frac{u^3}{s^2} \right)  \nonumber \\
& & \ \ \  + i f^{\mu \nu} k_{\nu} n^{\alpha} ( k \cdot n)^2 e^3 f^2 \left( \frac{22}{9} su^3 - \frac{110}{9} u^4 + 22 \frac{u^5}{s} - \frac{154}{9} \frac{u^6}{s^2} + \frac{44}{9} \frac{u^7}{s^3} \right) \nonumber \\
& & \ \ \ + i f^{\mu \nu} k_{\nu} n^{\alpha} \xi ( k \cdot n ) e^3 f^2 \left( -2su^2 + 8u^3 -10 \frac{u^4}{s} + 4 \frac{u^5}{s^2} \right) + ie^2 f^{\mu \nu} k_{\nu} n^{\alpha} ( \sigma f ) \left( \frac{1}{2} su - \frac{3}{2} u^2 + \frac{u^3}{s} \right) \nonumber \\
& & \ \ \ + i n^{\mu} k^{\alpha} ( k \cdot n)^2 e^2 f^2 \left( \frac{4}{3} u^3 - \frac{16}{3} \frac{u^4}{s} + \frac{20}{3} \frac{u^5}{s^2} - \frac{8}{3} \frac{u^6}{s^3} \right) + i n^{\mu} k^{\alpha} \xi ( k \cdot n) e^2 f^2 \left( - \frac{4}{3} u^2 + 4 \frac{u^3}{s} - \frac{8}{3} \frac{u^4}{s^2} \right) \nonumber \\
 & & \ \ \ + e n^{\mu} f^{\alpha \beta} k_{\beta} \left( \frac{1}{3} s - \frac{u^2}{s} + \frac{4}{3} \frac{u^3}{s^2} \right) \nonumber \\
 & & \ \ \ + i n^{\mu} f^{\alpha \beta} k_{\beta} ( k \cdot n)^2 e^3 f^2 \left( - \frac{20}{9} s u^3 + \frac{100}{9} u^4 -20 \frac{u^5}{s} + \frac{140}{9} \frac{u^6}{s^2} - \frac{40}{9} \frac{u^7}{s^3} \right)  \nonumber \\
 & & \ \ \ + i n^{\mu} f^{\alpha \beta} k_{\beta} \xi ( k \cdot n) e^3 f^2 \left( 2 su^2 - 8 u^3 + 10 \frac{u^4}{s} - 4 \frac{u^5}{s^2} \right) \nonumber \\
 & & \ \ \   + n^{\mu} n^{\alpha} ( k \cdot n ) e^2 f^2 \left( - \frac{4}{3} su + 8 u^2 - \frac{40}{3} \frac{u^3}{s} + \frac{20}{3} \frac{u^4}{s^2} \right) \nonumber \\
 & & \ \ \ + i n^{\mu} n^{\alpha} ( k \cdot n) ^3 e^4 f^4 \left( \frac{10}{9} s^2 u^3 - \frac{70}{9} s u^4 + \frac{190}{9} u^5 - \frac{250}{9} \frac{u^6}{s} + \frac{160}{9} \frac{u^7}{s^2} - \frac{40}{9} \frac{u^8}{s^3} \right) \nonumber \\
 & & \ \ \ + n^{\mu} n^{\alpha} \xi e^2 f^2 \left( \frac{2}{3} s - 4u + 4 \frac{u^2}{s} \right) \nonumber \\
& & \ \ \ + i n^{\mu} n^{\alpha} \xi ( k \cdot n) ^2 e^4 f^4 \left( - \frac{8}{9} s^2 u^2 + \frac{16}{3} s u^3 - \frac{104}{9} u^4 + \frac{32}{3} \frac{u^5}{s} - \frac{32}{9} \frac{u^6}{s^2} \right) \nonumber \\
& & \ \ \ \left.  +  i n^{\mu} n^{\alpha} ( k \cdot n) ( \sigma f) e^3 f^2 \left( \frac{1}{6} s^2 u - \frac{5}{6} s u^2 + \frac{4}{3} u^3 - \frac{2}{3} \frac{u^4}{s} \right) \right] ,
\end{eqnarray}
\begin{eqnarray}
& & \langle x(s)| \hat{\Pi}^{\mu} (s) \hat{\Pi}^{\alpha} (u) \exp \left[ -i k_{\delta} \hat{x}^{\delta} (u) \right] - \hat{\Pi}^{\alpha} (u) \exp \left[ -i k_{\delta} \hat{x}^{\delta} (u) \right] \hat{\Pi}^{\mu} (0) | x(0) \rangle \nonumber \\
&=& \langle x(s)| \exp \left[ -i k_{\delta} \hat{x}^{\delta} (u) \right] |x (0) \rangle \nonumber \\
& & \times \left[ k^{\mu} k^{\alpha} \frac{u}{s} + e k^{\mu} f^{\alpha \beta} k_{\beta} \left( -u + \frac{u^2}{s} \right) + k^{\mu} n^{\alpha} ( k \cdot n) e^2 f^2 \left( \frac{1}{3} su - u^2 + \frac{2}{3} \frac{u^3}{s} \right) \right] \nonumber \\
&+& i \Omega \langle x(s)| \exp \left[ -i k_{\delta} \hat{x}^{\delta} (u) \right] |x (0)  \rangle \nonumber \\
& & \times \left[ e f^{\mu \alpha} ( k \cdot n) \left( - \frac{1}{3} s + u \right) + i  e k^{\mu} f^{\alpha \beta} k_{\beta} ( k \cdot n ) \left( - \frac{4}{3} u^2 + \frac{8}{3} \frac{u^3}{s} - \frac{4}{3} \frac{u^4}{s^2} \right) + i e \xi k^{\mu} f^{\alpha \beta} k_{\beta} \left( u - \frac{u^2}{s} \right) \right. \nonumber \\
& & \ \ \ + i k^{\mu} n^{\alpha} ( k \cdot n )^2 e^2 f^2 \left( s u^2 - 4 u^3 + 5 \frac{u^4}{s} - 2 \frac{u^5}{s^2} \right) + i k^{\mu} n^{\alpha} \xi ( k \cdot n) e^2 f^2 \left( - \frac{2}{3} su + 2 u^2 - \frac{4}{3} \frac{u^3}{s} \right)  \nonumber \\
& & \ \ \ +  i k^{\mu} n^{\alpha} e ( \sigma f ) \left( \frac{1}{4} s - \frac{1}{2} u \right) + e f^{\mu \nu} k_{\nu} n^{\alpha} \left( \frac{1}{3} s - u \right) + i n^{\mu} k^{\alpha} ( k \cdot n)^2 e^2 f^2 \left( - \frac{2}{3} u^3 + \frac{4}{3} \frac{u^4}{s} - \frac{2}{3} \frac{u^5}{s^2} \right) \nonumber \\
 & & \ \ \ + i n^{\mu} k^{\alpha} e ( \sigma f ) \left( - \frac{1}{2} u \right) + e n^{\mu} f^{\alpha \beta} k_{\beta} \left( - \frac{1}{3} s + \frac{u^2}{s} \right) \nonumber \\
 & & \ \ \ + i n^{\mu} f^{\alpha \beta} k_{\beta} ( k \cdot n )^2 e^3 f^2 \left( \frac{2}{3} s u^3 - 2 u^4 + 2 \frac{u^5}{s} - \frac{2}{3} \frac{u^6}{s^2} \right) \nonumber \\
& & \ \ \ + i n^{\mu} f^{\alpha \beta} k_{\beta} e^2 ( \sigma f ) \left( \frac{1}{2} su - \frac{1}{2} u^2 \right) + n^{\mu} n^{\alpha} ( k \cdot n ) e^2 f^2 \left( \frac{2}{3} su - 2 u^2 + \frac{4}{3} \frac{u^3}{s} \right) \nonumber \\
& & \ \ \ + i n^{\mu} n^{\alpha} ( k \cdot n)^3 e^4 f^4 \left( - \frac{2}{9} s^2 u^3 + \frac{10}{9} s u^4 - 2 u^5 + \frac{14}{9} \frac{u^6}{s} - \frac{4}{9} \frac{u^7}{s^2} \right) \nonumber \\
& & \ \ \ \left. + i n^{\mu} n^{\alpha} e^3 f^2 ( \sigma f ) ( k \cdot n) \left( - \frac{1}{6} s^2 u + \frac{1}{2} su^2 - \frac{1}{3} u^3 \right) \right] .
\end{eqnarray}

\section{Permutations of Operators} \label{permutation}
We give some technical details relevant for permutations of operators in this section. The basic commutation relations are those among $\hat{x}^{\mu} (s)$, $\hat{x}^{\mu} (u)$ and $\hat{x}^{\mu} (0)$. 
It is written as 
\begin{eqnarray}
& & \left[ \hat{x}^{\mu} (0) , \hat{x}^{\alpha} (s) \right] \nonumber \\
&=& -2is \eta ^{\mu \alpha} + 2ies^2 f^{\mu \alpha} - \frac{4}{3} i n^{\mu} n^{\alpha} e^2 f^2 s^3  + i \Omega  es ^2 f^{\mu \alpha} \left( \frac{4}{3} \xi (s) + \frac{2}{3} \xi (0) \right) \nonumber \\
& & - \frac{2}{3} i \Omega e s^2 n^{\alpha} {f^{\mu}}_{\nu} [ \hat{x}^{\nu} (s) - \hat{x}^{\nu} (0) ] + i \Omega n^{\mu} n^{\alpha} e^2 f^2 s^3 \left( - \frac{4}{3} \xi (s) - \frac{4}{3} \xi (0) \right) \label{commutationrelationxx} 
\end{eqnarray}
for $\hat{x}^{\mu} (s)$ and $\hat{x}^{\alpha} (0)$.
Its derivation is as follows. The canonical commutation relation is written as
\begin{eqnarray}
\left[ \hat{x}^{\mu} (0),  \hat{\Pi}^{\nu} (0) \right] = \left[ \hat{x}^{\mu} (s),  \hat{\Pi}^{\nu} (s) \right] = -i \eta ^{\mu \nu} \label{canonicalcommutation}
\end{eqnarray}
and the expression of $\hat{x}^{\mu} (0)$ in terms of $\hat{\Pi}^{\mu} (s)$
is obtained from Eq.~(\ref{pis}) as
\begin{eqnarray}
 & &  \hat{x}^{\mu} (0) \nonumber \\
 & & = -2s \hat{\Pi}^{\mu} (s) + \hat{x}^{\mu} (s) + 2 e s^2 {f^{\mu}}_{\nu} \hat{\Pi}^{\nu} (s) + n^{\mu} e^2 f^2 s^2 \left( - \frac{2}{3} \xi (s) + \frac{2}{3} \xi (0) \right) \nonumber \\
& & + \Omega e s^2 {f^{\mu}}_{\nu} \hat{\Pi} ^{\nu} (s) \left( \frac{4}{3} \xi (s) + \frac{2}{3} \xi (0) \right) + \Omega n^{\mu} e^2 f^2 s^2 \left( - \frac{5}{6} \xi ^2 (s) + \frac{1}{3} \xi (s) \xi (0) + \frac{1}{2} \xi ^2 (0) \right) \nonumber \\
& & + \frac{1}{2} \Omega e s^2 n^{\mu} (\sigma f) .  \label{x0}
\end{eqnarray}
Let us first consider the commutation relation $\left[ \xi (0) , \hat{x}^{\alpha} (s) \right]$.
From Eq.~(\ref{x0}), we obtain 
\begin{eqnarray}
 \xi (0) = -2s n_{\mu} \hat{\Pi}^{\mu} (s) + \xi (s).
\end{eqnarray}
We then easily derive the following relation:
\begin{eqnarray}
 \left[ \xi (0) , \hat{x}^{\alpha} (s) \right] &=& \left[ -2s n_{\mu} \hat{\Pi}^{\mu} (s)  , \hat{x}^{\alpha} (s)\right] \nonumber \\
&=& -2s n_{\mu} i \eta ^{\mu \alpha} \nonumber \\
&=& -2is n^{\alpha} . \label{xi0xscommutation}
\end{eqnarray}
Combining Eqs.~(\ref{canonicalcommutation})-(\ref{xi0xscommutation}),
we obtain Eq.~(\ref{commutationrelationxx}) easily.

The following commutation relations, which are frequently used,
also follow immediately:
\begin{eqnarray}
& & \left[ \hat{x}^{\alpha} (0), {f^{\mu}}_{\nu} \hat{x}^{\nu} (s) \right] \nonumber \\
&=& 2is f^{\alpha \mu} -2i n^{\alpha} n^{\mu} e f^2 s^2 + i \Omega n^{\alpha} n^{\mu} e f^2 s^2 \left( - \frac{4}{3} \xi (s) - \frac{2}{3} \xi (0) \right) , 
\end{eqnarray}
\begin{eqnarray}
& & \left[ \hat{x}^{\alpha} (s) , {f^{\mu}}_{\nu} \hat{x}^{\nu} (0) \right] \nonumber \\
&=& -2is f^{\alpha \mu} -2i n^{\alpha} n^{\mu} e f^2 s^2 + i \Omega n^{\alpha} n^{\mu} e f^2 s^2 \left( - \frac{2}{3} \xi (s) - \frac{4}{3} \xi (0) \right) , 
\end{eqnarray}
\begin{eqnarray}
\left[ {f^{\alpha}}_{\beta} \hat{x}^{\beta} (s) , {f^{\mu}}_{\nu} \hat{x}^{\nu} (0) \right] = -2i n^{\alpha} n^{\mu} f^2 s ,  
\end{eqnarray}
\begin{eqnarray}
& & \left[ \xi (s) , \xi (0) \right] = 0 , \\
& & \left[ \hat{x}^{\mu} (0) , \xi (s) \right] = -2is n^{\mu} , \\
& & \left[ \xi (0) , \hat{x}^{\mu} (s) \right] = -2is n^{\mu} , \\
& & \left[ \hat{x}^{\mu} (s) , \hat{\Pi}^{\nu} (s) \right] = \left[ \hat{x}^{\mu} (0) , \hat{\Pi}^{\nu} (0) \right] = \left[ \hat{x}^{\mu} (u) , \hat{\Pi}^{\nu} (u) \right] = -i \eta ^{\mu \nu} .  \label{canonical} 
\end{eqnarray}

The commutation relations between $\hat{x}^{\alpha} (u)$ and $\hat{x}^{\beta} (0)$ or $\hat{x}^{\beta} (s)$ are derived by Eqs.~(\ref{xt}) and (\ref{commutationrelationxx}) as
\begin{eqnarray}
& & \left[ \hat{x}^{\alpha} (u), \hat{x}^{\beta} (0) \right] \nonumber \\
& & = 2i u \eta ^{\alpha \beta} + 2ie u^2 f^{\alpha \beta} + \frac{4}{3} i n^{\alpha} n^{\beta} e^2 f^2 u^3 + i \Omega e f^{\alpha \beta} \left[ \frac{2}{3} \frac{u^3}{s} \xi (s) + \left( 2 u^2 - \frac{2}{3} \frac{u^3}{s} \right) \xi (0) \right] \nonumber \\
 & & \ \ \ + i \Omega e n^{\beta} {f^{\alpha}}_{\nu} [ \hat{x}^{\nu} (s) - \hat{x}^{\nu} (0) ] \left( \frac{2}{3} \frac{u^3}{s} \right) \nonumber \\
 & & \ \ \ + i \Omega n^{\alpha} n^{\beta} e^2 f^2 \left[ \left( - \frac{2}{3} u^3  + 2 \frac{u^4}{s} \right) \xi (s) + \left( \frac{10}{3} u^3 - 2 \frac{u^4}{s} \right) \xi (0) \right] , 
\end{eqnarray}
\begin{eqnarray}
& & \left[ \hat{x}^{\alpha} (s) , \hat{x}^{\beta} (u) \right] \nonumber \\
&=& 2i (s-u ) \eta ^{\alpha \beta} + 2ie ( s^2 - 2su + u^2 ) f^{\alpha \beta} + i n^{\alpha} n^{\beta} e^2 f^2 \left( \frac{4}{3} s^3 - 4 s^2 u + 4 s u^2 - \frac{4}{3} u^3 \right) \nonumber \\
& & + i \Omega e f^{\alpha \beta} \left[ \left( \frac{4}{3} s^2 - 2su + \frac{2}{3} \frac{u^3}{s} \right) \xi (s) + \left( \frac{2}{3} s^2 - 2 su + 2 u^2 - \frac{2}{3} \frac{u^3}{s} \right) \xi (0) \right] \nonumber \\
& & + i \Omega e n^{\alpha} {f^{\beta}}_{\nu} [ \hat{x}^{\nu} (s) - \hat{x}^{\nu} (0) ] \left( \frac{2}{3} s^2 -2 su + 2u^2 - \frac{2}{3} \frac{u^3}{s} \right) \nonumber \\
& & + i \Omega n^{\alpha} n^{\beta} e^2 f^2 \left[ \left( \frac{4}{3} s^3 - 2 s^2 u - 2 s u^2 + \frac{14}{3} u^3 - 2 \frac{u^4}{s} \right) \xi (s) \right. \nonumber \\
& & \hspace{0.5cm} \left.  + \left( \frac{4}{3} s^3 - 6 s^2 u + 10 s u^2 - \frac{22}{3} u^3 + 2 \frac{u^4}{s} \right) \xi (0) \right].
\end{eqnarray}

\section{$x$-dependence of Transformation Amplitudes} \label{renormalization}

Here we discuss the $x$-dependence of the results.
Note that the calculations of the amplitudes in Eqs.~(\ref{amplitudepis})
 - (\ref{amplitudepitexppi0}) are calculated of $x^{\mu}$
in the neighborhood of each point under the assumption
that the wavelength of the external wave field is
much longer than the Compton wavelength of the electron.
Then $x$ appears explicitly only in the form of $\xi = n_{\mu} x^{\mu}$
and it turns out in addition that $\xi$ occurs only as a combination of
$f(0) (1 + \Omega \xi)$. For example, the amplitude in
Eq.~(\ref{amplitudeexp}) is written as 
\begin{eqnarray}
& & \langle x(s)| \exp \left[ -i k_{\mu} \hat{x}^{\mu} (u) \right] |x (0) \rangle \nonumber \\
& & \simeq \langle x (s)| x (0) \rangle \times \exp \left( -i k_{\mu} x^{\mu} \right)  \exp \left[ i (k )^2 \left( u- \frac{u^2}{s} \right) \right] \nonumber \\
& &  \times \exp \left[  i ( k \cdot n)^2 e^2 f^2 (0) (1 +\Omega \xi)^2 \left( - \frac{1}{3} su^2 + \frac{2}{3} u^3 - \frac{1}{3} \frac{u^4}{s} \right) \right] \nonumber \\
& & \times \biggl\{ 1 + \Omega \left[ i ( k \cdot n )^3 e^2 f^2 (0) \left( - \frac{2}{3} s^3 u + \frac{14}{9} s^2 u^2 - \frac{4}{9} s u^3 - \frac{10}{9} u^4 + \frac{2}{3} \frac{u^5}{s} \right) \right. \nonumber \\
& & \left. \left. + ie ( \sigma f(0) ) ( k \cdot n) \left( \frac{1}{2} su - \frac{1}{2} u^2 \right) \right] \right\} 
\end{eqnarray}
and that in Eq.~(\ref{amplitudepitexp}) is given as
\begin{eqnarray}
& & \langle x(s)| \hat{\Pi}^{\alpha} (u) \exp \left[ -i k_{\mu} \hat{x}^{\mu} (u) \right] |x (0)\rangle  \nonumber \\
& & \simeq \langle x(s) | \exp \left[ -i k_{\mu} \hat{x}^{\mu} (u) \right] | x (0) \rangle  \nonumber \\
& & \ \ \ \times \left[ \frac{u}{s} k^{\alpha} + e f ^{\alpha \beta} (0)  k_{\beta} (1 + \Omega \xi)   \left( \frac{u^2}{s} - u \right) \right. + n^{\alpha} ( k \cdot n ) e^2 f^2 (0) ( 1 + \Omega \xi )^2 \left( \frac{1}{3} su - u^2 + \frac{2}{3} \frac{u^3}{s} \right) \nonumber \\
& & \ \ \ + \Omega e f^{\alpha \beta} (0) k_{\beta} ( k \cdot n ) \left( \frac{4}{3} u^2 - \frac{8}{3} \frac{u^3}{s} + \frac{4}{3} \frac{u^4}{s^2} \right) + \Omega n^{\alpha} ( k \cdot n)^2 e^2 f^2 (0) \left( - su^2 + 4 u^3 - 5 \frac{u^4}{s} + 2 \frac{u^5}{s^2} \right) \nonumber \\
& & \ \ \ \left. + \Omega e ( \sigma f (0)  ) n^{\alpha} \left( - \frac{1}{4} s + \frac{1}{2} u \right) \right] .
\end{eqnarray}
Note that the terms proportional to $\Omega$ in these equations
are of higher order and that $f(0)$ in these terms can be replaced with
$f(0) (1 + \Omega \xi)$. Considering $f(0)(1 + \Omega \xi) \approx f(x)$
in the same approximation,
we may conclude that all the explicit $x$-dependence can be included
in the amplitude of the external field and hence that 
the current term depends on the field strength $f$ and its gradient $\Omega$
at each point.
We can then assume that $x^{\mu} = 0$ at any points and
the terms that contain $\xi$ disappear in our results.

\section{Furry's Theorem in Proper-Time Method} \label{furry}
It is well known as Furry's theorem in QED that
all loop diagrams with an odd number of vertices vanish.
The same reasoning applies to our theory and we find that the terms in the induced electromagnetic
current that include
odd numbers of the external electromagnetic 
fields should be dropped in our case.
To understand this, we consider the charge conjugation
of the electron propagator with the external electromagnetic 
fields. 
\begin{figure}[htbp]
\begin{center}
   \includegraphics[width=70mm,clip]{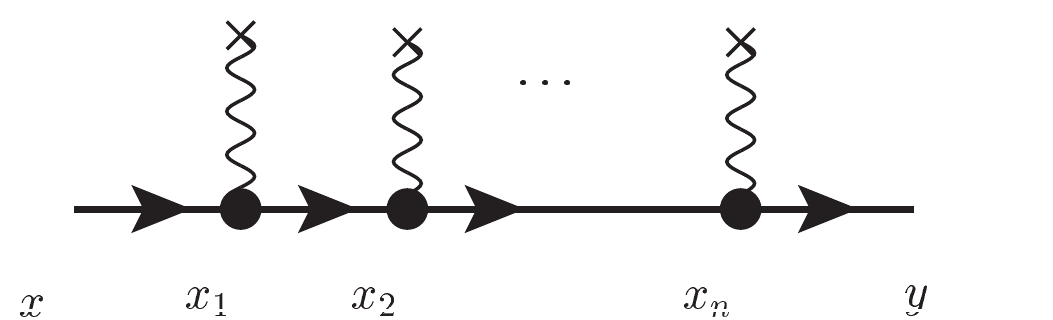}
  \end{center}
  \caption{Electron propagator with external electromagnetic fields is shown.}
  \label{1external}
\end{figure}

The propagator with $n$ external fields $S_{n, A}$ is represented as 
\begin{eqnarray}
& & S_{n, A} (y-x) = S (y - x_1 ) [ - e \gamma ^{\mu} A_{\mu} (x_1) ] S (x_1 - x_2 )  \cdots [ - e \gamma ^{\mu} A_{\mu} (x_n) ] S (x_n - x ) , 
\end{eqnarray}
where $S (y-x)$ is the electron free propagator.
Because the charge conjugation of the free propagator is
\begin{eqnarray}
S^c (y-x) &=& \mathcal{C} S ( y- x ) \mathcal{C} ^{\dagger} \nonumber \\
&=& C S^{T} (x - y) C^{-1} ,
\end{eqnarray}
where $C$ is the matrix, which is $C = i \gamma ^2 \gamma ^0$ for the Dirac representation
and the charge conjugation of the electromagnetic field $A_{\mu}$ is
\begin{eqnarray}
 A_{\mu} ^c  = \mathcal{C} A_{\mu} \mathcal{C} ^{\dagger} =  -A_{\mu} ,
\end{eqnarray} 
the charge conjugation of $S_{n,A} (y-x)$ is 
\begin{eqnarray}
S_{n, A} ^c ( y-x) &=& \mathcal{C} S_{n, A} ( y- x ) \mathcal{C} ^{\dagger} \nonumber \\
&=& \mathcal{C} S ( y- x_1 ) \mathcal{C} ^{\dagger} \mathcal{C} [ -e \gamma ^{\mu} A_{\mu} ( x_1 ) ] \mathcal{C}^{\dagger} \cdots \mathcal{C} ^{\dagger} \mathcal{C} S ( x_n - x ) \mathcal{C} ^{\dagger} \nonumber \\
&=& C S^{T} (x_1 - y) C^{-1} \left\{ -e \gamma ^{\mu} [ - A_{\mu} (x_1 ) ] \right\} C  \cdots C S^{T} ( x - x_n ) \mathcal{C} ^{\dagger} C^{-1} \nonumber \\
&=& C S^{T} ( x_1 - y ) ( -e ) ( - \gamma ^{\mu T} ) [ - A_{\mu} ( x _1 ) ]  \cdots S^{T} ( x - x_n ) C^{-1} \nonumber \\
&=& C \left\{ S ( x- x_n ) [ - e \gamma ^{\mu} A_{\mu} ( x_1 ) ] \cdots  S ( x_1 - y ) \right\} ^T C^{-1} \nonumber \\
&=& C S_{A} ^T ( x- y ) C^{-1} .  \label{procc1}
\end{eqnarray}
Another expression of charge conjugation is
\begin{eqnarray}
S_{n, A} ^c ( y-x ) &=& \mathcal{C} S_{n, A} ( y-x ) \mathcal{C} ^{\dagger} \nonumber \\
&=& \mathcal{C} S ( y- x_1 ) \mathcal{C} ^{\dagger} \mathcal{C} ( -e \gamma ^{\mu} A_{\mu} ( x_1 ) ) \cdots \mathcal{C}^{\dagger} \mathcal{C} S ( x_n - x ) \mathcal{C} ^{\dagger} \nonumber \\
&=& S (y - x_1 ) ( -e \gamma ^{\mu} ) ( - A_{\mu} (x_1 ) )  \cdots S ( x_n - x ) \nonumber \\
&=& S _{n, -A} ( y-x) \label{procc2}
\end{eqnarray}
because of the property of the electron propagator that it does not change
by the charge conjugation $\mathcal{C} S \mathcal{C}^{\dagger} = S$. 
From Eq.~(\ref{procc2}), we conclude that the sign of the propagator changes when the propagator contains 
odd numbers of electromagnetic fields
\begin{eqnarray}
S_{\mathrm{odd} , A} ^c ( y-x) = - S_{ \mathrm{odd} , A} ( y-x )
\end{eqnarray}
and that the sign of the propagator does not change when the propagator contains 
even numbers of electromagnetic fields
\begin{eqnarray}
S_{\mathrm{even} , A} ^c ( y-x) = S_{ \mathrm{even} , A} ( y-x ) . 
\end{eqnarray}

As shown in Eq.~(\ref{currentterm}), the induced electromagnetic current
is represented by the propagator
\begin{eqnarray}
j^{\mu} (x) = \frac{\partial \mathcal{L} [A,a] (x)}{\partial a_{\mu}} = ie \mathrm{tr} \left[ \gamma ^{\mu} G (x,x) \right] . 
\end{eqnarray}
There are two ways to obtain the charge conjugation of the propagator. One is to extract the matrix $C$
\begin{eqnarray}
j^{c \mu} _{A} (x) &=& ie \mathrm{tr} \left[ \gamma ^{\mu} G^c _A (x,x) \right] = ie \mathrm{tr} \left[ \gamma ^{\mu} C G^T _A (x,x) C^{-1} \right] \nonumber \\
&=& ie \mathrm{tr} \left[ - \gamma ^{\mu T} G^T _A (x,x)  \right] = -  ie \mathrm{tr} \left[ \gamma ^{\mu T} G^T _A (x,x)  \right] \nonumber \\
&=& -ie \mathrm{tr} \left[ \gamma ^{\mu} G_A (x,x) \right] = - j^{\mu} _{A} (x) . 
\end{eqnarray}
The other is to change the sign of the electromagnetic field 
\begin{eqnarray}
j^{c \mu} _{A} (x) &=& ie \mathrm{tr} \left[ \gamma ^{\mu} G^c _A (x,x) \right] = ie \mathrm{tr} \left[ \gamma ^{\mu} G _{-A} (x,x) \right] \nonumber \\
&=& j^{\mu} _{-A} (x) . 
\end{eqnarray}
Comparing these two expression, we obtain 
\begin{eqnarray}
- j^{\mu} _A (x) = j^{\mu} _{-A} (x) . 
\end{eqnarray}
Thus, the induced electromagnetic current should contain only those terms with
odd numbers of external electromagnetic fields.
Since it is represented as $\langle j_{\mu} \rangle = {\Pi _{\mu}}^{\nu} b_{\nu}$ with the probe photon $b_\nu$ and the polarization tensor ${\Pi _{\mu}}^{\nu}$, the number of the external fields in ${\Pi_{\mu}}^{\nu}$ should be even.

\section{Expression of the Induced Electromagnetic Current} \label{current_expression}

The induced electromagnetic current $\langle j^{\mu} \rangle$ as given in Eq.~(\ref{current}) is
given as follows:
\begin{eqnarray}
& &  \langle j^{\mu} \rangle \simeq \frac{e}{2} \int ^{\infty} _0 ds \int ^s _0 du \ e^{-im^2 s} \biggl[  A^{\mu} + B^{\mu} +  C^{\mu} +  D^{\mu} + E^{\mu} +  F^{\mu} + G^{\mu} + H^{\mu} \biggr] , \label{currentlong}
\end{eqnarray}
in which the terms $A^{\mu}, B^{\mu}, \cdots , H^{\mu}$ are expressed as follows:
\begin{eqnarray}
& & A^{\mu}  = \mathrm{tr} \left[ i e b_{\alpha} k^{\alpha}  \langle  x(s)|  \hat{\Pi}^{\mu} (s) \exp \left[ -i k_{\delta} \hat{x}^{\delta} (u) \right] +  \exp \left[ -i k_\delta \hat{x}^{\delta} (u) \right] \hat{\Pi}^{\mu} (0) |x (0) \rangle   \right] , \\
& & B^{\mu} =  \mathrm{tr} \left[  \left( - \frac{ie}{2} \right)  \langle x(s) | \hat{\Pi}^{\mu} (s) \exp \left[ -i k_{\delta} \hat{x}^{\delta} (u) \right] + \exp \left[ -i k_{\delta} \hat{x}^{\delta} (u) \right] \hat{\Pi}^{\mu} (0) |x (0) \rangle ( \sigma g)   \right] , \\
& & C^{\mu} = \mathrm{tr} \left[ ( -2ie ) b_{\alpha}  \langle x(s)| \hat{\Pi}^{\mu} (s) \hat{\Pi}^{\alpha} (u) \exp \left[ -i k_{\delta} \hat{x}^{\delta} (u) \right] + \hat{\Pi}^{\alpha} (u)  \exp \left[ -i k_{\delta} \hat{x}^{\delta} (u) \right] \hat{\Pi}^{\mu} (0) |x (0) \rangle   \right] , \\
& & D^{\mu} =  \mathrm{tr} \left[ ( -2e b_{\alpha} ) \sigma ^{\mu \nu}  \langle x(s)| \hat{\Pi}_{\nu} (s) \hat{\Pi}^{\alpha} (u) \exp \left[ -i k_{\delta} \hat{x}^{\delta} (u) \right] - \hat{\Pi}^{\alpha} (u) \exp \left[ -i k_{\delta} \hat{x}^{\delta} (u)  \right] \hat{\Pi}_{\nu} (0)  |x (0) \rangle \right] , \hspace{0.8cm} \\
& & E^{\mu} = \mathrm{tr} \biggl[ \left( - \frac{e}{2} \right) \sigma ^{\mu \nu} \langle x(s)| \hat{\Pi}_{\nu} (s)  \exp \left[ -i k_{\delta} \hat{x}^{\delta} (u) \right] - \exp \left[ -i k_{\delta} \hat{x}^{\delta} (u) \right] \hat{\Pi}_{\nu} (0) |x (0) \rangle  \nonumber \\
& & \hspace{3cm} \times \left.  \left\{ ( \sigma g ) +  \frac{e^2 u^2}{4} ( \sigma f )( \sigma g)( \sigma f)  \right\}   \right] , \\
& & F^{\mu} = \mathrm{tr} \left[ \left( - \frac{ie^2 u}{4} \right) \sigma ^{\mu \nu}  \langle  x(s)| \hat{\Pi}_{\nu} (s) \exp \left[ -i k_{\delta} \hat{x}^{\delta} (u) \right] - \exp \left[ -i k_{\delta} \hat{x}^{\delta} (u) \right] \hat{\Pi}_{\nu} (0) |x (0) \rangle \right.  \nonumber \\
& &  \hspace{3cm} \times  \left\{ ( \sigma f ) ( \sigma g) - ( \sigma g) ( \sigma f) \right\}  \biggr] , \\
& & G^{\mu} = \mathrm{tr} \left[ \left( \frac{\Omega e^2 u}{4} \right) \sigma ^{\mu \nu} n_{\nu}  \langle x(s)| \exp \left[ -i k_{\delta} \hat{x}^{\delta} (u) \right] |x (0) \rangle  \left\{ ( \sigma f )( \sigma g) - ( \sigma g)( \sigma f) \right\}  \right] , \\
& & H^{\mu} =  \mathrm{tr} \left[ \left( - \frac{i \Omega e^3 u^2}{4}  \right) \sigma ^{\mu \nu} n_{\nu}  \langle x(s)| \exp \left[ -i k_{\delta} \hat{x}^{\delta} (u) \right] |x (0) \rangle   ( \sigma f )( \sigma g )( \sigma f ) \right] .
\end{eqnarray}
They are further decomposed: e.g., $A^{\mu}$ is written as the sum of $A_i ^{\mu}$ as $A^{\mu} = \sum _{i=1} ^{5} A_i ^{\mu}$. The same notation is used for $B^{\mu}, C^{\mu}, \cdots , F^{\mu}$. All these components are explicitly written as follows:
\begin{eqnarray}
 A_1 ^{\mu} &=&   K (u) ie ( b \cdot k) k^{\mu} \left( 2 \frac{u}{s} -1 \right) - c.t. , \\
 A_2 ^{\mu} &=&  K (u) ie^3 f^2 ( b \cdot k) (k \cdot n) n^{\mu} \left( \frac{2}{3} su - 2 u^2 + \frac{4}{3} \frac{u^3}{s} \right) , \\
 A_3 ^{\mu} &=&  K (u) i \Omega e^3 f^2 ( b \cdot k) ( k \cdot n)^2 n^{\mu} \left( - \frac{4}{3} su^2 + \frac{16}{3} u^3 - \frac{20}{3} \frac{u^4}{s} + \frac{8}{3} \frac{u^5}{s^2} \right) , \\
 A_4 ^{\mu} &=&  K(u) \Omega e^3 f^2 ( b \cdot k) ( k \cdot n)^3 k^{\mu} \left( - \frac{2}{3} s^3 u + \frac{26}{9} s^2 u^2 - \frac{32}{9} s u^3  - \frac{2}{9} u^4 + \frac{26}{9} \frac{u^5}{s} - \frac{4}{3} \frac{u^6}{s^2} \right) ,  \hspace{1.2cm} \\
 A_5 ^{\mu} &=& K(u) \Omega e^5 f^4 ( b \cdot k ) ( k \cdot n)^4 n^{\mu} \left( \frac{4}{9} s^4 u^2 - \frac{64}{27} s^3 u^3 + \frac{116}{27} s^2 u^4 - \frac{20}{9} s u^5  - \frac{56}{27} u^6 + \frac{76}{27} \frac{u^7}{s} - \frac{8}{9} \frac{u^8}{s^2} \right) , \hspace{1.2cm} \\
B_1 ^{\mu} &=&   \mathrm{tr} \left[ ( \sigma f) ( \sigma g ) \right] L(u) e^3 f^{\mu \nu} k_{\nu} \left( - \frac{su}{2} + \frac{u^2}{2} \right) , \\
B_2 ^{\mu} &=&   \mathrm{tr} \left[ ( \sigma f) ( \sigma g ) \right] L (u) i \Omega e^5 f^2 (k \cdot n)^3 f^{\mu \nu} k_{\nu} \left( \frac{1}{3} s^4 u^2 - \frac{10}{9} s^3 u^3 + s^2 u^4 + \frac{1}{3} s u^5 - \frac{8}{9} u^6 + \frac{1}{3} \frac{u^7}{s} \right) , \\
 B_3 ^{\mu} &=&  \mathrm{tr} \left[ ( \sigma f) ( \sigma g ) \right] L (u) \Omega e^3 ( k \cdot n) f^{\mu \nu} k_{\nu} \left( \frac{1}{2} s u^2 - u^3 + \frac{1}{2} \frac{u^4}{s} \right) , \\
 B_4 ^{\mu} &=&  \mathrm{tr} \left[ ( \sigma f) ( \sigma g ) \right] L (u) \Omega e^3 ( k \cdot n) f^{\mu \nu} k_{\nu} \left( \frac{1}{3} s u^2 - \frac{2}{3} u^3 + \frac{1}{3} \frac{u^4}{s} \right) , 
\end{eqnarray}
\begin{eqnarray}
  C_1 ^{\mu} &=& K(u) i e ( b \cdot k) k^{\mu} \left( 2 \frac{u}{s} - 4 \frac{u^2}{s^2} \right) - c.t. , \\
 C_2 ^{\mu} &=& K(u) i e^3 ( bfk ) f^{\mu \nu} k_{\nu} \left( 4 u^2 - 8 \frac{u^3}{s} + 4 \frac{u^4}{s^2} \right) ,  \\
 C_3 ^{\mu} &=& K(u) i e^3 f^2 ( b \cdot n) ( k \cdot n) k^{\mu} \left( \frac{2}{3} su - \frac{10}{3} u^2 + \frac{16}{3} \frac{u^3}{s} - \frac{8}{3} \frac{u^4}{s^2} \right) , \\
 C_4 ^{\mu} &=& K (u) i e^3 f^2 ( b \cdot k) ( k \cdot n)  n^{\mu} \left( - \frac{4}{3} u^2 + 4 \frac{u^3}{s} - \frac{8}{3} \frac{u^4}{s^2} \right) , \\
 C_5 ^{\mu} &=& K (u) e^3 f^2 ( b \cdot n) n^{\mu} \left( \frac{2}{3} s - 4 u + 4 \frac{u^2}{s} \right) , \\
 C_6 ^{\mu} &=& K(u) i e^5 f^4 ( b \cdot n) ( k \cdot n)^2 n^{\mu} \left( - \frac{4}{9} s^2 u^2 + \frac{8}{3} s u^3 - \frac{52}{9} u^4 + \frac{16}{3} \frac{u^5}{s} - \frac{16}{9} \frac{u^6}{s^2}\right) , \\
 C_7 ^{\mu} &=& K (u) e b^{\mu} \frac{2}{s} - c.t. , \\
 C_8 ^{\mu} &=& K(u) \Omega e^3 f^2 ( b \cdot k) ( k \cdot n)^3 k^{\mu} \left( \frac{4}{3} s^2 u^2 - \frac{52}{9} s u^3 + \frac{64}{9} u^4 + \frac{4}{9} \frac{u^5}{s} - \frac{52}{9} \frac{u^6}{s^2} + \frac{8}{3} \frac{u^7}{s^3} \right) , \\
 C_9 ^{\mu} &=& K (u) \Omega e^5 f^2 ( bfk ) ( k \cdot n)^3 f^{\mu \nu} k_{\nu} \nonumber \\
 & & \ \ \  \times \left( \frac{8}{3} s^3 u^3 - \frac{104}{9} s^2 u^4 + \frac{152}{9} s u^5 - \frac{16}{3} u^6 - \frac{88}{9} \frac{u^7}{s} + \frac{88}{9} \frac{u^8}{s^2} - \frac{8}{3} \frac{u^9}{s^3} \right) , \\
 C_{10} ^{\mu} &=& K (u) \Omega e^5 f^4 ( b \cdot n) ( k \cdot n)^4 k^{\mu} \nonumber \\
 & & \ \ \ \times \left( \frac{4}{9} s^4 u^2 - \frac{88}{27} s^3 u^3 + \frac{244}{27} s^2 u^4 - \frac{292}{27} s u^5  + \frac{64}{27} u^6 + \frac{188}{27} \frac{u^7}{s} - \frac{176}{27} \frac{u^8}{s^2} + \frac{16}{9} \frac{u^9}{s^3} \right) ,  \hspace{1.2cm} \\
 C_{11} ^{\mu} &=& K (u) \Omega e^5 f^4 ( b \cdot k) ( k \cdot n)^4 n^{\mu}  \nonumber \\
 & & \times \left( - \frac{8}{9} s^3 u^3 + \frac{128}{27} s^2 u^4 - \frac{232}{27} s u^5 + \frac{40}{9} u^6  + \frac{112}{27} \frac{u^7}{s} - \frac{152}{27} \frac{u^8}{s^2} + \frac{16}{9} \frac{u^9}{s^3} \right) , \\
 C_{12} ^{\mu} &=& K(u) i \Omega e^5 f^4 ( b \cdot n) ( k \cdot n)^3 n^{\mu} \nonumber \\
 & & \times \left( - \frac{4}{9} s^4 u + \frac{100}{27} s^3 u^2 - \frac{248}{27} s^2 u^3 + \frac{196}{27} s u^4 + \frac{28}{9} u^5 - \frac{64}{9} \frac{u^6}{s} + \frac{8}{3} \frac{u^7}{s^2} \right) , \\
 C_{13} ^{\mu} &=& K(u) \Omega e^7 f^6 ( b \cdot n) ( k \cdot n)^5 n^{\mu} \left( - \frac{8}{27} s^5 u^3 + \frac{200}{81} s^4 u^4 - \frac{664}{81} s^3 u^5 \right. \nonumber \\
 & & \ \ \ \left. + \frac{1072}{81} s^2 u^6 - \frac{712}{81} s u^7 - \frac{248}{81} u^8 + \frac{728}{81} \frac{u^9}{s} - \frac{448}{81} \frac{u^{10}}{s^2} + \frac{32}{27} \frac{u^{11}}{s^3} \right) , \\
 C_{14} ^{\mu} &=& K(u) i \Omega e^3 f^2 ( k \cdot n)^3 b^{\mu} \left( - \frac{4}{3} s^2 u + \frac{28}{9} s u^2 - \frac{8}{9} u^3 - \frac{20}{9} \frac{u^4}{s} + \frac{4}{3} \frac{u^5}{s^2} \right) , \\
 C_{15} ^{\mu} &=& K(u) i \Omega e^3 ( bfk ) ( k \cdot n ) f^{\mu \nu} k_{\nu}  \left( - 8 u^3  + 24 \frac{u^4}{s} -24 \frac{u^5}{s^2} +8 \frac{u^6}{s^3} \right)  , \\
C_{16} ^{\mu} &=& K(u) i \Omega e^3 f^2 ( b \cdot n) ( k \cdot n)^2 k^{\mu}  \left( -2 su^2 + 12 u^3 - 26 \frac{u^4}{s} + 24 \frac{u^5}{s^2} - 8 \frac{u^6}{s^3} \right)  , \\
 C_{17} ^{\mu} &=& K(u) i \Omega e^3 f^2 ( b \cdot k) ( k \cdot n)^2 n^{\mu} \left( \frac{8}{3} u^3 - \frac{32}{3} \frac{u^4}{s} + \frac{40}{3} \frac{u^5}{s^2} - \frac{16}{3} \frac{u^6}{s^3} \right) ,
  \end{eqnarray}
\begin{eqnarray}
 C_{18} ^{\mu} &=& K (u) \Omega e^3 f^2 ( b \cdot n) ( k \cdot n) n^{\mu} \left( - \frac{8}{3} su + 16 u^2 - \frac{80}{3} \frac{u^3}{s} + \frac{40}{3} \frac{u^4}{s^2} \right)  , \\
 C_{19} ^{\mu} &=& K (u) i \Omega e^5 f^4 ( b \cdot n) ( k \cdot n)^3 n^{\mu} \left( \frac{20}{9} s^2 u^3 - \frac{140}{9} su^4 + \frac{380}{9} u^5  - \frac{500}{9} \frac{u^6}{s} + \frac{320}{9} \frac{u^7}{s^2} - \frac{80}{9} \frac{u^8}{s^3} \right) , \\
 D_1 ^{\mu} &=&  \mathrm{tr} \left[ \sigma ^{\mu \nu} ( \sigma f ) \right] L(u) i e^3 ( bfk) k_{\nu} ( -su + u^2) , \\
 D_2 ^{\mu} &=&  \mathrm{tr} \left[ \sigma ^{\mu \nu} ( \sigma f ) \right] L(u) \Omega e^5 f^2 ( bfk) ( k \cdot n)^3 k_{\nu} \nonumber \\
 & & \ \ \ \times \left( - \frac{2}{3} s^4 u^2 + \frac{20}{9} s^3 u^3 - 2 s^2 u^4 - \frac{2}{3} s u^5 + \frac{16}{9} u^6 - \frac{2}{3} \frac{u^7}{s} \right) , \\
 D_3 ^{\mu} &=&  \mathrm{tr} \left[ \sigma ^{\mu \nu} ( \sigma f ) \right] L (u) \Omega e^3 ( k \cdot n) {f_{\nu}}^{\alpha} b_{\alpha} \left( \frac{1}{3} s^2 - su \right) , \\
 D_4 ^{\mu} &=&  \mathrm{tr} \left[ \sigma ^{\mu \nu} ( \sigma f ) \right] L(u) i \Omega e^3 ( bfk) ( k \cdot n) k_{\nu}  \left( \frac{4}{3} s u^2 - \frac{8}{3} u^3 + \frac{4}{3} \frac{u^4}{s} \right) , \\
 D_5 ^{\mu} &=&  \mathrm{tr} \left[ \sigma ^{\mu \nu} ( \sigma f ) \right] L(u) \Omega e^3 ( b \cdot n) {f_{\nu}}^{\lambda} k_{\lambda} \left( - \frac{1}{3} s^2 + su \right)  , \\
 D_6 ^{\mu} &=&  \mathrm{tr} \left[ \sigma ^{\mu \nu} ( \sigma f ) \right] L(u) \Omega e^3 ( bfk ) n_{\nu}  \left( \frac{1}{3} s^2 - u^2 \right) , \\
 D_7 ^{\mu} &=&  \mathrm{tr} \left[ \sigma ^{\mu \nu} ( \sigma f ) \right] L(u) i \Omega e^5 f^2 ( bfk ) ( k \cdot n)^2  n_{\nu} \left( - \frac{2}{3} s^2 u^3 + 2 s u^4 - 2 u^5 + \frac{2}{3} \frac{u^6}{s} \right) , \\
  D_8 ^{\mu} &=& \mathrm{tr} \left[ \sigma ^{\mu \nu} ( \sigma f ) \right] L(u) i \Omega e^3 ( bfk) ( k \cdot n) k_{\nu}  \left( su^2 - 2 u^3 + \frac{u^4}{s} \right) , \\
 D_9 ^{\mu} &=&  \mathrm{tr} \left[ \sigma ^{\mu \nu} ( \sigma f ) \right] L(u) \Omega e^3 ( bfk) n_{\nu} ( su - u^2 )  , \\
 E_1 ^{\mu} &=&  \mathrm{tr} \left[ \sigma ^{\mu \nu} \left(  ( \sigma g)  +  \frac{e^2 u^2}{4} ( \sigma f )( \sigma g)( \sigma f)  \right) \right] L(u)  \left( - \frac{e}{2} k_{\nu} \right) - c.t.  , \\
 E_2 ^{\mu} &=&  \mathrm{tr} \left[ \sigma ^{\mu \nu} \left( ( \sigma g) + \frac{e^2 u^2}{4} ( \sigma f) ( \sigma g) ( \sigma f) \right) \right] L(u) \Omega e^3 f^2 ( k \cdot n)^2 n_{\nu} \left( - \frac{1}{3} s u^2 + \frac{2}{3} u^3 - \frac{1}{3} \frac{u^4}{s} \right) , \hspace{1.2cm} \\
E_3 ^{\mu} &=&  \mathrm{tr} \left[ \sigma ^{\mu \nu} \left( ( \sigma g) + \frac{e^2 u^2}{4} ( \sigma f) ( \sigma g) ( \sigma f) \right) \right] L(u) i \Omega e^3 f^2 ( k \cdot n)^3 k_{\nu} \nonumber \\
& & \times \left( \frac{1}{3} s^3 u - \frac{7}{9} s^2 u^2 + \frac{2}{9} s u^3 + \frac{5}{9} u^4 - \frac{1}{3} \frac{u^5}{s} \right) , \\
 F_1 ^{\mu} &=&  \mathrm{tr} [ \sigma ^{\mu \nu} ( \sigma f ) ( \sigma g) ( \sigma f) ] L(u) e^3 k_{\nu} \frac{su}{8}  , \\
& & \nonumber \\
 F_2 ^{\mu} &=&  \mathrm{tr} [ \sigma ^{\mu \nu} ( \sigma f) ( \sigma g ) ( \sigma f )] L(u) i \Omega e^5 f^2 ( k \cdot n)^3 n_{\nu} \nonumber \\
& & \times \left( - \frac{1}{12} s^4 u^2 + \frac{7}{36} s^3 u^3 - \frac{1}{18} s^2 u^4 - \frac{5}{36} s u^5 + \frac{1}{12} u^6 \right) , \\
 F_3 ^{\mu} &=&  \mathrm{tr} [ \sigma ^{\mu \nu} ( \sigma f) ( \sigma g) ( \sigma f) ] L(u) \Omega e^5 f^2 ( k \cdot n)^2 k_{\nu} \left( \frac{1}{12} s^2 u^3 - \frac{1}{6} s u^4 + \frac{1}{12} u^5 \right) , \\
 F_4 ^{\mu} &=&  \mathrm{tr} [ \sigma ^{\mu \nu} ( \sigma f) ( \sigma g) ( \sigma f) ] L(u) \Omega e^3 ( k \cdot n) k_{\nu} \left( - \frac{1}{8} s u^2 + \frac{1}{8} u^3 \right)  , \\
 F_5 ^{\mu} &=&  \mathrm{tr} [ \sigma ^{\mu \nu} ( \sigma f) ( \sigma g) ( \sigma f) ] L(u) i \Omega e^3 n_{\nu} \frac{su}{8} , 
\end{eqnarray}
\begin{eqnarray}
G ^{\mu} &=& F_5 ^{\mu} , \\
H ^{\mu} &=&  \mathrm{tr} [ \sigma ^{\mu \nu} ( \sigma f) ( \sigma g) ( \sigma f) ] L(u) \left( - i \Omega e^3 n_{\nu}  \frac{u^2}{4} \right) ,  \label{fin} 
\end{eqnarray}
where we employ the following abbreviations: $k \cdot n = k_{\mu} n^{\mu}$ and $bfk = b_{\alpha} f^{\alpha \beta} k_{\beta}$. In the above equations, $K(u)$ and $L(u)$ are defined as
\begin{eqnarray}
K(u) &=& \frac{1}{4 i \pi ^2 s^2} \exp ( -ikx ) \exp \left\{ i ( k)^2 \left( u- \frac{u^2}{s} \right) + i ( k \cdot n )^2 e^2 f^2 \left( - \frac{1}{3} su^2 + \frac{2}{3} u^3 - \frac{1}{3} \frac{u^4}{s} \right) \right\} ,  \hspace{1.2cm} \label{termK} \\
L(u) &=& K(u) /4 ,
\end{eqnarray}
where $\exp (-ikx) = \exp ( -i k_{\mu} x^{\mu})$. The counter terms that originate from renormalization are denoted by $c.t.$ in some equations. For the crossed-field, i.e., the long wavelength limit ($\Omega \rightarrow 0$) of the external plane-wave, the above expression is reduced to
\begin{eqnarray}
 \langle j^{\mu} \rangle |_{\Omega =0 } &\simeq& \frac{e}{2} \int ^{\infty} _0 ds \int ^s _0 du \ e^{-im^2 s} \biggl[ \sum ^2 _{i=1} A_i ^{\mu} + B_1 ^{\mu} + \sum ^{7} _{i=1} C_i ^{\mu} +  D_1 ^{\mu} + E_1 ^{\mu} + F_1 ^{\mu} \biggr] . \label{current_crossed}
\end{eqnarray}


\begin{thebibliography}{10}

\bibitem{Bialynicki-Birula1983}
I.~{Bialynicki-Birula},
\newblock In B.~{Jancewicz} and J.~{Lukierski}, editors, {\it Quantum Theory Of Particles and Fields}, pages 31--48. World Scientific (1983).

\bibitem{1987PhRvL..59.1065A}
I.~{Affleck} and L.~{Kruglyak}, Phys. Rev. Lett. {\bf 59}, 1065 (1987).

\bibitem{mcgill}
S.~A. Olausen and V.~M. Kaspi, Astrophys.\ J. Suppl. Ser. {\bf 212}, 6 (2014).

\bibitem{2008A&ARv..15..225M}
S.~{Mereghetti}, Astron. Astrophys. Rev. {\bf 15}, 225 (2008).

\bibitem{2002PhRvD..66b3002H}
J.~S. {Heyl} and N.~J. {Shaviv}, Phys. Rev. D {\bf 66}, 023002 (2002).

\bibitem{2015MNRAS.454.3254T}
R.~{Taverna}, R.~{Turolla}, D.~{Gonzalez Caniulef}, S.~{Zane}, F.~{Muleri}, and
P.~{Soffitta}, Mon.\ Not.\ R.\ Astron.\ Soc. {\bf 454}, 3254 (2015).

\bibitem{2017MNRAS.465..492M}
R.~P. {Mignani}, V.~{Testa}, D.~{Gonz{\'a}lez Caniulef}, R.~{Taverna}, R.~{Turolla}, S.~{Zane}, and K.~{Wu}, Mon.\ Not.\ R.\ Astron.\ Soc. {\bf 465}, 492 (2017).

\bibitem{1979PhRvD..19.3565M}
P.~{M\'esz\'aros} and J.~{Ventura}, Phys. \ Rev. \ D {\bf 19}, 3565 (1979).

\bibitem{2003PhRvL..91g1101L}
D.~{Lai} and W.~C. {Ho}, Phys. \ Rev. \ Lett. {\bf 91}, 071101 (2003).

\bibitem{2017ApJ...850..185Y}
A.~{Yatabe} and S.~{Yamada}, Astrophys. J, {\bf 850}, 185, (2017).

\bibitem{hercules}
V.~Yanovsky, V.~Chvykov, G.~Kalinchenko, P.~Rousseau, T.~Planchon, T.~Matsuoka, A.~Maksimchuk, J.~Nees, G.~Cheriaux, G.~Mourou, and K.~Krushelnick, Optics Express {\bf 16}, 2109 (2008).

\bibitem{2006OptCo.267..318H}
T.~{Heinzl}, B.~{Liesfeld}, K.-U. {Amthor}, H.~{Schwoerer}, R.~{Sauerbrey}, and A.~{Wipf}, Opt. Commun. {\bf 267}, 318 (2006).

\bibitem{2014PhRvD..89l5003D}
V.~{Dinu}, T.~{Heinzl}, A.~{Ilderton}, M.~{Marklund}, and G.~{Torgrimsson}, Phys. \ Rev. \ D {\bf 89}, 125003 (2014).

\bibitem{2014PhRvD..90d5025D}
V.~{Dinu}, T.~{Heinzl}, A.~{Ilderton}, M.~{Marklund}, and G.~{Torgrimsson}, Phys. Rev. D {\bf 90}, 045025 (2014).

\bibitem{karbstein1}
F.~Karbstein and R.~Shaisultanov, Phys.\ Rev.\ D {\bf 91}, 085027 (2015).

\bibitem{KingHeinzl2016}
B.~{King} and T.~{Heinzl}, High Power Laser Science and Engineering {\bf 4}, e5 (2016).

\bibitem{1952PhDT........21T}
J.~S. {Toll}, \newblock {\it {The Dispersion Relation for Light and its Application to Problems Involving Electron Pairs.}}, \newblock PhD thesis, PRINCETON UNIVERSITY. (1952).

\bibitem{Baier1967a}
R.~{Baier} and P.~{Breitenlohner}, Acta Phys. Austriaca {\bf 25}, 212 (1967).

\bibitem{1971PhRvD...3..618B}
E.~{Brezin} and C.~{Itzykson}, Phys. \ Rev. \ D {\bf 3}, 618 (1971).

\bibitem{Adler71}
S.~L. Adler, Ann.\ Phys. {\bf 67}, 599 (1971).

\bibitem{tsaierber74}
W.~Tsai and T.~Erber, Phys.\ Rev.\ D {\bf 10}, 492 (1974).

\bibitem{tsaierber75}
W.~Tsai and T.~Erber, Phys.\ Rev.\ D {\bf 12}, 1132 (1975).

\bibitem{kohriyamada}
K.~Kohri and S.~Yamada, Phys.\ Rev.\ D {\bf 65}, 043006 (2002).

\bibitem{2007NuPhB.778..219S}
G.~M. {Shore}, Nucl. Phys. B {\bf 778}, 219--258 (2007).

\bibitem{hattoriitakura1}
K.~Hattori and K.~Itakura, Ann.\ Phys. {\bf 330}, 23 (2013).

\bibitem{hattoriitakura2}
K.~Hattori and K.~Itakura, Ann.\ Phys. {\bf 334}, 58 (2013).

\bibitem{kishikawa}
K.~Ishikawa, D.~Kimura, K.~Shigaki, and A.~Tsujii, Int.\ J.\ Mod.\ Phys.\ A {\bf 28}, 1350100 (2013).

\bibitem{karbstein2013}
F.~Karbstein, Phys.\ Rev.\ D {\bf 88}, 085033 (2013).

\bibitem{1970PhRvD...2.2341B}
Z.~{Bialynicka-Birula} and I.~{Bialynicki-Birula}, Phys. Rev. D {\bf 2}, 2341 (1970).

\bibitem{batalin_shabad_1971}
I.~A. {Batalin} and A.~E. {Shabad}, Sov. Phys. -JETP {\bf 33}, 483 (1971).

\bibitem{urrutia1978vacuum}
L.~F. Urrutia, Phys. Rev. D {\bf 17}, 1977 (1978).

\bibitem{artimovich_1990}
G.~K. {Artimovich}, Sov. Phys. -JETP {\bf 70}, 787 (1990).
	
\bibitem{DitGies}
W.~Dittrich and H.~Gies, \newblock {\it Probing the Quantum Vacuum}, \newblock Number 166 in Springer Tracts in Modern Physics.  (Springer, 2000).

\bibitem{2000NuPhB.585..407S}
C.~{Schubert}, Nucl. Phys. B {\bf 585}, 407 (2000).
	
\bibitem{Baier1967b}
R.~{Baier} and P.~{Breitenlohner}, Nuovo Cimento B {\bf 47}, 117 (1967).

\bibitem{narozhnyi69}
N.~B. Narozhny\u{\i}, Sov.\ Phys.\ -JETP {\bf 28}, 371 (1969).

\bibitem{ritus72}
V.~I. Ritus, Ann.\ Phys. {\bf 69}, 555 (1972).

\bibitem{heinzl}
T.~Heinzl and O.~Schr{\"{o}}der, J.\ Phys. A {\bf 39}, 11623 (2006).

\bibitem{1975JPhA....8.1638B}
W.~{Becker} and H.~{Mitter}, J. Phys. A {\bf 8}, 1638 (1975).

\bibitem{Mitter1975}
H.~{Mitter}, Acta Phys. Austriaca Suppl. {\bf 14}, 397 (1975).

\bibitem{baier75}
V.~N. Ba\u{\i}er, A.~I. Mil'shte\u{\i}n, and V.~M. Strakhovenko, Sov.\ Phys. -JETP {\bf 42}, 961 (1975).

\bibitem{1988JPhA...21..693A}
I.~{Affleck}, J. Phys. A {\bf 21}, 693 (1988).

\bibitem{meuren}
S.~Meuren, C.~H. Keitel, and A.~Di Piazza, Phys.\ Rev.\ D {\bf 88}, 013007 (2013).

\bibitem{schwinger51}
J.~Schwinger, Phys.\ Rev. {\bf 82}, 664 (1951).

\bibitem{kogut1970quantum}
J.~B. Kogut and D.~E. Soper, Phys.\ Rev. D {\bf 1}, 2901 (1970).
	
\bibitem{neville1971quantum}
R.~A. Neville and F.~Rohrlich, Phys. Rev. D {\bf 3}, 1692 (1971).

\bibitem{volkov1935class}
D.~M. Volkov, Z.\ Phys. {\bf 94}, 250 (1935).
	
\bibitem{gusynin1999derivative}
V.~P. Gusynin and I.~A. Shovkovy, Journal of Mathematical Physics {\bf 40}, 5406 (1999).

\bibitem{zel1967quasienergy}
Ya.~B. Zel'Dovich, Sov.\ Phys.\ -JETP {\bf 24}, 1006 (1967).
	
\bibitem{HH98}
J.~S. Heyl and L.~Hernquist, Phys.\ Rev.\ D {\bf 58}, 043005 (1998).

\bibitem{HH99}
J.~S. Heyl and L.~Hernquist, Phys.\ Rev.\ D {\bf 59}, 045005 (1999).

\bibitem{HH05}
J.~S. Heyl and L.~Hernquist, Astrophys. \ J. {\bf 618}, 463 (2005).

\end{thebibliography}
\end{document}